\documentclass[aps,prd,onecolumn,showpacs,nofootinbib,amsmath,amssymb,floatfix,superscriptaddress,showkeys]{revtex4}
\pdfoutput=1
\usepackage{bm}
\usepackage{graphicx}
\usepackage{amssymb,amsmath}
\usepackage{physics}
\usepackage{multirow}
\usepackage{units,changes}
\usepackage{color,url}
\usepackage{tabu}
\usepackage{array}
\usepackage[colorlinks=true,urlcolor=blue,anchorcolor=blue
,citecolor=blue,filecolor=blue,linkcolor=blue,menucolor=blue
,linktocpage=true,pdfproducer=medialab,pdfa=true]{hyperref}
\usepackage{colordvi}
\def\beqa{\begin{eqnarray}}
\def\eeqa{\end{eqnarray}}

\newcommand{\degree}{$^\circ$}
\usepackage{soul}
\usepackage{hhline} 

\newcommand{\mev}{\ensuremath{\,\mathrm{MeV}}}
\newcommand{\gev}{\ensuremath{\,\mathrm{GeV}}}
\newcommand{\tev}{\ensuremath{\,\mathrm{TeV}}}
\newcommand{\pc}{\ensuremath{\,\mathrm{pc}}}
\newcommand{\kpc}{\ensuremath{\,\mathrm{kpc}}}

\newcommand{\fermi}{Fermi}

\newcommand{\sv}{\ensuremath{\langle\sigma v_{\rm rel.}\rangle}}
\newcommand{\dnde}{\ensuremath{\frac{dN_\gamma}{dE_\gamma}}} 
\newcommand{\avgdnde}{\ensuremath{\langle dN_\gamma/dE_\gamma\rangle[r]}}

\begin{document}

\title{Searching Accretion-Enhanced Dark Matter Annihilation Signals in the Galactic Centre}
\def\slash#1{#1\!\!\!/}

\author{Meiwen Yang}
\affiliation{Department of Physics and Institute of Theoretical Physics, Nanjing Normal University, Nanjing, 210023, China}
\affiliation{Key Laboratory of Dark Matter and Space Astronomy, Purple Mountain Observatory, Chinese Academy of Sciences, Nanjing 210023, China}

\author{Zhi-Qi Guo}
\affiliation{Department of Physics and Institute of Theoretical Physics, Nanjing Normal University, Nanjing, 210023, China}
\affiliation{Key Laboratory of Dark Matter and Space Astronomy, Purple Mountain Observatory, Chinese Academy of Sciences, Nanjing 210023, China}

\author{Xiao-Yi Luo}
\affiliation{Department of Physics and Institute of Theoretical Physics, Nanjing Normal University, Nanjing, 210023, China}

\author{Zhao-Qiang Shen}\email{zqshen@pmo.ac.cn}
\affiliation{Key Laboratory of Dark Matter and Space Astronomy, Purple Mountain Observatory, Chinese Academy of Sciences, Nanjing 210023, China}

\author{\\Zi-Qing Xia}
\affiliation{Key Laboratory of Dark Matter and Space Astronomy, Purple Mountain Observatory, Chinese Academy of Sciences, Nanjing 210023, China}

\author{Chih-Ting Lu}\email{ctlu@njnu.edu.cn}
\affiliation{Department of Physics and Institute of Theoretical Physics, Nanjing Normal University, Nanjing, 210023, China}

\author{Yue-Lin Sming Tsai}\email{smingtsai@pmo.ac.cn}
\affiliation{Key Laboratory of Dark Matter and Space Astronomy, Purple Mountain Observatory, Chinese Academy of Sciences, Nanjing 210023, China}
\affiliation{School of Astronomy and Space Science, University of Science and Technology of China, Hefei 230026, China}

\author{Yi-Zhong Fan}
\affiliation{Key Laboratory of Dark Matter and Space Astronomy, Purple Mountain Observatory, Chinese Academy of Sciences, Nanjing 210023, China}
\affiliation{School of Astronomy and Space Science, University of Science and Technology of China, Hefei 230026, China}

\date{\today}

\begin{abstract}
This study reanalyzes the detection prospects of dark matter (DM) annihilation signals in the Galactic Center, 
focusing on velocity-dependent dynamics within a spike density near the supermassive black hole (Sgr~A$^{\star}$). 
We investigate three annihilation processes---$p$-wave, resonance, and forbidden annihilation---under semi-relativistic velocities, leveraging gamma-ray data from Fermi and DAMPE telescopes. 
Our analysis integrates a fermionic DM model with an electroweak axion-like particle (ALP) portal, exploring annihilation into two or four photons. 
Employing a comprehensive six-dimensional integration, we precisely calculate DM-induced gamma-ray fluxes near Sgr~A$^{\star}$, incorporating velocity and positional dependencies in the annihilation cross-section and photon yield spectra. 
Our findings highlight scenarios of resonance and forbidden annihilation, where the larger ALP-DM-DM coupling constant $C_{a\chi\chi}$ can affect spike density, potentially yielding detectable gamma-ray line spectra within Fermi and DAMPE energy resolution. 
We set upper limits for $C_{a\chi\chi}$ across these scenarios, offering insights into the detectability and spectral characteristics of DM annihilation signals from the Galactic Center.
\end{abstract}

\maketitle

\section{Introduction}

Dark matter (DM) is inferred from gravitational effects, yet understanding its non-gravitational interactions remains a fundamental challenge in modern physics. 
Investigating interactions between the dark sector and the standard model (SM) sector is crucial for unraveling its nature. 
The weakly interacting massive particle, a potential candidate for DM, is expected to yield detectable signals in DM experiments. 
However, recent results from collider experiments~\cite{ATLAS:2021ldb,CMS:2017dmg,BaBar:2017tiz}, DM direct detection experiments~\cite{XENON:2023cxc,LZ:2022lsv,PandaX-4T:2021bab}, and DM indirect detection experiments~\cite{Foster:2022nva,Albert:2014hwa,MAGIC:2022acl}, have not definitively detected it.
Fortunately, measuring the density of the DM relics from cosmic microwave background radiation sets constraints on DM interaction rates, crucially limiting the DM annihilation rate required to match the relic density reported by Planck~\cite{Planck:2018vyg}. 
The simplest DM particle models, featuring only one DM particle and one mediator particle, often fail to reproduce correct relic densities with velocity-independent ($s$-wave) DM annihilation cross-sections, 
thus prompting interest in velocity-dependent DM annihilation cross-sections like new mediator resonance~\cite{Ibe:2008ye, Guo:2009aj,Croon:2020ntf,Ding:2021sbj,Binder:2022pmf,Belanger:2024bro}, $p$-wave secluded DM~\cite{Diamanti:2013bia,Bondarenko:2019vrb,Siegert:2024hmr} and forbidden DM~\cite{Griest:1990kh,DAgnolo:2015ujb,Delgado:2016umt,DAgnolo:2020mpt,Hara:2021lrj,Wojcik:2021xki}.

Velocity-dependently annihilating DM particles, while currently evading stringent limits, are experimentally challenging to detect due to the velocity suppression. 
The environment around supermassive black hole (SMBH) at galactic center (GC) provides an ideal region for detecting such signals. 
The gigantic gravitational potential of SMBH, coupled with their accretion, accumulates significant numbers of DM particles, forming a spike~\cite{Gondolo:1999ef}. 
These DM particles within the spike move at semi-relativistic speeds, around half the speed of light ($v \approx 0.5c$), significantly enhancing their annihilation cross-sections and detection rates. 
In general, near SMBH, $p$-wave annihilation cross-sections are boosted by a factor of $v^2$, while resonance annihilation is enhanced at velocities near the pole. 
In the forbidden scenario, the annihilation channel opens only if the total energy exceeds the final state masses. 
Consequently, phase space for both resonance and forbidden scenarios can be constrained, resulting in either continuum or line-shaped photon spectra, depending on model parameters~\cite{Shelton:2015aqa, Johnson:2019hsm}.

The Fermi Large Area Telescope (Fermi-LAT)~\cite{2021AjelloApJS}, referred to as Fermi hereafter, and the Dark Matter Particle Explorer (DAMPE)~\cite{DAMPE:2017cev} play crucial roles in monitoring gamma-ray emissions, advancing indirect detection of DM. Fermi's nearly 16 years of data, ranging from $500\mev$ to $1\tev$, provide the largest exposure for searching gamma-ray continuum spectra. In contrast, DAMPE offers higher energy resolution ($\lesssim 1.0\%$ at 100 GeV) compared to Fermi, 
reducing the systematic uncertainty from the background
and improving the capability to detect sharp gamma-ray features across the GeV to TeV range.

In this study, we revisit the accretion-enhanced annihilation process in a DM halo spike, with a focus on a fermionic DM model incorporating an electroweak axion-like particle (ALP) portal~\cite{Peccei:1977hh, Peccei:1977ur, Weinberg:1977ma, Wilczek:1977pj,Ghosh:2023tyz}.
We explore three specific annihilation channels (new mediator resonance, $p$-wave secluded DM and forbidden DM) based on mass conditions, analyzing the photon yield spectrum and annihilation cross-section relative to the position and velocity of DM particles with respect to the SMBH. 
For gamma-ray continuum spectra, we exclusively utilize Fermi data to establish upper limits on the ALP-DM-DM coupling $C_{a\chi\chi}$ across three scenarios. 
Additionally, following the methodology developed in Ref.~\cite{Cheng:2023chi}, we combine publicly available 6 years of DAMPE and 15 years of Fermi-LAT data from Sgr A$^{\star}$ to search for spectral line signals.

The rest of the paper is organized as follows. 
In Sec.~\ref{sec:model}, we provide a brief overview of the axion-portal DM model, 
focusing on the properties of the ALP and DM particles.
Sec.~\ref{sec:halo} covers the density profile of the DM spike and outlines the formalism used for calculating velocity-averaged annihilation cross-sections and photon yield spectra across three annihilation mechanisms.
In Sec.~\ref{sec:GaFlux}, we conduct a comprehensive six-dimensional integration to precisely compute DM-induced gamma-ray fluxes near Sgr A$^{\star}$, considering the velocity and positional dependencies of the annihilation cross-section and photon yield spectra.
Our numerical analysis methodology is detailed in Sec.~\ref{sec:analysis}, and the obtained results are presented in Sec.~\ref{sec:result}.
Finally, we conclude and discuss our findings in Sec.~\ref{sec:conclusion}.

\section{The DM model with electroweak ALP portal}
\label{sec:model}
\begin{figure}[tb]
\centering{\includegraphics[width=0.38\textwidth]{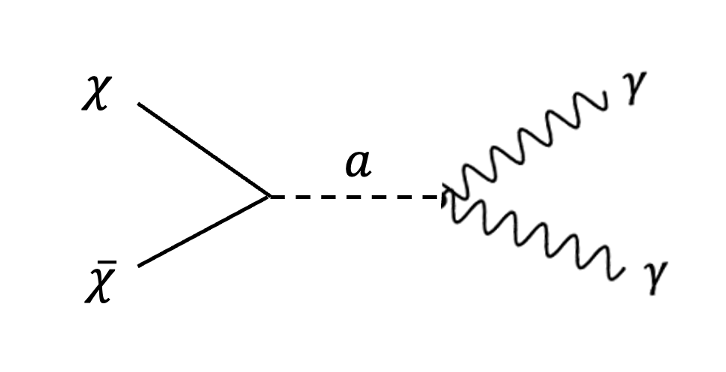}}
\centering{\includegraphics[width=0.38\textwidth]{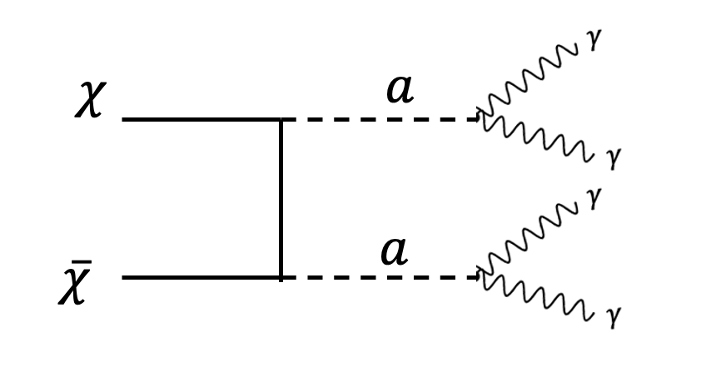}}
\caption{The DM annihilation processes to two photons (left panel) and four photons (right panel) in the model with electroweak ALP portal.
}
\label{fig:anni}
\end{figure}

Considering DM particles to be Dirac fermions $\chi$ with the mass of $m_{\chi}$, their annihilations or decays are via the mediator, pseudoscalar ALP $a$, and successively to photons as shown in Fig.~\ref{fig:anni}. 
We take the ALP mass $m_a$ as a free parameter and it could originate either from explicit PQ breaking~\cite{Wilczek:1977pj} or a confining gauge interaction in the hidden sector~\cite{Allen:2024ndv}. 
Taking that $m_\chi$ and $m_a$ to be well below the radial-mode masses of the symmetry-breaking sector, 
the effective field theory (EFT) is valid. 
In the ALP EFT framework, the ALP-DM Lagrangian can be written as~\cite{Bharucha:2022lty}
\begin{eqnarray}
\label{eq:DM-ALP}
\mathcal{L}_{\rm DM-ALP} &=&C_{a\chi\chi}m_{\chi}a\bar{\chi}i\gamma_{5}\chi, 
\end{eqnarray}
where $C_{a\chi\chi}$ is the inverse mass dimensional coupling constant of the interaction between $\chi$ and $a$.
The ALP could also couple to the SM fields. 
Focusing on electroweak ALP portal in this study, only the EFT of ALP couplings to the SM gauge bosons ($W$, $Z$, and $\gamma$) is considered, and the relevant Lagrangian for the ALP-SM couplings is~\cite{Allen:2024ndv} 
\begin{align}
\label{eq:EFT1}
\mathcal{L}_{\rm gauge-ALP} =& - \frac{C_{BB}}{4}a B_{\mu\nu} \widetilde{B}^{\mu\nu} \nonumber
	 -   \frac{C_{WW}}{4} aW_{\mu\nu}^i \widetilde{W}^{i;\mu\nu},
\end{align}
where $C_{BB}$ and $C_{WW}$ are the inverse mass dimensional coupling constants, $W_{\mu\nu}$ and $B_{\mu\nu}$ are the field strength tensors of SM ${\rm SU}(2)_{\rm L}$ and ${\rm U}(1)_{\rm Y}$ gauge groups before electroweak symmetry breaking (EWSB), respectively.  Here $\widetilde{W}_{\mu\nu}\equiv \epsilon_{\rho\sigma \mu\nu} W^{\rho\sigma}/2$ is the corresponding dual field strengths, and $i=1, 2, 3$ are the $SU(2)_L$ indices.

Including DM-ALP and the EFT of ALP couplings to the SM gauge bosons ($W$, $Z$, and $\gamma$),
the full Lagrangian after EWSB can be written as  
\begin{eqnarray}
\label{eq:fullL}
    {\cal L} = 
    &&-\frac{1}{2}\partial_{\mu}a\partial^{\mu}a-
    \bar{\chi}(i{\:/\!\!\!\! {\partial}}-m_{\chi})\chi
    +\frac{1}{2}m^{2}_{a}a^{2}-C_{a\chi\chi}m_{\chi}a\bar{\chi}i\gamma_{5}\chi \nonumber \\
    &&-\frac{C_{a\gamma\gamma}}{4}\,a F_{\mu\nu}\tilde{F}^{\mu\nu} - \frac{C_{aZZ}}{4}\,a Z_{\mu\nu}\tilde{Z}^{\mu\nu}
    -\frac{C_{a\gamma Z}}{2} \,aF_{\mu\nu}\tilde{Z}^{\mu\nu}-\frac{C_{aWW}}{2}\,a W_{\mu\nu}^+\tilde{W}^{-\mu\nu}.    
\end{eqnarray}
Those effective couplings ($C_{a\gamma\gamma}$, $C_{aZZ}$, $C_{aWW}$, and $C_{a\gamma Z}$) can be expanded from $C_{BB}$, $C_{WW}$, and the Weinberg angle $\theta_W$ as 
\begin{equation}
\begin{aligned}
   &C_{a\gamma\gamma} = C_{BB}\cos^2\theta_{\rm W} + C_{WW}\sin^2\theta_{\rm W},\\
   &C_{aZZ} = C_{BB}\sin^2\theta_{\rm W} + C_{WW}\cos^2\theta_{\rm W},\\
   &C_{a\gamma Z}= \left(C_{WW}-C_{BB}\right)\sin\theta_{\rm W}\cos\theta_{\rm W},\\
   &C_{aWW} =C_{WW}.
	\label{eq:couplingsEWSB}
\end{aligned}
\end{equation}
The annihilation rates of DM to photons are determined by the couplings $C_{a\gamma\gamma}$ and $C_{a\chi\chi}$. For simplicity, we use the following assumptions for the model parameters in our numerical studies without loss of generality. First, we set $C_{WW}$ = $C_{BB}$ to reduce one free model parameter. 
Then, the electroweak ALP strongly suffers from collider constraints~\cite{Allen:2024ndv}: $C_{a\gamma\gamma}\lesssim 2.5\times10^{-5} \; \rm GeV^{-1}$ for $m_a\sim 1$ TeV from the LHC ALP to diphoton searches~\cite{Jaeckel:2012yz}, and $C_{a\gamma\gamma}\lesssim 1.0\times10^{-6} \; \rm GeV^{-1}$ for $m_a\sim 4$ GeV from the BaBar $B\to K\gamma\gamma$ searches~\cite{BaBar:2021ich}. 

Generally speaking, DM annihilations to photons in this model can be dominant by the $s$-wave contribution via $a$-resonance, or $p$-wave contribution ($\chi\overline{\chi}\to a a$). 
In addition, some different mass combinations between $\chi$ and $a$ can alter the velocity dependence of the DM annihilation cross-section.  
To highlight various velocity-dependent features of the annihilation cross section, we set $C_{a\gamma\gamma}=10^{-9} \gev^{-1}$ for $m_a>500 \; \rm MeV$ in numerical calculations when $m_\chi>m_a$ and $m_\chi\lesssim m_a$, thereby neglecting the $s$-channel contribution. In these cases, the main processes involve $p$-wave annihilation (velocity-squared suppression) and forbidden annihilation (velocity-exponential suppression). When $2 m_\chi\approx m_a$, the decay channel $a\to \chi\chi $ opens, we assign $C_{a\gamma\gamma}=2.5\times 10^{-5} \gev^{-1}$ to increase the efficiency of DM annihilation to photons.

If DM particles can be accelerated and annihilated to photons as presented in Fig.~\ref{fig:anni}, 
three possible annihilation mechanisms based on the relationships between $m_\chi$ and $m_a$ appear as what follows,     
\begin{itemize}
    \item $m_\chi > m_a$:\\
    The dominant DM annihilation channel is $\chi\overline{\chi}\to aa$ which is $p$-wave. 
    The photon energy spectra are a box-shape due to Lorentz boost, especially for $m_\chi\gg m_a$.   
    \item Resonance annihilation ($2 m_\chi\approx m_a$):\\
The annihilation cross-section of $\chi\overline{\chi}\to \gamma\gamma$ can undergo a large enhancement through an $a$-resonance. 
In the rest frame, its photon spectrum shapes as a monochromatic line, but this feature can become distorted under a high acceleration of DM particles. 
To characterize the resonance nature, we utilize $R_{a\chi}=\sqrt{1-4 m_\chi^2/m_a^2}$ instead of $m_a$ itself.
    \item Forbidden annihilation ($m_\chi\lesssim m_a$):\\
The annihilation process is also $\chi\overline{\chi}\to aa$, albeit with $m_\chi\lesssim m_a$. 
Essentially, forbidden annihilations cease when the kinetic energies of DM particles fall below threshold energies. 
The shape of the energy spectrum varies, either monochromatic or continuum, depending on the mass splitting $\Delta m\equiv m_a - m_\chi$. 
If $\Delta m$ exceeds a certain threshold, the photon spectrum will appear monochromatic (line-like); otherwise, it remains a continuum spectrum.
We will see later in Sec.~\ref{sec:GaFlux} that the shape of photon energy spectra from the GC is dependent on both the size of $\Delta m$ and $C_{a\chi\chi}$.
As a result, this requires different strategies for searching for forbidden annihilation signals. 
\end{itemize}


\section{DM annihilation in the spike halo}
\label{sec:halo}

Without the presence of a black hole, the DM density profile can be assumed as the generalized Navarro-Frenk-White (NFW) profile~\cite{Navarro:1996gj} obtained by $N$-body simulation, parameterized as
\begin{equation}
    \rho_{\rm NFW}(r)=\rho_s(r/r_s)^{-\gamma}(1+r/r_s)^{\gamma-3},
    \label{eq:gNFW}
\end{equation}
where $\rho_s\approx 0.0156 \; \rm M_{\odot}/\rm pc^3$ and $r_s=1.5\times 10^4 \; \rm pc$ are the scaled density and scaled radius~\cite{McMillan:2016jtx}. 
Here, we take the slope of the density profile $\gamma=0.8$ based on the S-star combined constraints $\gamma<0.83$~\cite{Shen:2023kkm}. 
The DM-induced gravitational potential with positive definition can be written as~\cite{Yang:2022hkm}
\begin{equation}
    \Phi(r)=4 \pi{\rm G}\left[\frac{1}{r}\int_0^r\rho_{\rm NFW}(r^\prime)r^{\prime 2} {\rm d} r^\prime + \int_r^{\infty} \rho_{\rm NFW}(r^\prime)r^{\prime} {\rm d} r^\prime\right].
\end{equation}
By considering an isotropic and spherically symmetric DM distribution, 
we can then use the Eddington inversion method to obtain the DM phase-space distribution function as~\cite{2018Anatomy}
\begin{equation}\label{Eddington_formula}
    f_0(\epsilon)=\frac{1}{\sqrt{8}\pi^{2}}\int^{\varepsilon}_{0}
                \frac{\mathrm{d}\Phi}{\sqrt{\varepsilon-\Phi}}
                \frac{\mathrm{d}^{2}\rho_{\rm NFW}}{\mathrm{d}\Phi^{2}},
\end{equation}
where $\varepsilon\equiv \Phi(r)-v^{2}/2$ is the relative energy per unit mass with the DM velocity $v=\left|\mathbf{v}\right|$.

Imagining a point-mass black hole with the adiabatic growth at galactic center as the gravitational potential near the point mass changes, each particle responds to the change by altering its energy $\epsilon$, but conserved the radial action $I_r$ and angular momentum $L$,
\begin{equation}
\begin{aligned}
    I_r(\epsilon,L)&=\oint v_r dr=2 \int_{r_{\rm min}}^{r_{\rm max}}   dr 
    \sqrt{2\Phi(r)-2\epsilon
    -\left(\frac{L}{r}\right)^2
    },  
\end{aligned}
\label{eq:Ir0}
\end{equation}
where $r_{\rm min}$ and $r_{\rm max}$ are the two roots of the radial velocity. 
When the black hole governs the gravitational potential, we can analytically obtain the integration in Eq.~\eqref{eq:Ir0} 
based on the gravitational potential of a point-mass black hole $\Phi_{\rm BH}(r)= {\rm G} M_{\rm BH}/r$, 
\begin{equation}
    I_r'(\epsilon',L') = 2 \pi \times
    \left(-L'+ \frac{{\rm G} M_{\rm BH}}{\sqrt{2\epsilon'}} \right).
    \label{eq:Irp}
\end{equation}
Given the conservation relations $L' = L$ and $I_r'(\epsilon',L') = I_r(\epsilon,L)$, 
we can obtain $\epsilon$ as a function of $\epsilon'$ and $L'$, 
then the final distribution in phase space can be written as $f'(\epsilon', L')=f_0\left[\epsilon(\epsilon', L')\right]$.


We can utilize the parametrization of $I_r$ provided in~\cite{Gondolo:1999ef}, which relies on the power-law profile and exhibits uncertainties below $8\%$, to establish a connection with our expression in Eq.~\eqref{eq:Irp}, allowing us to derive the relationship among $\epsilon$, $\epsilon'$, and $L'$. 
Once we obtain $\epsilon(\epsilon', L')$ and $f'(\epsilon', L')=f_0\left[\epsilon(\epsilon', L')\right]$, 
the spike halo profile can be calculated by integrating the final-state phase space, 
\begin{equation}\label{eq:density}
    \rho_{\rm spike}(r)=\int^{\epsilon'_m}_0 d\epsilon' 
    \int_{L'_{\rm min} }^{L'_{\rm max}} 
    dL' 
    \frac{4 \pi L'}{r^2 v_{\rm BH}} 
    f'(\epsilon',L'),~~{\rm with}~~
    v_{\rm BH}\equiv \sqrt{\frac{2 {\rm G} M_{\rm BH}}{r}- 2\epsilon'- \left(\frac{L'}{r}\right)^2}. 
\end{equation}
The upper and lower limits of the integrals are 
\begin{equation}
 \epsilon'_m=\frac{{\rm G} M_{\rm BH}}{r}\left(1-\frac{2R_{\rm s}}{r}\right), \quad
 L'_{\rm min}=\sqrt{2}cR_{\rm s},
 \quad {\rm and}~~L'_{\rm max}=\sqrt{2r^2\left(\frac{{\rm G} M_{\rm BH}}{r}-\epsilon'\right)},
\end{equation}
where $R_{\rm s}$ and $c$ are the Schwarzschild radius and the speed of light, respectively. 
When employing the non-relativistic approximation with a capture radius of approximately $2 R_{\rm s}$, 
the deviation in spike density from the relativistic numerical results is only about $15\%$~\cite{Sadeghian:2013laa}. 
Although the typical non-relativistic approximation uses a capture radius of $4 R_{\rm s}$~\cite{Gondolo:1999ef}, 
we choose to reduce it to $2 R_{\rm s}$ for a complementary purpose.

\begin{figure*}[ht]
\centering
\includegraphics[width=8.5cm]{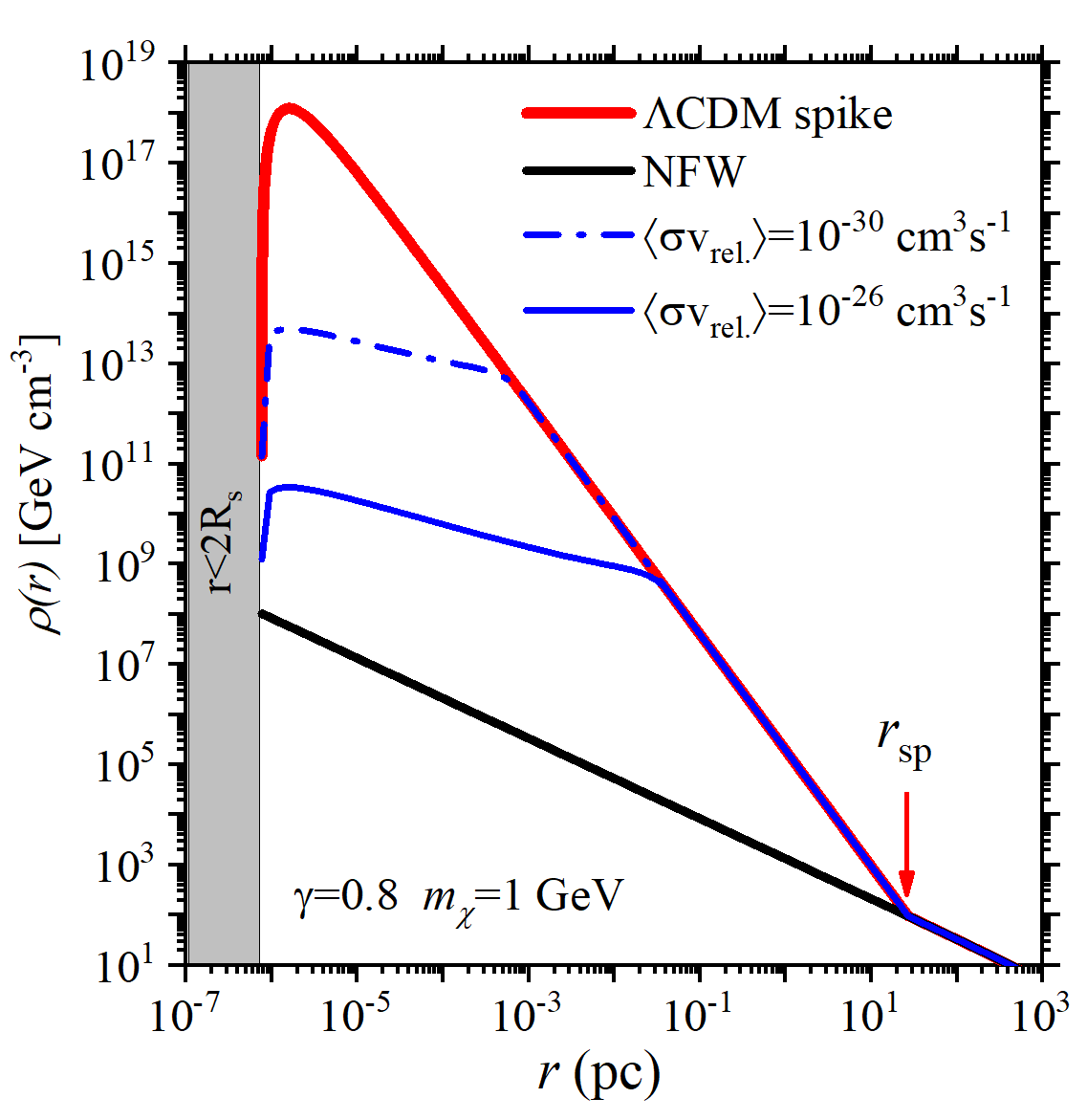}
\includegraphics[width=8.5cm]{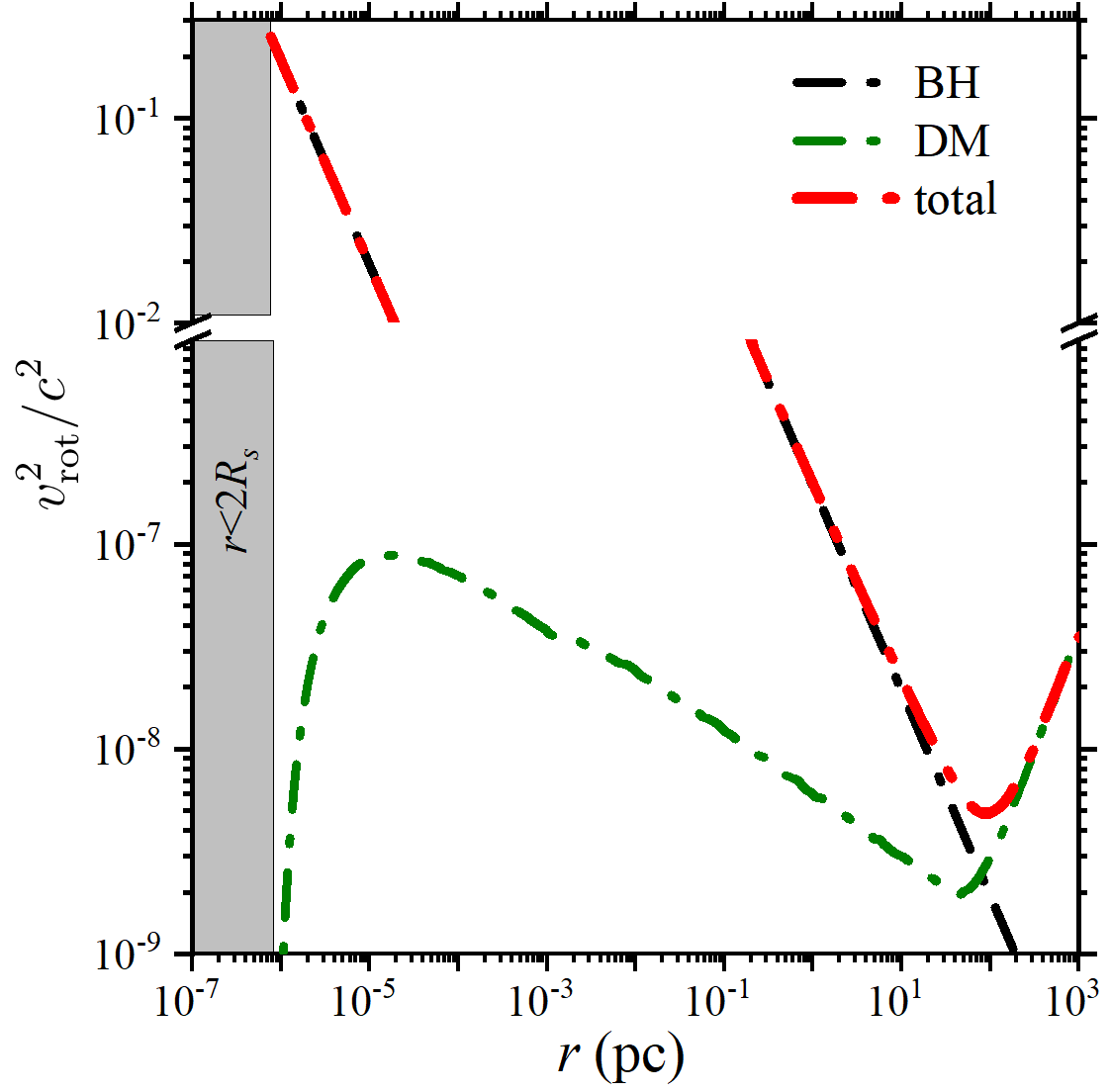}
 \caption{
The halo density profile (left panel) and the rotation velocity squared (right panel) as a function of radius. 
 The gray region $r<2R_s$ represents a no-DM particle zone. 
 The spike halo profile (red solid line) represents the collisionless DM paradigm. 
 The blue lines represent velocity-averaged annihilation cross-sections with $m_\chi=1\gev$ for 
 $\sv=10^{-26}$~cm$^3$s$^{-1}$ (solid) and $\sv=10^{-30}$~cm$^3$s$^{-1}$ (dot-dashed). 
 The spike halo profile matches the NFW profile (depicted as a black solid line) at $r\ge r_{\rm sp}$. 
As a comparison in the right panel, the total rotation velocity squared  (illustrated as a red dot-dashed line) contains the contributions from both BH (black dot-dashed line) and DM (green dot-dashed line).}
 \label{fig:rho}
\end{figure*}

Considering DM particles annihilating within the innermost region of the spike, the slope of the spike density can be weakened and 
their annihilation time is $T_{\rm ann}=m_\chi/\left[\rho_{\rm spike}(r) \langle\sigma v\rangle\right]$.
By taking the time scale equal to the galactic age $T_{\rm ann} \approx 10^{10}~{\rm yr}$, 
we can obtain the annihilation radius $ r_{\rm ann}$. 
Namely, if a particle moves with an orbital radius smaller than $r_{\rm ann}$ where DM density is higher, 
leading to a shorter annihilation time than $T_{\rm ann}$ and large DM particle annihilation within the range of $r<r_{\rm ann}$~\cite{Shapiro:2016ypb}. 
Thus, the DM phase distribution and total DM density profile can be modified as
\begin{equation}\label{eq:DF}
    f=
    \left\{
        \begin{array}{ll}    
        f'(\epsilon',L'),& \quad  0\leq \epsilon' \leq \frac{{\rm G}M_{\rm BH}}{r_{\rm ann}}, \\
                       0,& \quad \epsilon' > \frac{{\rm G}M_{\rm BH}}{r_{\rm ann}}, 
        \end{array}
    \right.
\end{equation}
and 
\begin{equation}
    \rho(r)=\left\{
    \begin{aligned}
    &0, \quad r<2R_{\rm s},\\
    &\rho_{\rm ann}(r), \quad  2R_{\rm s}<r<r_{\rm ann},\\
    &\rho_{\rm spike}(r), \quad r_{\rm ann}<r<r_{\rm sp},\\
    &\rho_{\rm NFW}(r), \quad r>r_{\rm sp}.
    \end{aligned}
    \right
.
\end{equation}
Gravitational accretion by the BH forms a dense DM spike in the central region, while DM annihilation reduces its density. 
The spike radius $r_{\rm sp}$ corresponds to the crossover where spike density becomes smaller than the NFW density $r_{\rm sp}\approx 27$~pc, 
while $r_{\rm ann}$ depends on the annihilation cross-section. 
Note that the spike can also be smoothed by the DM self-interaction, namely  $\chi\bar{\chi}\to\chi\bar{\chi}$, $\chi\chi\to\chi\chi$ and   $\bar{\chi}\bar{\chi}\to\bar{\chi}\bar{\chi}$, 
and their cross-sections are expressed in Appendix~\ref{app:self_interaction}.
In our interesting parameter spaces (taking $a$-resonance as an example, $C_{a\chi\chi}  \le 10^{-2} /\gev$ and $1\gev \le m_\chi \le 200\gev$), 
when we take the non-relativistic limit $\sqrt{s}\approx 2m_\chi$, the predicted DM self-interaction cross section per unit mass $\sigma_{\rm self}/m_\chi$ is $\mathcal{O}(10^{-8})$~cm$^{2}/$g, which is much smaller than the required value of the strong self-interaction DM (SIDM) scenario $\sigma_{\rm self}/m_\chi\approx 0.1~{\rm cm}^2/\rm{g}$.~\footnote{Unlike our electroweak ALP portal model, Ref.~\cite{Cheng:2022esn} shows a SIDM scenario with a predicted value  
$\langle \sigma_{\rm self} v\rangle/m_\chi\approx 1.5~{\rm cm}^2/\rm{g} \cdot km/s$. 
This value is roughly the same order of the normal SIDM value $\sigma_{\rm self}/m_\chi\approx 0.1~{\rm cm}^2/\rm{g}$. 
Because of the SIDM scenario in their model, the spike density can be weakened by $\mathcal{O}(10^{7})$ near the BH.}
As a result, such a small value of $\sigma_{\rm self}/m_\chi$ in this work cannot form the SIDM halo~\cite{Shapiro:2014oha}.

In Fig.~\ref{fig:rho}, the left panel illustrates how DM annihilation alters the spike halo profile slope~\footnote{The slope of the spike density from Eq.~\eqref{eq:density}, under adiabatic conditions and no annihilation, can be parameterized as $\gamma_{\rm sp}=(9-2\gamma)/(4-\gamma)$~\cite{Gondolo:1999ef}.}. 
Larger annihilation cross-sections lead to larger $r_{\rm ann}$ and can significantly weaken the spike.
With the DM mass of $m_\chi=1\gev$, a lower velocity-averaged annihilation cross-section $\sv=10^{-30}$~cm$^3$\,s$^{-1}$ (blue dashed line) can reduce the collisionless $\Lambda$CDM spike (red solid line) by approximately three orders of magnitude. 
A Larger cross-section $\sv=10^{-26}$~cm$^3$\,s$^{-1}$ (blue solid line) weakens it further, approaching the NFW profile (black solid line). 
The weak cusp profile coincides with the spike profile beyond the annihilation radius ($r_{\rm ann} = 10^{-3}~\rm pc$ and $r_{\rm ann} = 10^{-1}~\rm pc$ for $\sv = 10^{-30}~\mathrm{cm^3 \, s^{-1}}$ and $\sv = 10^{-26}~\mathrm{cm^3 \, s^{-1}}$, respectively).
All cases align with the NFW profile for $r\ge r_{\rm sp}$ (approximately $27$ pc), with the gray region ($r\le 2R_s$) representing a no-DM particle zone. 
At $r\approx 2 R_s$, the spike profile density exceeds the NFW density by $\sim 10$ orders of magnitude. 
The weak cusp density ($\sv = 10^{-30} \; \mathrm{cm^3 \, s^{-1}}$) surpasses the NFW density by six orders, while for $\sv = 10^{-26} \; \mathrm{cm^3 \, s^{-1}}$, it is only three orders higher.

In the right panel of Fig.~\ref{fig:rho}, the rotation velocity squared
\begin{equation}
    v_{\rm rot}^2=\frac{{\rm G} M(r)}{r}, \qquad {\rm where} \qquad M(r)=M_{\rm BH}+4 \pi\int_0^r \rho(r^\prime){r^\prime}^2 dr^\prime.
\end{equation}
The individual contributions to $v_{\rm rot}^2$ due to gravity are BH (black line), DM (green line), and the sum (red dot-dashed line). 
Within the innermost region, BH gravity dominates and accelerates DM particles to half the speed of light at $2R_s$. 
When $r>r_{\rm sp}$, the gravity of the DM halo dominates, leading the density profile to revert to the NFW.

\begin{figure*}[ht]
\centering
\includegraphics[width=8.5cm]{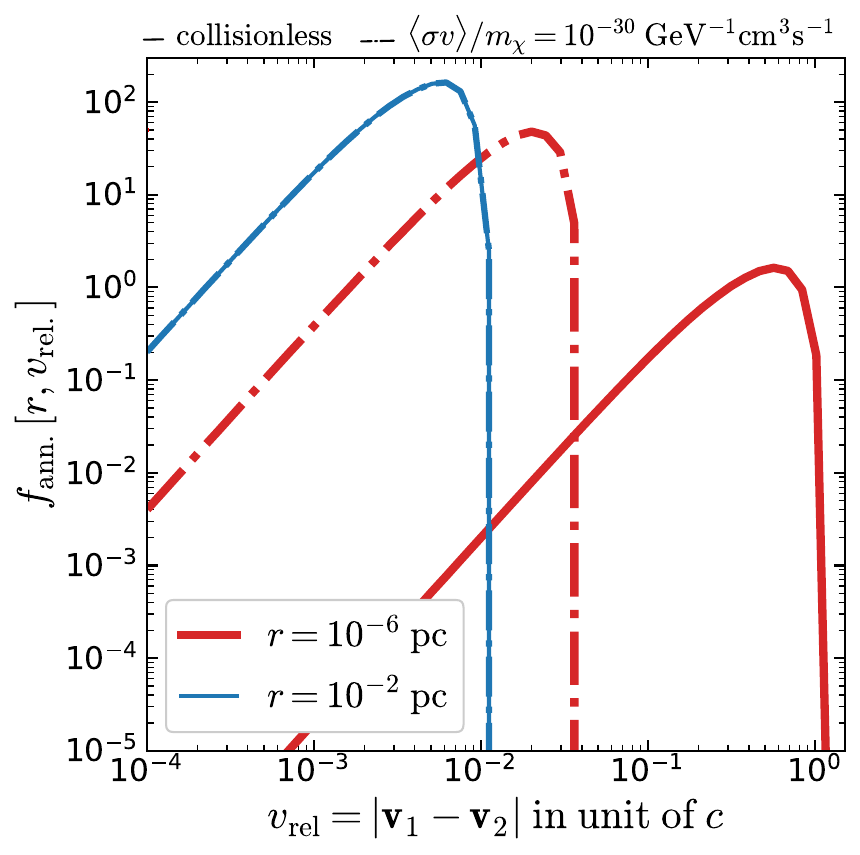}
 \caption{The distributions of relative velocities at $r=10^{-6}\pc$ (red lines) and 
 $r=10^{-2}\pc$ (blue lines). 
 The solid lines represent the velocity distribution of the collisionless DM, while 
 the dash-dotted lines are based on $\sv/m_\chi=10^{-30}$~cm$^3$s$^{-1}\gev^{-1}$.  
Since annihilations mostly occur within $r \leq r_{\rm ann}\approx 10^{-3}\pc$, the velocity distribution $f_{\rm ann.}$ remains unchanged at $r=10^{-2}\pc$, resulting in overlapping blue solid and dash-dotted lines. Note that $v_{\rm rel.}=|\mathbf{v_1}-\mathbf{v_2}|$ is not physical velocity and it does not require to be smaller than $c$. 
 }
 \label{eq:f_v_r}
\end{figure*}

To describe a DM particle, its velocity distribution can only be computed numerically using Eq.~\eqref{eq:DF}. 
In an NFW-like halo, the velocity distribution of DM is expected to depend on $r$. 
However, once the DM annihilation occurs near a massive BH, 
the velocity distribution of two DM particles also depends on the angle $\theta$ between them. 
For two DM particles (denoted as suffixes $1$ and $2$) annihilate at $r$ with two different velocity vectors $\mathbf{v_1}$ and $\mathbf{v_2}$, 
the total distribution is described by $f_1(\mathbf{v_1}, r)f_2(\mathbf{v_2}, r)$ and  it is more useful to define the relative velocity $\mathbf{v_{\rm rel.}}=\mathbf{v_2}-\mathbf{v_1}$ and 
the velocity at the center-of-mass frame $\mathbf{v_{\rm CM}}=(\mathbf{v_2}+\mathbf{v_1})/2$. 
Therefore, we can integrate over $\mathbf{v_{\rm CM}}$ to have a three-dimensional velocity distribution for DM annihilations,   
\begin{equation}
    f_{\rm ann.}[r, \mathbf{v_{\rm rel.}}]\propto \int d^3 \mathbf{v_{\rm CM}} f_1(\mathbf{v_1}, r)f_2(\mathbf{v_2}, r).
    \label{eq:fv_3D}
\end{equation}
Here, we use notations $v_{\rm rel.}=|\mathbf{v_{\rm rel.}}|$ and $v_{\rm CM}=|\mathbf{v_{\rm CM}}|$. 
Taking axial symmetric gravitational potential, 
we can again rewrite three-dimensional velocity distribution to be one-dimensional velocity distribution~\cite{2018Anatomy}, 
\begin{eqnarray}
\label{eq:f_v_r_1D}
    f_{\rm ann.}[r, v_{\rm rel.}]  =  \frac{v_{\rm rel.}^2}{N_0} \times 
    &&\int_0^{v_{\rm esc}} dv_{\rm CM} v_{\rm CM}^2 
    \int_{-1}^1 d\cos\alpha
    \int_0^{2 \pi} d\phi \times \nonumber\\ 
    && \int_{-\mu_0}^{\mu_0} d\cos\theta \times
    f_1(v_{\rm rel.}, v_{\rm CM} ,L_1,r)\times f_2(v_{\rm rel.}, v_{\rm CM},L_2,r), \\
    {\rm with}\quad 
    \mu_0  \equiv&&\frac{4 v_{\rm esc}^2- 4 v_{\rm CM}^2-v_{\rm rel.}^2}{4 v_{\rm rel.}v_{\rm CM}},
    \quad \cos\alpha \equiv  
    \frac{\mathbf{v_{\rm CM}}\cdot \mathbf{r}} {v_{\rm CM} \times r} \quad {\rm and} \quad v_{\rm esc}=\sqrt{2 \Phi_{\rm total}(r)},\nonumber
\end{eqnarray}
where the graph of the potential function $\Phi_{\rm total}(r)$ is displayed in Appendix~\ref{app:Potential profile}.

The angular momentum $L_1$ and $L_2$ are expressed as a function of angles, shown in Appendix~\ref{app:phase_space}. 
The normalization factor $N_0$ ensures $\int f_{\rm ann.}[r, v_{\rm rel.}] dv_{\rm rel.}=1$.
The azimuth angle, denoted by $\phi$, represents the angle in the polar coordinate, with the direction of $\mathbf{v}_{\rm CM}$ serving as the $z$-axis.

In Fig.~\ref{eq:f_v_r}, we present $f_{\rm ann.}[r, v_{\rm rel.}]$ for two distances, $r=10^{-6}\pc$ (red lines) and $r=10^{-2}\pc$ (blue lines). 
We also compare scenarios for the collisionless DM (solid lines) and $\sv/m_\chi=10^{-30}$ cm$^3$s$^{-1}\gev^{-1}$ (dash-dotted lines). 
The peak of $f_{\rm ann.}[r, v_{\rm rel.}]$ occurs around $\sqrt{2}v_{\rm rot}$. 
Near the BH around $r\approx 10^{-6}\pc$, DM particles can achieve higher velocities (also see the right panel of Fig.~\ref{fig:rho}), 
the peak velocity approaches $0.5 c$. 
Moving away from the BH to $r\approx 10^{-2}\pc$, the peak velocity decreases to be non-relativistic $\approx 5\times 10^{-3}c$.

For the benchmark (dash-dotted lines) shown in Fig.~\ref{eq:f_v_r}, annihilation mainly occurs within $r \leq r_{\rm ann} \approx 10^{-3} \text{ pc}$, where the velocities exceed $v_{\rm esc}[r_{\rm ann}]$ (the escape velocity at $r_{\rm ann}$) and significantly enhance the velocity-dependent annihilation rate. Thus, at a smaller radius $r = 10^{-6} \text{ pc}$, DM particles with $v_{\rm esc}[r_{\rm ann}] \leq v < v_{\rm esc}[10^{-6} \text{ pc}]$ can fully annihilate, leading to a truncation of the high-velocity tail of the velocity distribution, as illustrated by the leftward shift of the peak position in Fig.~\ref{eq:f_v_r}.
In contrast, the high-speed tail of collisionless DM particles can still be truncated at $v_{\rm esc}$, but there is no annihilation to consume high-speed DM particles. 
Thus, weak DM annihilation facilitates the accumulation of high-speed DM particles. 
However, there is almost no difference made by DM annihilation in $r>r_{\rm ann}$, as seen from the overlapping blue dash-dotted and blue solid lines ($r=10^{-2}\pc$) in Fig.~\ref{eq:f_v_r}.

Upon determining the DM velocity distribution, we can define the velocity-averaged cross-section $\sv$ and photon-yield spectrum $\avgdnde$ for DM annihilations at radius $r$ as
\begin{eqnarray}
    \sv[r] &=& \int \sigma[v_{\rm rel.}]\times v_{\rm rel.}\times f_{\rm ann.}[r, v_{\rm rel.}] d v_{\rm rel.}, ~~{\rm and}\nonumber\\
    \avgdnde &=& \int \dnde[v_{\rm rel.}]\times f_{\rm ann.}[r, v_{\rm rel.}] d v_{\rm rel.}, 
\label{eq:avgXS}
\end{eqnarray}
respectively. 
The photon spectrum $\dnde$ represents the photon energy distribution per DM annihilation. 
In the case of non-relativistic DM annihilation, specifically $\chi\overline{\chi}\to \gamma\gamma$, the spectrum takes the form of a delta function centered at $m_\chi$. 
However, for $\chi\overline{\chi}\to a a \to 4 \gamma$, the spectrum becomes the box-like shape with a Lorentz boost factor of $m_\chi/m_a$. 
By integrating over all $E_\gamma$, we obtain two photons for $\chi\overline{\chi}\to \gamma\gamma$ and four photons for $\chi\overline{\chi}\to a a \to 4 \gamma$. 
In the case of relativistic DM annihilation near a BH, the central energy $\sqrt{s}$ of DM particles varies with $r$. 
As a result, the shape of $\avgdnde$ also relies on $r$.

In the following subsections, we will explore the impact of the spike profile, focusing on $\sv$ and $\avgdnde$. 
We will investigate $p$-wave annihilation for scenario $m_\chi>m_a$ in Sec.~\ref{sec:pwave}, resonance annihilation in Sec.~\ref{sec:resonance}, and forbidden annihilation in Sec.~\ref{sec:forbidden}.

\subsection{Scenario $m_\chi>m_a$}
\label{sec:pwave}

\begin{figure*}[ht]
\centering
\includegraphics[width=8.5cm]{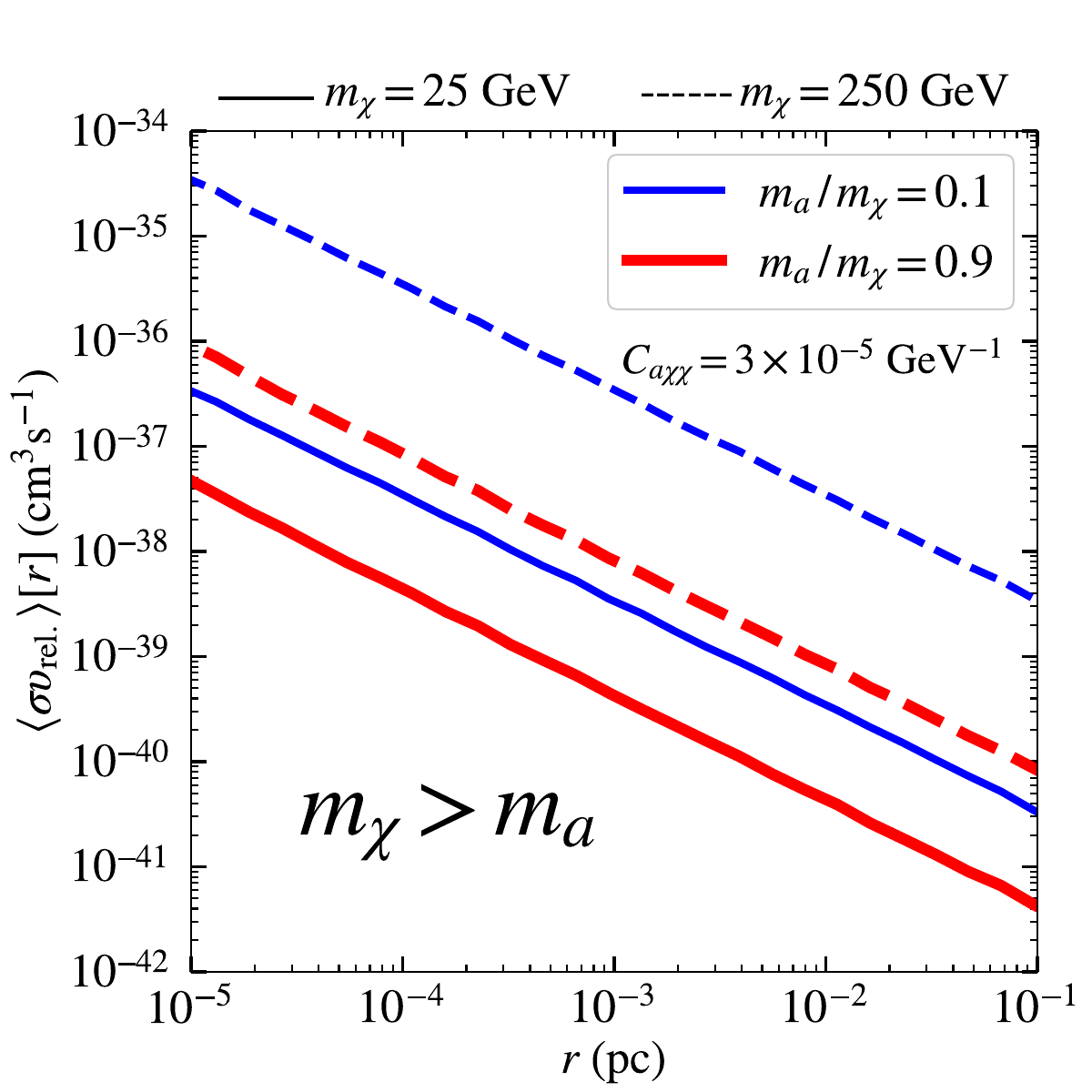}
\includegraphics[width=8.5cm]{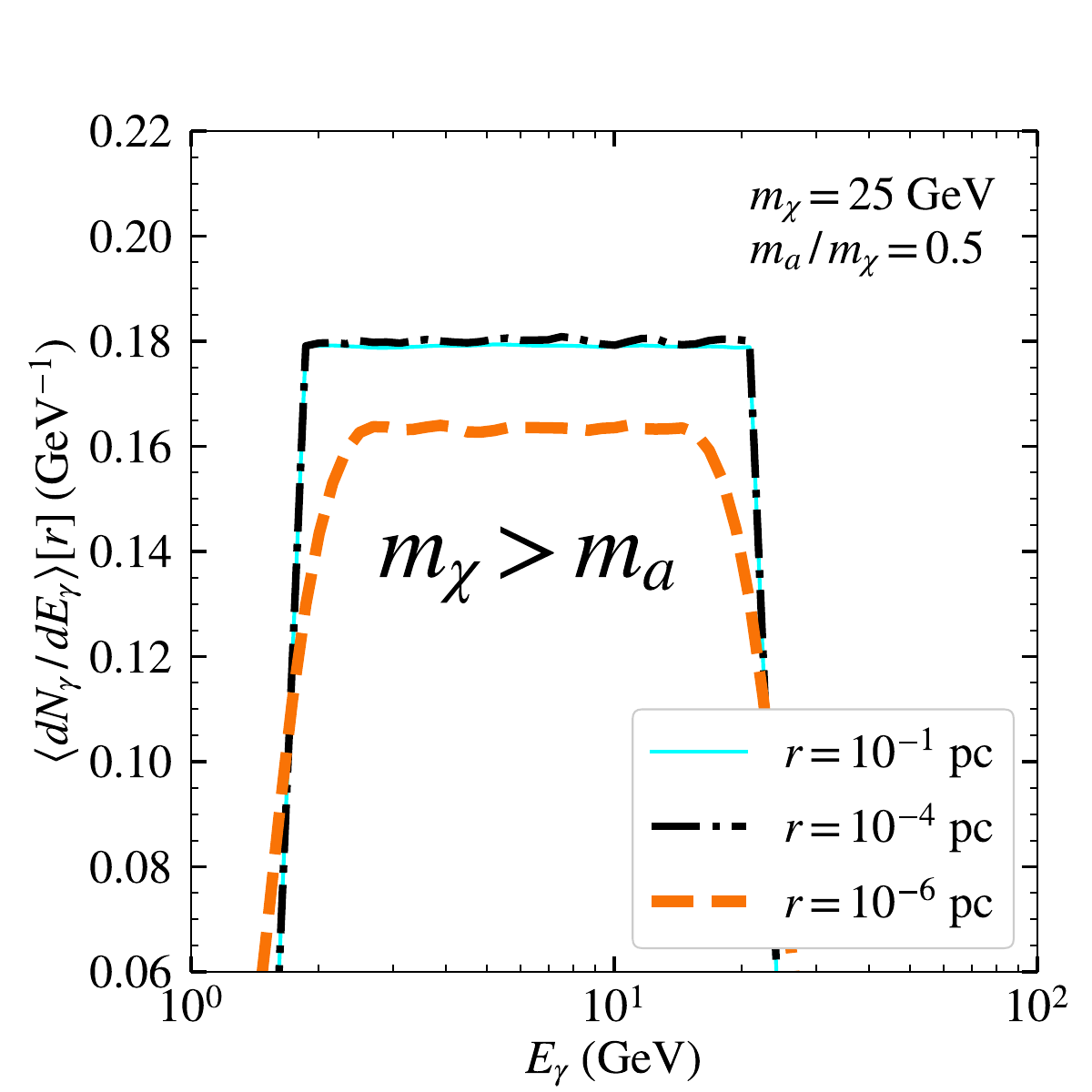}
 \caption{
The left panel: the radial distribution of $\sv$ in the $m_\chi>m_a$ scenario as a function of $r$ for $m_a/m_\chi=0.1$ (blue lines) and $m_a/m_\chi=0.9$ (red lines). 
The solid and dash lines represent $m_\chi=25\gev$ and $m_\chi=250\ \gev$, respectively, by taking $C_{a\chi\chi}=3\times 10^{-5}/\gev$ as the benchmark parameter. 
The right panel: the photon energy distribution at $r=10^{-6}\pc$ (orange dashed line), 
 $r=10^{-4}\pc$ (black dash-dotted line), and $r= 10^{-1}\pc$ (cyan solid line) with benchmark $m_\chi=25\gev$ and $m_a/m_\chi=0.5$. 
 }
 \label{fig:sigma_pwave}
\end{figure*}

With the interaction given in Eq.~\eqref{eq:DM-ALP}, the integrated cross-section $\sigma(\chi\bar{\chi}\to a a)$~\cite{Bharucha:2022lty} can be expressed by 
\begin{eqnarray}
\label{eq:xsecxxaa}
\sigma(\chi\bar{\chi}\to a a)&=&
\frac{C_{a\chi\chi}^4 m_\chi^2}{64\pi }\times 
\frac{1-\beta_\chi^2}{\beta_\chi^2}\times
\left[ 
F_a
\ln\left(\frac{\beta_a^2+2 \beta_a \beta_\chi+1}{\beta_a^2-2 \beta_a \beta_\chi+1}\right)
-\beta_a\beta_\chi F_{a\chi}\right],\quad {\rm with}\\
\beta_\chi&=& \sqrt{1-\frac{4m_\chi^2}{s}}, \quad {\rm and}\quad \beta_a=\sqrt{1-\frac{4m_a^2}{s}}, \nonumber 
\end{eqnarray}
where $s\equiv E_{\rm cm}^2$ is the center-of-mass energy squared of DM annihilation.
Here, functions $F_a$ and $F_{a\chi}$ are defined as
\begin{eqnarray}
F_a&\equiv&\frac{ (1-\beta_a^2)^2+ 2(1+\beta_a^2)^2}{4 (1+\beta_a^2)},\quad {\rm and} \nonumber\\
F_{a\chi} &\equiv& \left(
\frac{\beta_a^2+8\beta_\chi^2-8\frac{1-\beta_\chi^2}{1-\beta_a^2}-5}
{\beta_a^2+4 \beta_\chi^2+4\frac{1-\beta_\chi^2}{1-\beta_a^2}-3}
\right).
\end{eqnarray}
By expanding the cross-section in terms of the DM velocity $\beta_\chi$, 
we can find that $p$-wave contribution dominates the annihilation cross-section
\begin{equation}
    \sigma v_{\rm rel.}(\chi\bar{\chi}\rightarrow a a)\approx
    \frac{C_{a \chi\chi}^4 m_\chi^5 (m_\chi^2-m_a^2)^{5/2}}{3\pi\times (2m_\chi^2-m_a^2)^4} \beta_\chi^2 .
    \label{eq:av_sv}
\end{equation}
Using the relative velocity $v_{\rm rel}=2\beta_\chi$, we then perform the integral of velocities to obtain $\sv$ by Eq.~\eqref{eq:avgXS}.

In Fig.~\ref{fig:sigma_pwave}, the left panel shows results of $\sv$ with $C_{a\chi\chi}=3\times 10^{-5}/\gev$ for $p$-wave annihilation, 
while the right panel presents $\avgdnde$. 
From the right panel of Fig.~\ref{fig:rho}, we see that $v_{\rm rot} \propto r^{-1/2}$ for $r<100\pc$, 
indicating that $v_{\rm rot}$ decreases with radius. 
Consequently, the $p$-wave feature, where $\sigma v_{\rm rel.} \propto v_{\rm rel.}^2$, results in $\sv \propto r^{-1}$.
In addition, $\sv$ increases with $m_\chi^2$ for $m_\chi\gg m_a$ as presented in Eq.~\eqref{eq:av_sv}, 
leading to dash lines ($m_\chi=250 \gev$) larger than solid lines ($m_\chi=25 \gev$) by two orders of magnitude. 
Similarly, those red lines ($m_a/m_\chi=0.9$) are one order of magnitude smaller than the blue lines ($m_a/m_\chi=0.1$), because 
a larger value of $m_a/m_\chi$ corresponds to a smaller $\sv$.

In the right panel of Fig.~\ref{fig:sigma_pwave}, we display $\avgdnde$ versus $E_\gamma$ for $r=0.1\pc$ (cyan solid line), $r=10^{-4}\pc$ (black dash-dotted line), and $r=10^{-6}\pc$ (orange dashed line). 
With $m_a/m_\chi=0.5$, the produced ALP speed is approximately $0.9c$. 
As previously mentioned, $\dnde$ exhibits a box-shaped spectrum, but its width varies with the central energy of annihilating DM particles. 
Therefore, even after convolving with the velocity distribution, the spectrum broadens as the particles approach the BH.

\subsection{Resonance annihilation}
\label{sec:resonance}

\begin{figure*}[ht]
\centering
\includegraphics[width=8.5cm]{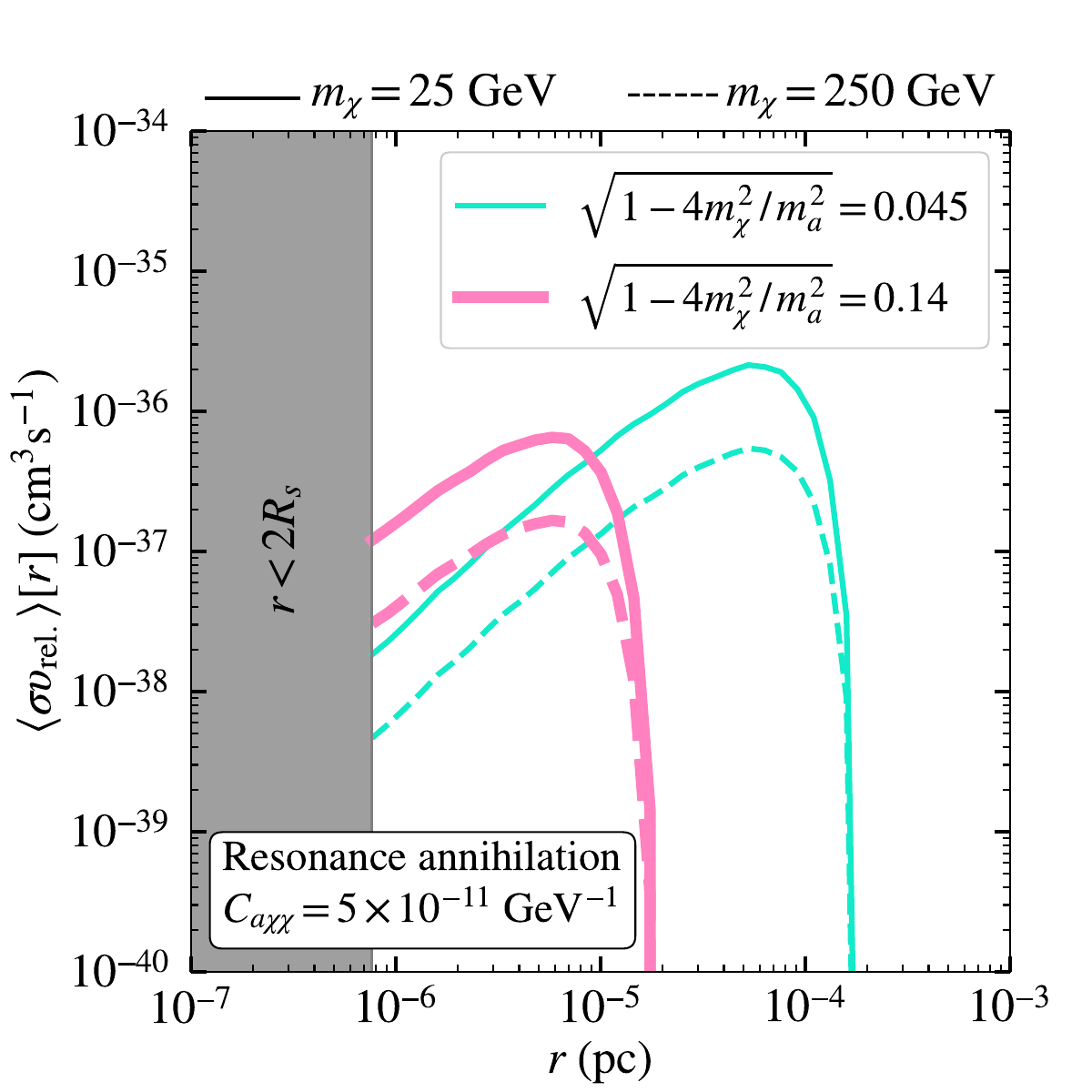}
\includegraphics[width=8.5cm]{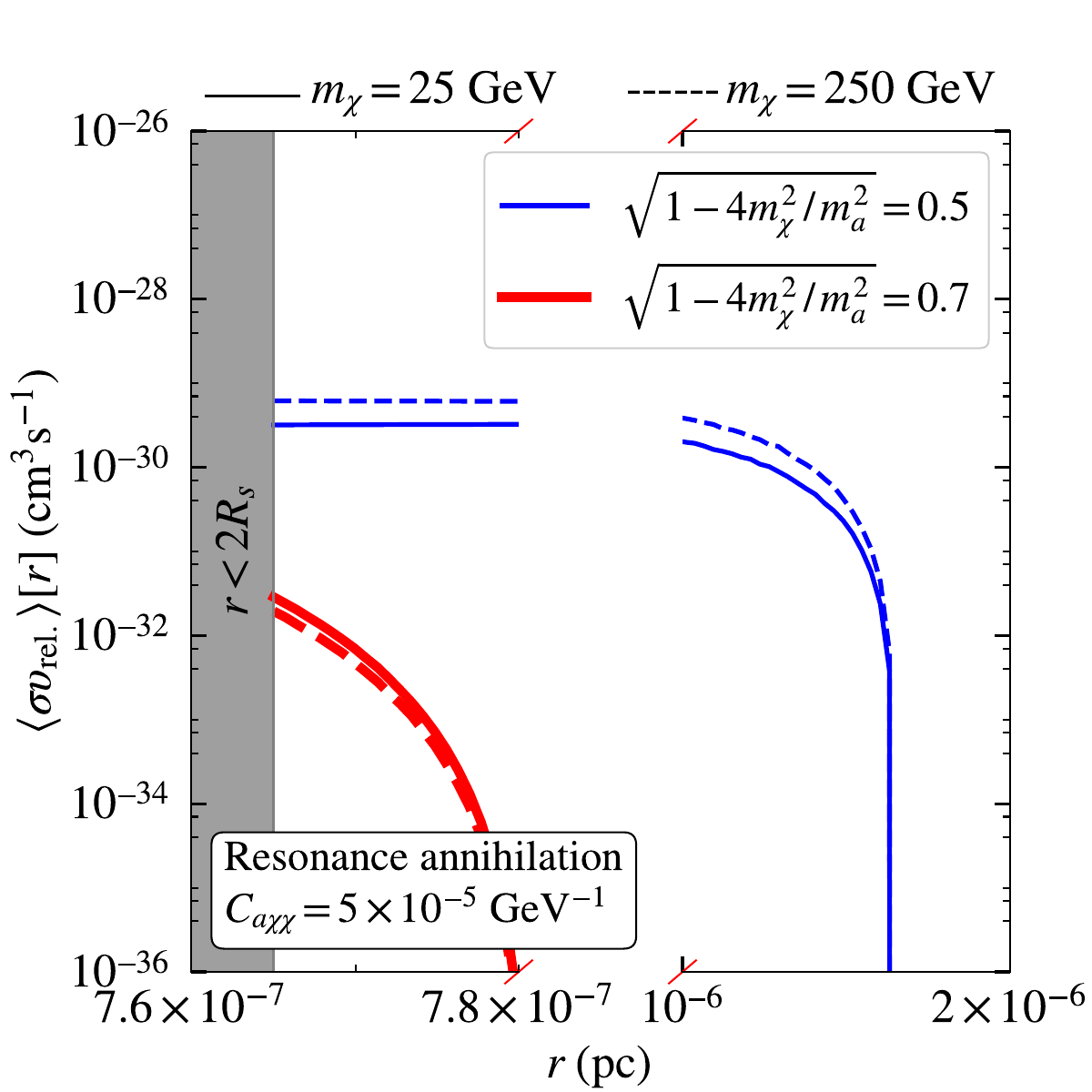}
\includegraphics[width=8.5cm]{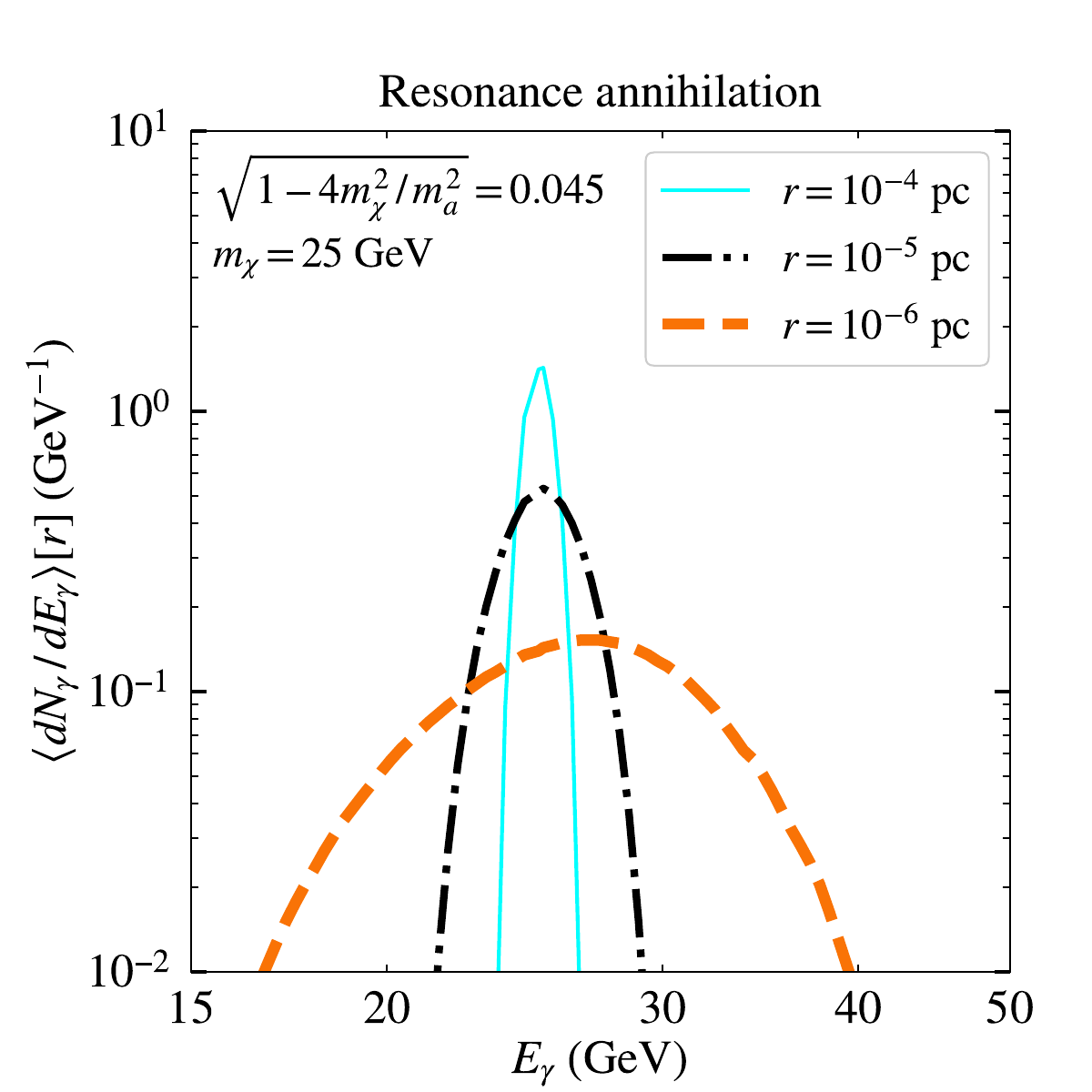}
\includegraphics[width=8.5cm]{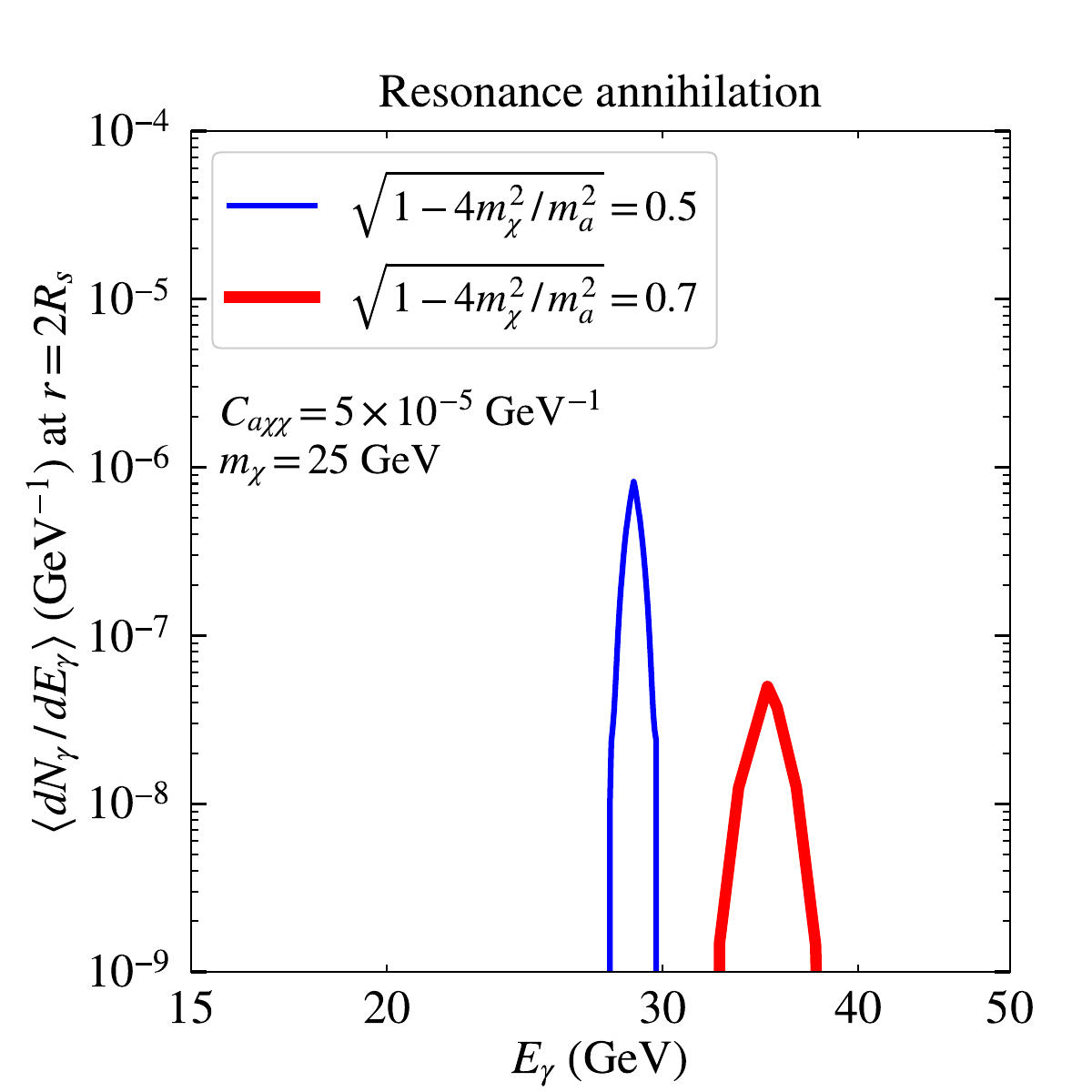}
 \caption{Upper panels: the velocity-averaged annihilation cross-sections of resonance scenario as a function of radius 
 for benchmarks in the upper left panel: \{$R_{a\chi}<0.2$, $C_{a\chi\chi}/\gev^{-1}=5\times 10^{-11}$\} and in the upper right panel: \{$R_{a\chi}>0.2$, $C_{a\chi\chi}/\gev^{-1}=5\times 10^{-5}$\}. 
 The solid and dash lines are $m_\chi=25\gev$ and $m_\chi=250\gev$, respectively. 
 Lower panels: the annihilation yield photon spectra in the resonance scenario. 
 The lower left panel corresponds to $m_\chi = 25 \gev$ and $R_{a\chi} = 0.045$, demonstrating the impact of different radii on the spectrum, 
 while the right panel corresponds to $m_\chi = 25 \gev$ and $r=2R_s$, by changing $R_{a\chi}$ on the spectrum. 
 }
 \label{sigma_ave}
\end{figure*}

The interactions in Eq.\eqref{eq:DM-ALP} and Eq.\eqref{eq:fullL} give the integrated cross-section $\sigma (\chi\overline{\chi}\to\gamma\gamma)$~\cite{Ibe:2008ye} as 
\begin{equation}
\label{eq:resonance Xsec}
    \sigma (\chi\overline{\chi}\to\gamma\gamma)=\frac{16 \pi}{s R_{a\chi}\beta_\chi}\frac{m_a^2 \Gamma_a^2}{(s-m_a^2)^2+m_a^2 \Gamma_a^2}B_i B_f,
\end{equation}
where $\Gamma_a=\Gamma_i+\Gamma_f$ is the total decay width of the mediator $a$, while $s\equiv E_{cm}^2$ is the center-of-mass energy squared of DM annihilation. The phase space factor $R_{a\chi}=\sqrt{1-4m_\chi^2/m_a^2}$ and $\beta_\chi=\sqrt{1-4m_\chi^2/s}$ are at the center of mass energy of the collision. 
The branching ratios of the mediator $a$ decaying into the initial/final-state particles are denoted as $B_{i,f}=\Gamma_{i,f}/\Gamma_{\rm tot.}$. In our study, $\Gamma_{i}=\Gamma(a\to\chi\overline{\chi})$ and $\Gamma_{f}=\Gamma(a\to\gamma\gamma)$. Each partial decay width of $a$ is calculated by closely following Refs.~\cite{Bao:2022onq,Allen:2024ndv}: 
\begin{equation}
    \begin{aligned}
    &\Gamma(a\to\chi\overline{\chi}) = \frac{(C_{a\chi\chi} m_{\chi})^2 m_a }{8\pi }\sqrt{1-\frac{4m_\chi^2}{m_a^2}},\\
    &\Gamma(a\to\gamma\gamma) = \frac{C_{a\gamma\gamma}^2m_a^3}{64\pi},\\
    &\Gamma(a\to\gamma Z) = \frac{C_{a\gamma Z}^2m_a^3}{32\pi}\left(1-\frac{m_Z^2}{m_a^2}\right)^3,\\
    &\Gamma(a\to Z Z) = \frac{C_{aZZ}^2m_a^3}{64\pi}\left(1-\frac{4m_Z^2}{m_a^2}\right)^{3/2},\\
    &\Gamma(a\to W^+ W^-) = \frac{C_{aWW}^2m_a^3}{32\pi}\left(1-\frac{4m_W^2}{m_a^2}\right)^{3/2},
   \end{aligned}
	\label{eq:decay}
\end{equation}
where $M_Z$ and $M_W$ are the masses of the $Z$ and $W$ bosons, respectively. 
With the assumption of $C_{BB}= C_{WW}$, the partial decay widths of $a\to\chi\overline{\chi}$, $a\to\gamma\gamma$, $a\to ZZ$, and $a\to W^+ W^-$ can be found in detail in Appendix~\ref{app:Gamma_a}. 
The decay mode $a\to\gamma Z$ vanished when $C_{BB}= C_{WW}$.

In the limit of $m_a \Gamma_a  \rightarrow 0$, the denominator in Eq.~\eqref{eq:resonance Xsec} can be written as 
\begin{equation}
\label{eq:limit1}
\displaystyle\lim_{m_a \Gamma_a  \rightarrow 0}\frac{1}{(s-m_a^2)^2+m_a^2\Gamma_a^2}=\delta(s-m_a^2)\frac{\pi}{m_a \Gamma_a}=
\delta(v^2_{\rm rel.}-4R^2_{a\chi})
\frac{\pi}{m_a \Gamma_a}\frac{(1-R_{a\chi}^2)^2}{m_\chi^2}. 
\end{equation} 
Finally, we use Eq.~\eqref{eq:limit1} to replace the denominator in Eq.~\eqref{eq:resonance Xsec} such that $\sigma v_{\rm rel}$ can be simplified as 
\begin{equation}
\label{eq:xsec_vrel_reson}
\sigma v_{\rm rel}=\frac{32\pi^2 m_a}{m_\chi^4 R_{a\chi}} \frac{\Gamma_{a\chi\overline{\chi}}\Gamma_{a\gamma\gamma}}{\Gamma_a} 
\delta(\beta_\chi^2 -R_{a\chi}^2)
(1-R_{a\chi}^2)^2(1-\beta_\chi^2).  
\end{equation}
Plugging Eq.~\eqref{eq:xsec_vrel_reson} into Eq.~\eqref{eq:avgXS}, 
we can obtain $\sv$ and confirm that it is mainly $s$-wave for $\beta_\chi\to 0$. 
Close to the BH, if $\beta_\chi$ approaches $R_{a\chi}$, we might observe a significant enhancement in the DM annihilation signal due to resonance.

In the upper two panels of Fig.~\ref{sigma_ave}, we depict $\sv$ as a function of $r$. 
In the left panels, we choose the non-relativistic resonance pole velocity, namely $R_{a\chi}<0.2$, and $C_{a\chi\chi}=5\times 10^{-11}/\gev$. 
The peak of $\sv$ shifts from $r\approx 10^{-5}\pc$ (pink lines) to $r\approx 10^{-4}\pc$ (aqua lines) when $R_{a\chi}$ changes from $0.14$ to $0.045$, respectively. 
When changing $m_\chi$ from $25\gev$ (solid lines) to $250\gev$ (dashed lines), $\sv$ can be altered by a factor of $4$, because $a\to ZZ$ and $a\to WW$ open only for $m_a>200 \gev$ and the decay branching ratio of $a\to\chi\bar{\chi}$ decreases by a factor of $4$. 

By contrast, the right panels of Fig.~\ref{sigma_ave} depict a scenario where $R_{a\chi}$ is approximately half the speed of light, 
resulting in resonant annihilation only near $2R_s$. 
The high values of $R_{a\chi}$ suggest that DM particles annihilate at small radii. 
For instance, for $R_{a\chi}=0.5$ (blue lines) or $R_{a\chi}=0.7$ (red lines), DM particles annihilate only at $r < 1.5 \times 10^{-6} \pc$ or $r < 7.8 \times 10^{-7} \pc$, respectively. 
Note that a larger $C_{a\chi\chi}=5\times 10^{-5}/\gev$ is required for $R_{a\chi}\sim\mathcal{O}(0.5c)$ to produce detectable photon fluxes. 
The case with $R_{a\chi}=0.5$ shows a smoother variation of $\sv$ with radius than $R_{a\chi}=0.7$ when $r<10^{-6} \pc$,
because the condition $R_{a\chi}=0.5$ results a larger cross-section after forming the spike to truncate  
the high-velocity tail in the velocity distribution within $r_{\rm ann}$ and suppress the resonance velocity phase space distribution. 
Moreover, in the upper right panel, the $\sv$ for the solid blue line ($m_\chi=25\gev$ and $R_{a\chi}=0.5$) is lower than the one  for the dashed blue line ($m_\chi=250\gev$ and $R_{a\chi}=0.5$), while the red lines show the opposite trend, attributed to modifications in the velocity distribution. 
The blue solid line, associated with a larger $r_{\rm ann}$, predicts a mild suppression in the velocity distribution, leading to a decrease in $\sv$.

In addition, $\sv[r]$ of the resonance scenario peaks at the radius where the pole velocity is matched with the mean of velocity distribution. 
For smaller $R_{a\chi}$ values (the upper left panel of Fig.~\ref{sigma_ave}), under the condition $T_{\rm ann}=m_\chi/\left[\rho_{\rm spike}(r) \langle\sigma v\rangle\right]\approx 10^{10}~{\rm yrs}$, two annihilation radii can be found: 
the smaller radius denoted as $r_{\rm ann}^{\rm min}$ and the larger radius denoted as $r_{\rm ann}^{\rm max}$. 
Only particles moving entirely within the region $r_{\rm ann}^{\rm min} \le r \le r_{\rm ann}^{\rm max}$ of the Milky Way are annihilated. 
Therefore, the distribution function in the resonance annihilation, based on Eq.~\ref{eq:DF}, has to be rewritten as 
\begin{equation}
    f=\left\{
        \begin{array}{ll}    
        f'(\epsilon',L'),& \quad  0\leq \epsilon' \leq \frac{{\rm G}M_{\rm BH}}{r_{\rm ann}} \quad {\rm or} \quad \epsilon' > \frac{{\rm G}M_{\rm BH}}{r_{\rm ann}^{min}}, \\
    0,& \frac{{\rm G}M_{\rm BH}}{r_{\rm ann}^{max}}< \epsilon' < \frac{{\rm G}M_{\rm BH}}{r_{\rm ann}^{min}}, 
        \end{array}
    \right.
\end{equation}

In the lower left panel of Fig.~\ref{sigma_ave}, the shape of $\avgdnde$ changes with radius. 
At low velocities, the photon spectrum appears as a monochromatic line, as seen at $r= 10^{-4}\pc$ (cyan solid line). 
However, as DM particles approach the BH and accelerate to higher velocities, like at $r=10^{-6}\pc$ (orange dashed line) and $r=10^{-5}\pc$ (black dash-dotted line), the spread of $\avgdnde$ widens. 
If not considering DM particles boosted by the BH, we would expect to observe a monochromatic line spectrum for the resonance annihilation scenario $\chi\overline{\chi}\to \gamma\gamma$. 
However, once we consider the effects of acceleration from the BH gravitational potential, the spectrum cannot be treated as a line.
In the right panels of Fig.~\ref{sigma_ave}, resonant annihilation mainly occurs near $2R_s$, highlighting the widest $\avgdnde$ at $r=2 R_s$. 
The peak of $\avgdnde$ occurs at $m_a/2$, shifting to higher energies with increasing $R_{a\chi}$. 
For parameters $C_{a\chi\chi} = 5\times 10^{-5}/\gev$, $m_\chi=25\gev$, and $R_{a\chi}=0.5$ (depicted by the blue line), 
the high-velocity tail of the velocity distribution disappears, resulting in a narrower spectral line.

\textit{After summing all of the $\avgdnde$ values at different radii along the line of sight, 
the overall spectrum for a small $R_{a\chi}$ may appear softer than what is typically expected for $s$-wave annihilation.} 
Still, as shown in the right panels of Fig.~\ref{sigma_ave}, the overall spectrum for a large $R_{a\chi}$ behaves like a spectral line.

\subsection{Forbidden annihilation}
\label{sec:forbidden}

\begin{figure*}[ht]
\centering
\includegraphics[width=8.5cm]{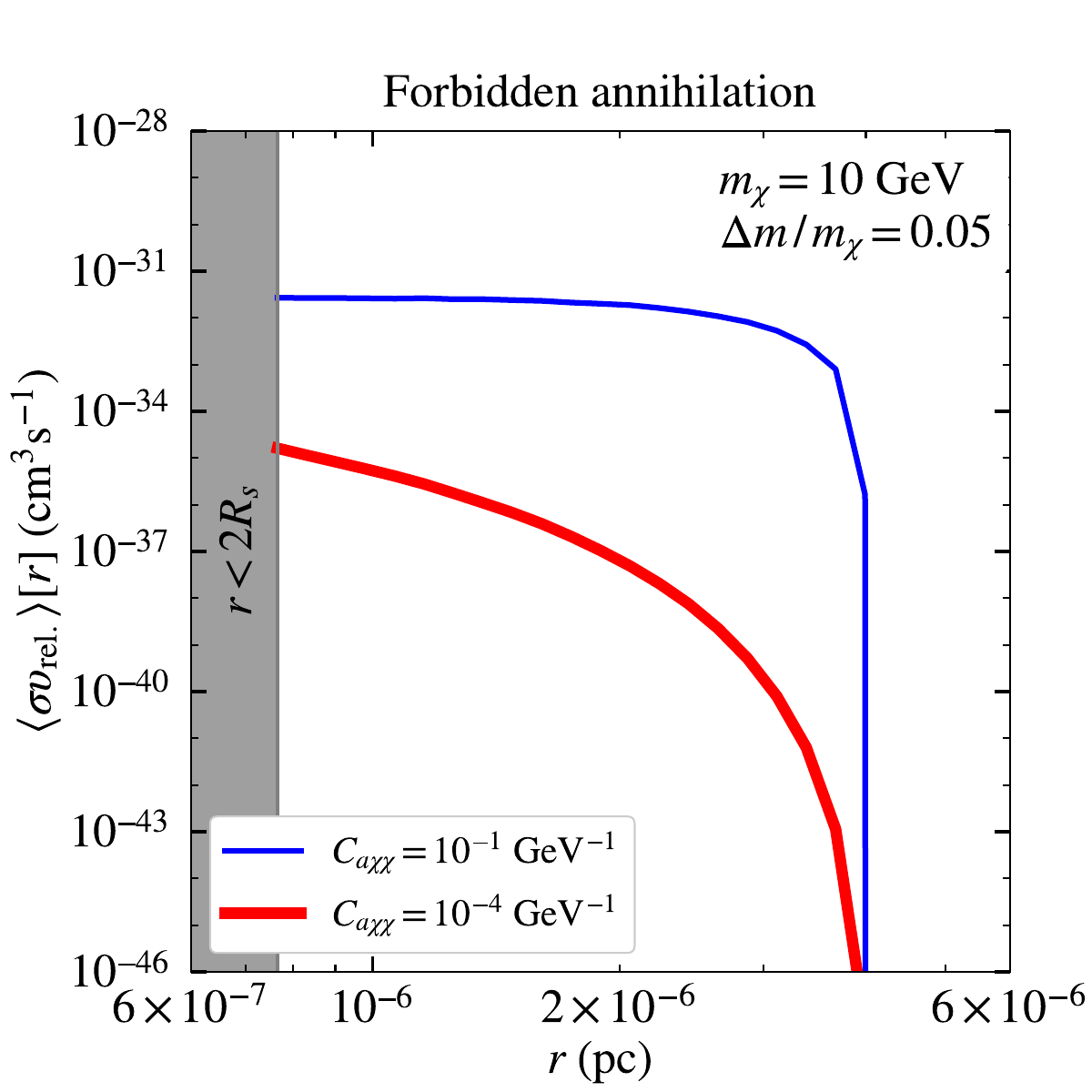}
\includegraphics[width=8.5cm]{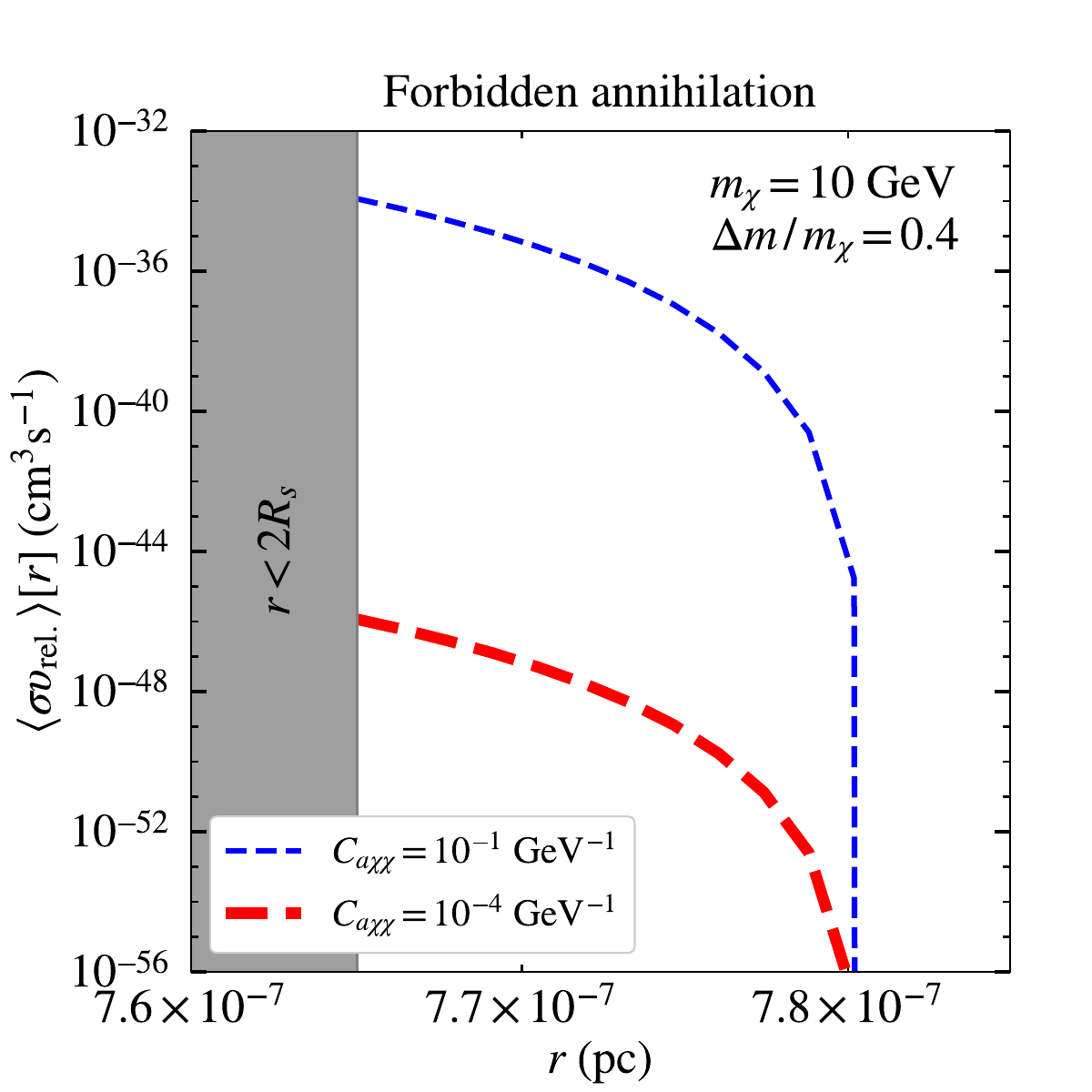}
\includegraphics[width=8.5cm]{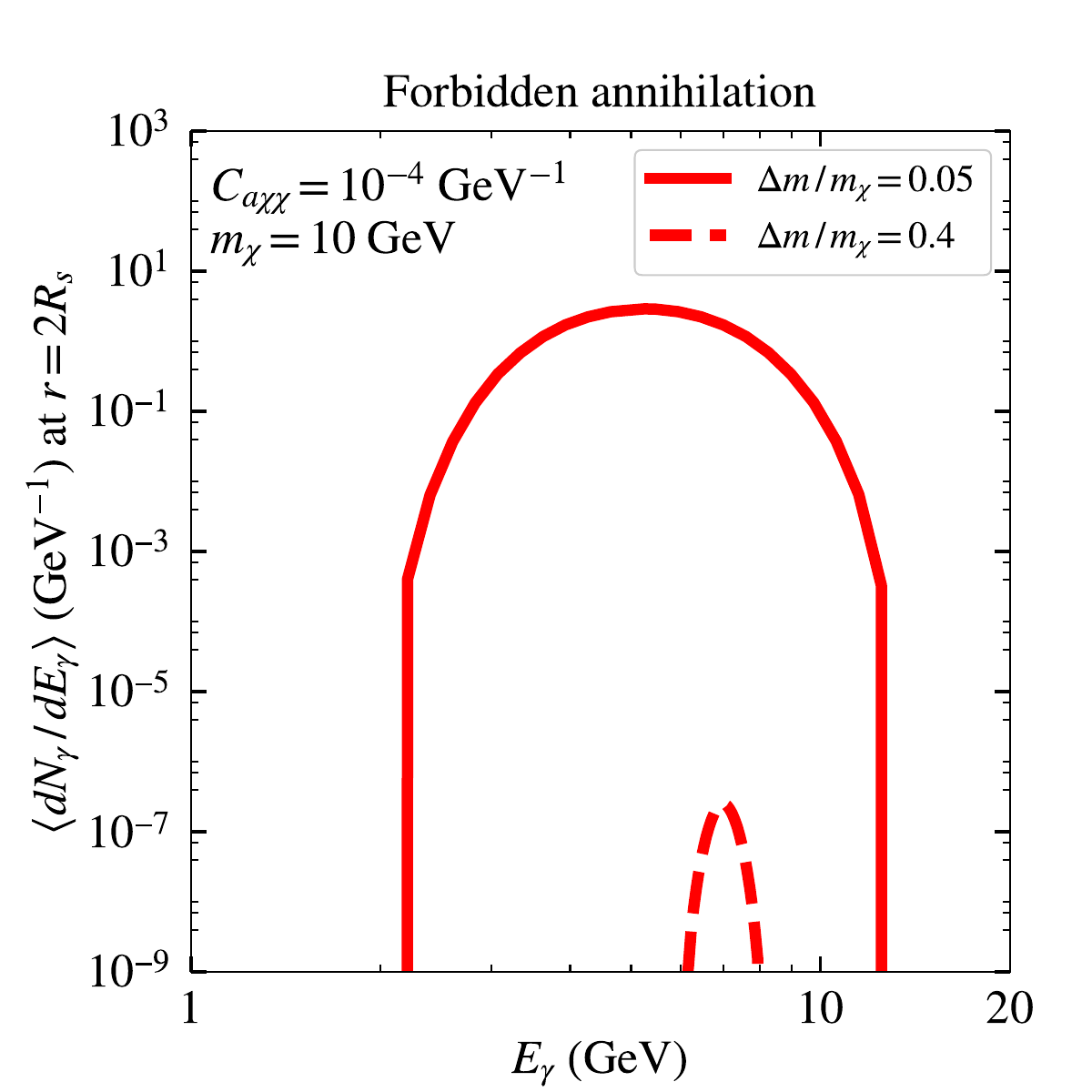}
\includegraphics[width=8.5cm]{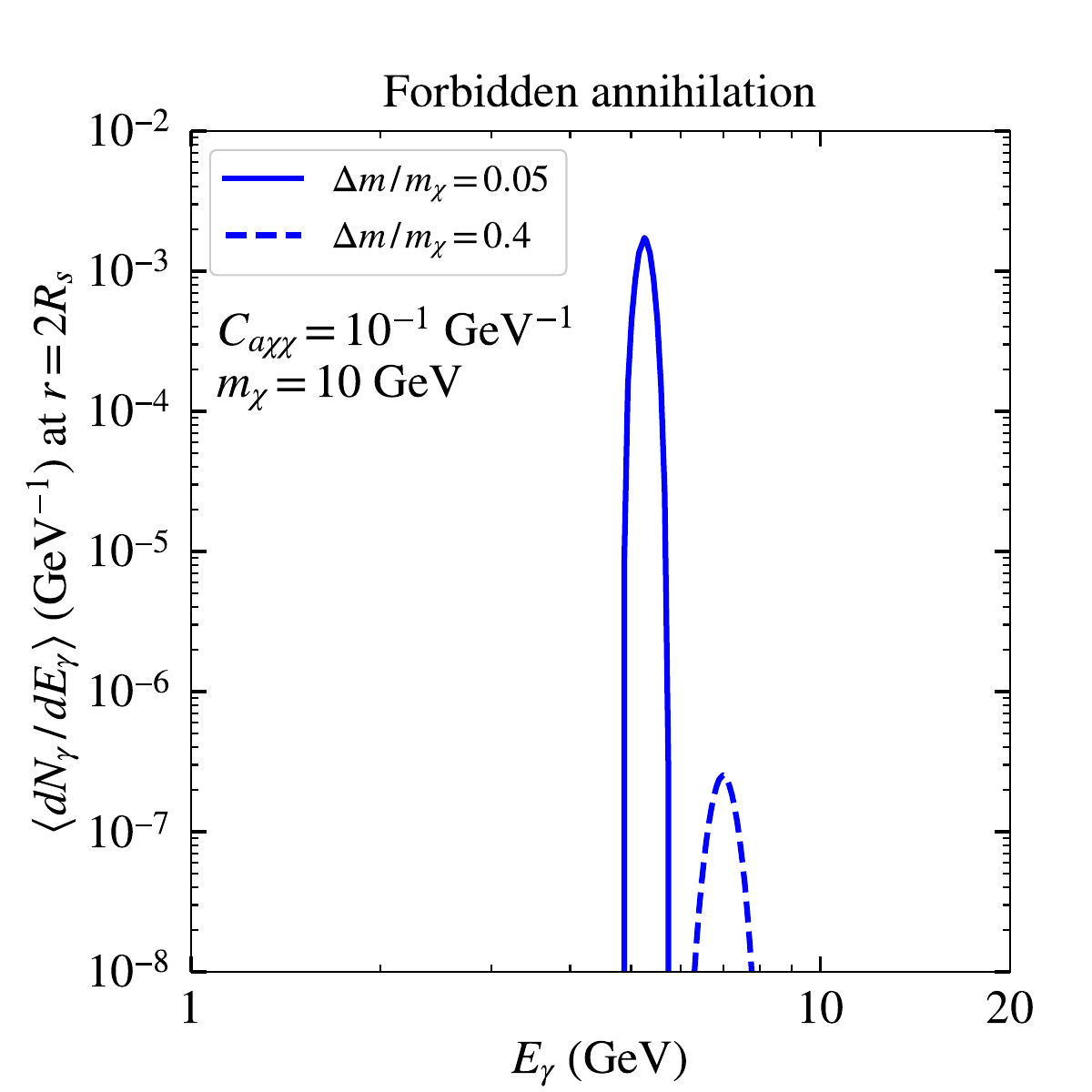}
 \caption{The velocity averaged annihilation cross-sections (upper panels) and photon spectra (lower panels). 
In all panels, solid and dashed lines represent $\Delta m/m_{\chi}=0.05$ and $\Delta m/m_{\chi}=0.4$, 
while red and blue lines indicate two benchmark couplings, $C_{a\chi\chi}=10^{-4}/\gev$ and $C_{a\chi\chi}=10^{-1}/\gev$. 
The spectra $\avgdnde$ in the lower right panel also depend on $C_{a\chi\chi}$, because $C_{a\chi\chi}=10^{-1}/\gev$ is sufficiently large to modify the halo profile. }
 \label{fig:sigma_forbidden}
\end{figure*}

In most instances,
the forbidden DM annihilation cannot be turned on in the present-day Universe, due to $m_\chi<m_a$. 
Only those DM particles are very close to the BH and can be accelerated to open the annihilations $\chi\overline{\chi}\to a a$. 
Because their $\sv$ are suppressed by narrow phase space, a larger $C_{a\chi\chi}$ than the values used in $m_\chi>m_a$ cases may be required to reach the sensitivities of gamma-ray telescopes.  
Furthermore, a large $C_{a\chi\chi}$ also implies a large $r_{\rm ann.}$ and less spike density, resulting in an efficient annihilation.

In the upper two panels of Fig.~\ref{fig:sigma_forbidden}, we compare the values of $\sv$ for $C_{a\chi\chi}=0.1/\gev$ (blue lines) and $C_{a\chi\chi}=10^{-4}/\gev$ (red lines) with $\Delta m/m_\chi=0.05$ in the upper left panel and $\Delta m/m_\chi=0.4$ in the upper right panel. 
Our benchmarks exhibit two interesting features. 
First, the spectra decrease sharply at $r\approx 4\times 10^{-6}\pc$ in the upper left panel and $r\approx 7.8\times 10^{-7}\pc$ in the upper right panel, corresponding to the values of $\Delta m/m_\chi$. 
Beyond this range, the DM particles cannot annihilate into ALPs because they are kinetically forbidden. 
A higher value of $\Delta m/m_\chi$ leads to a smaller value of $\sv$ with the same $C_{a\chi\chi}$ value. 
When comparing the $C_{a\chi\chi}=10^{-4}/\gev$ lines in the two upper panels, even a slight change in the $\Delta m/m_\chi$ value can result in a ten-order difference in $\sv$ at $r=2R_s$.       
Second, for $C_{a\chi\chi}= 0.1/\gev$ and $\Delta m/m_\chi=0.05$ illustrated by the solid blue line in the upper left panel, the value of $\sv$ only varies slightly within $r_{\rm ann}$. 
As discussed in Fig.~\ref{eq:f_v_r}, the velocity distributions for the region $r<r_{\rm ann}$ peaks near the escape velocity at $r_{\rm ann}$. 
Such a semi-relativistic velocity also largely enhances the annihilation cross-section, truncating the high-speed tail of the velocity distribution.

In the forbidden annihilation scenario, we only display the widest $\avgdnde$ at $r=2 R_s$ in Fig.~\ref{fig:sigma_forbidden}.  
Since the gravitational potential farther from the BH provides less acceleration to the DM particles, DM particle annihilation only happens at the region near $2R_s$. Hence, the distribution width of $\avgdnde$ narrows in $r\gg 2R_s$.

In the lower panels of Fig.~\ref{fig:sigma_forbidden}, larger values of $\Delta m/m_\chi$ require higher velocities for DM particle annihilation. 
This leads to a weaker Lorentz boost of ALPs and a narrower spectrum. In the lower left panel, with fixed $C_{a\chi\chi}=10^{-4}/\gev$, a larger $\Delta m/m_\chi$ (red dashed line) suppresses the spectrum due to kinematic constraints. 
For the annihilation $\chi\overline{\chi} \to aa \to 4\gamma$, the relative velocity $v_{\rm rel.}$ must exceed $2 \sqrt{1 - m_\chi^2/m_a^2}$. 
Increasing $\Delta m/m_\chi$ reduces phase space in the velocity distribution, resulting in fewer average photons per annihilation and a peak shift to higher energies.

In the lower right panel, for a higher value of $C_{a\chi\chi}=10^{-1}/\gev$, the spectral shapes of two cases are constrained to a narrow range of $E_\gamma$. 
The case with $\Delta m/m_\chi = 0.4$ (blue dashed line) shows no change in the distribution and aligns with $C_{a\chi\chi} = 10^{-4}/\gev$ (red dashed line). 
However, the spectrum of the compressed mass case $\Delta m/m_\chi = 0.05$ (blue solid line) almost behaves like a delta function.

\section{Gamma-Ray Flux from GC}
\label{sec:GaFlux}

\begin{figure*}[ht]
\centering
\includegraphics[width=10.0cm]{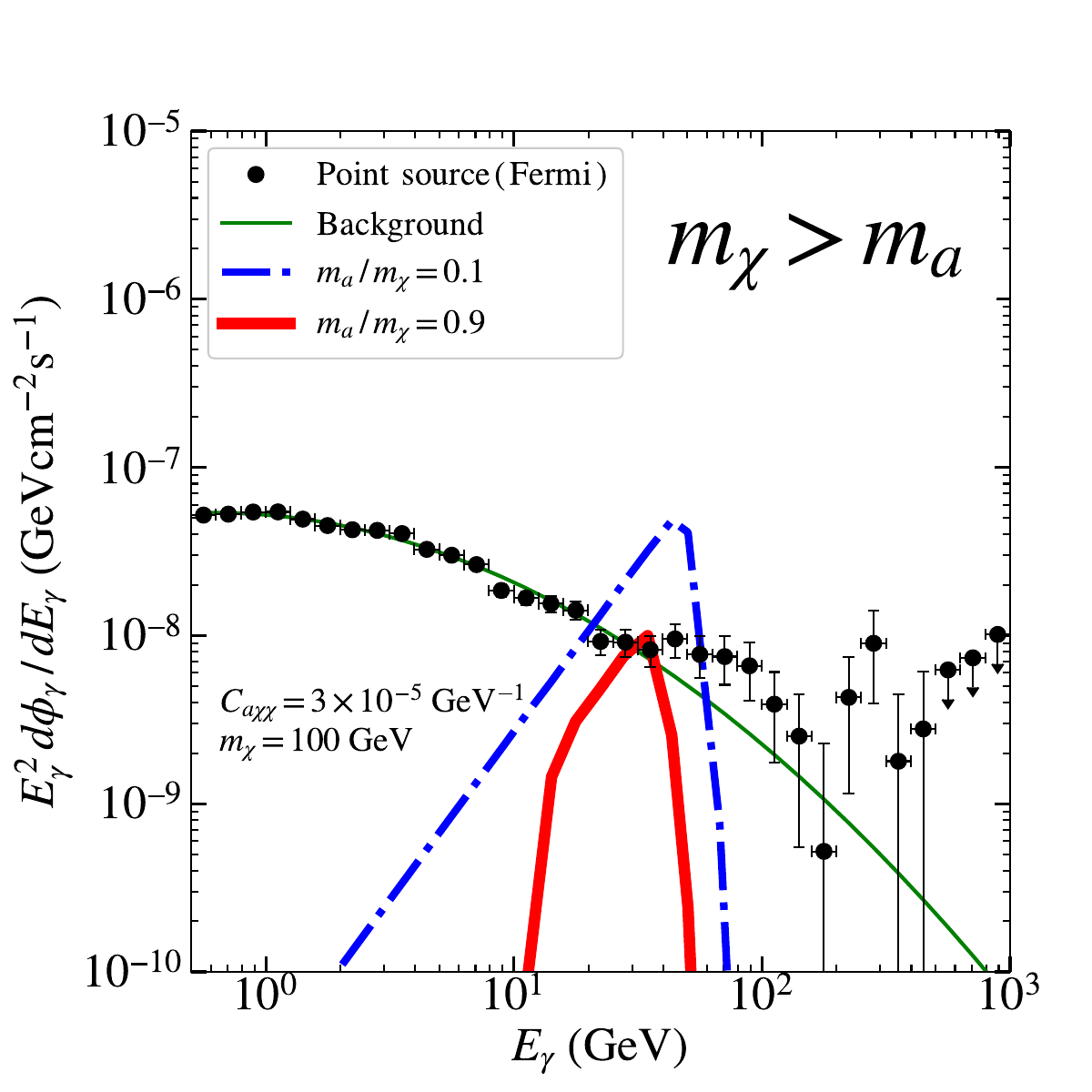}
 \caption{The photon energy spectra of $m_{\chi} > m_{a}$ scenario, based on $m_\chi=100\gev$ and $C_{a\chi\chi}/\gev^{-1}=3\times 10^{-5}$, 
 in comparison with Fermi-LAT spectral energy distribution (SED, black cycles). 
 The green lines denote the background. Two benchmarks are $m_a/m_\chi=0.1$ (blue dash-dotted line) and $m_a/m_\chi=0.1$ (red solid line). 
 }
 
 \label{fig:flux1_p_r}
\end{figure*}

The general form of gamma-ray differential flux density from DM annihilation near Sgr A$^{\star}$ is 
\begin{equation}\label{VD_Eq1}
\begin{aligned}
    \frac{d\phi_{\gamma}}{dE_{\gamma}}= 
    \frac{1}{4m_{\chi}^{2} D^2} &\times 
    \int_{2R_{\rm s}}^{0.4\kpc} \rho^2_\chi(r) r^2 dr
    \int_0^{2v_{\rm esc}} \frac{v_{\rm rel.}^2}{N_0} d{v}_{\rm rel.} 
    \int_0^{v_{\rm c}} v_{\rm CM}^2 dv_{\rm CM}  
    \int_{-1}^1 d\cos\alpha
    \int_0^{2 \pi} d\phi  \\ 
   \times&\int_{-\mu_0}^{\mu_0} d\cos\theta \times     
   \frac{dN_{\gamma}}{dE_{\gamma}} \times
    \sigma v_{\rm rel.}  \times 
    f_1(v_{\rm rel.}, v_{\rm CM} ,L_1,r)\times f_2(v_{\rm rel.}, v_{\rm CM},L_2,r),\\
    &{\rm with}\quad v_{\rm c}=
    \sqrt{1-\frac{s}{E_{\rm esc}^2}} 
    \quad {\rm and} \quad E_{\rm esc}=\frac{2m_\chi}{\sqrt{1-v_{\rm esc}^2}}.
\end{aligned}
\end{equation}
Here, $D=8.5\kpc$ is the distance from Sgr A$^{\star}$ to Earth.  
In this work, we treat Sgr A$^{\star}$ as a point source, thus we can simply integrate $r$ up to $0.4\kpc$, corresponding to the degree of $3$ of the region of interest. 
However, our three benchmark annihilation channels only trigger at a region near the Sgr A$^{\star}$. This paper utilizes Lepage’s VEGAS python library~\cite{vegas} for numerical integration.

In some studies, for example, Ref.~\cite{Yuan:2021mzi}, 
a large velocity-independent $\sv$ counter-intuitively results in a small $\frac{d\phi_{\gamma}}{dE_{\gamma}}$, 
owing to the spike density largely suppressed by large $\sv$, as shown in Fig.~\ref{fig:rho}.        
This does not appear in the case of our three annihilation scenarios. 
For $m_\chi > m_a$ annihilation, $\sv$ decreases as $r$ increases, which does not significantly reduce spike density,
especially under the simple perturbative limit $m_\chi C_{a\chi\chi}<4\pi$. 
Similarly, $\sv$ is generally small in the spike region for both the resonance and forbidden DM scenarios.   
Nevertheless, in a specific velocity range, resonance annihilation can be enhanced, or forbidden annihilation becomes available.

Fig.~\ref{fig:flux1_p_r} shows the predicted gamma-ray continuum spectra for the scenario $m_\chi>m_a$, 
and the ratio $m_a/m_\chi$ determines the shapes of the energy spectra. 
The spectrum represented by the dash-dotted  blue line ($m_a/m_\chi=0.1$) is broader than the one represented by the solid red line ($m_a/m_\chi=0.9$).
When comparing the log-parabola background spectrum (solid green lines) with the Fermi continuum spectrum data (black error bars) for $10\gev<E_\gamma<100\gev$, the background spectrum does not fit the data very well. 
This may require a contribution from some unknown sources. 
Considering DM annihilation as the candidate solution, only the scenario $m_\chi>m_a$ can accommodate this due to its smoother spectrum.

\begin{figure*}[ht]
\centering
\includegraphics[width=7.5cm]{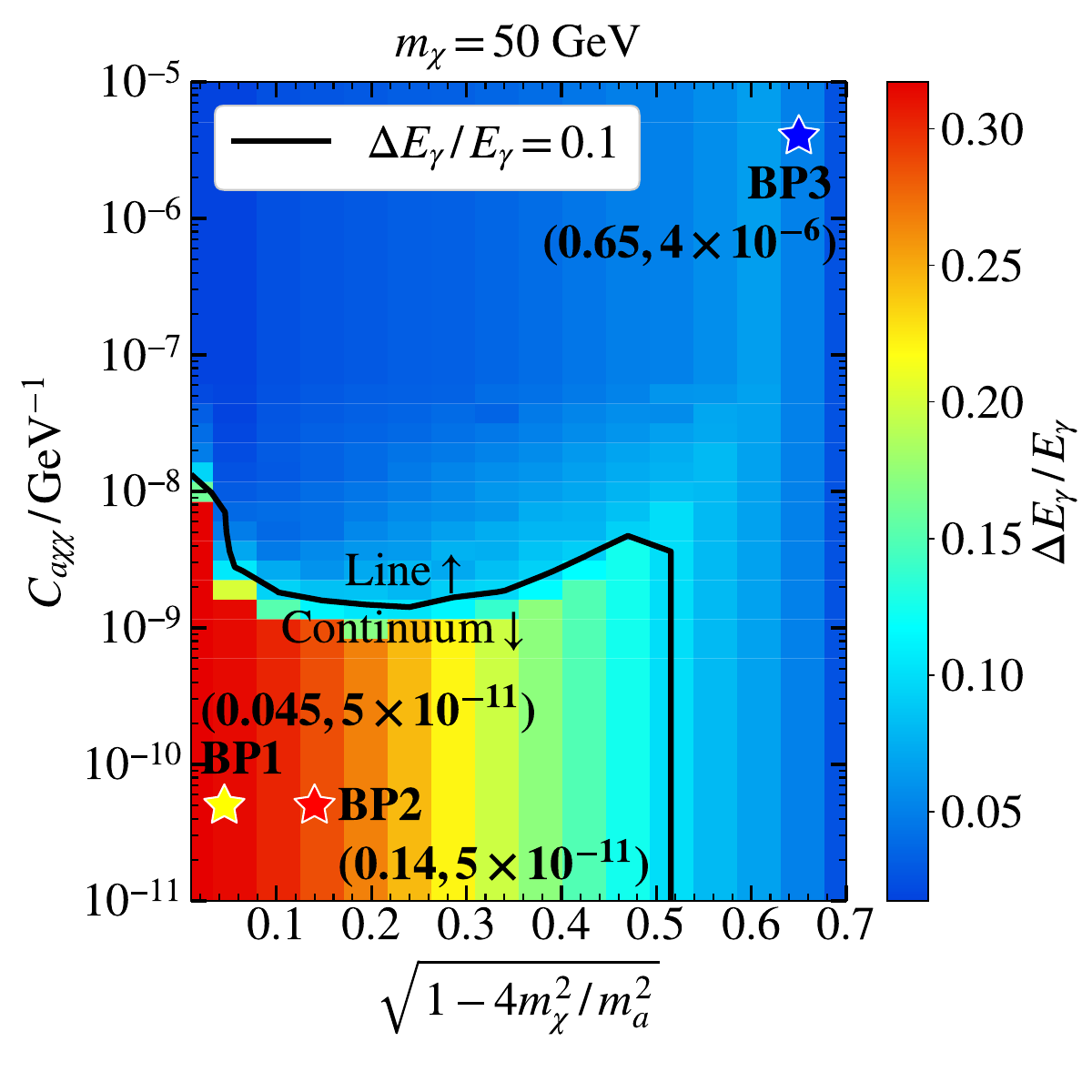}
\includegraphics[width=7.5cm]{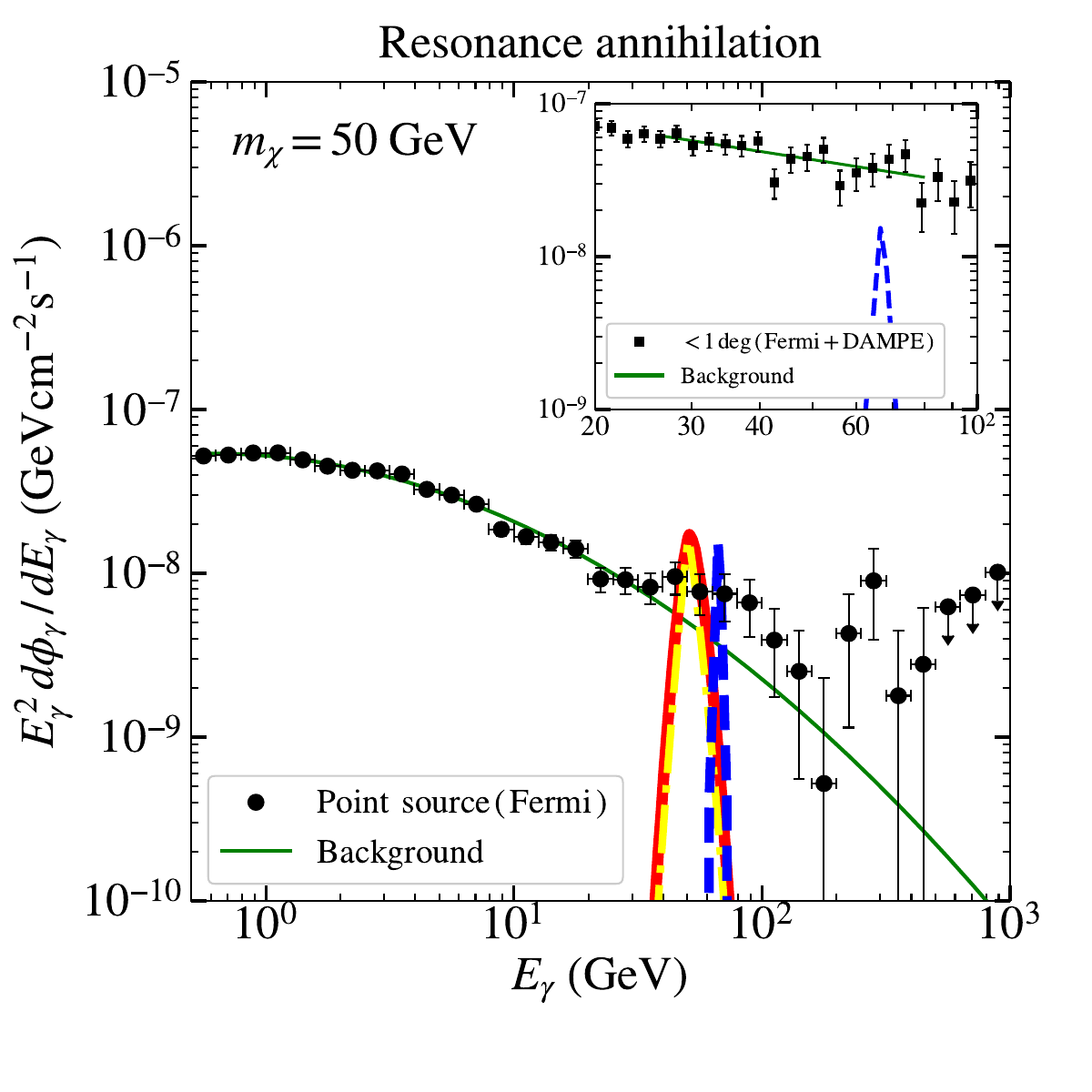}
\includegraphics[width=7.5cm]{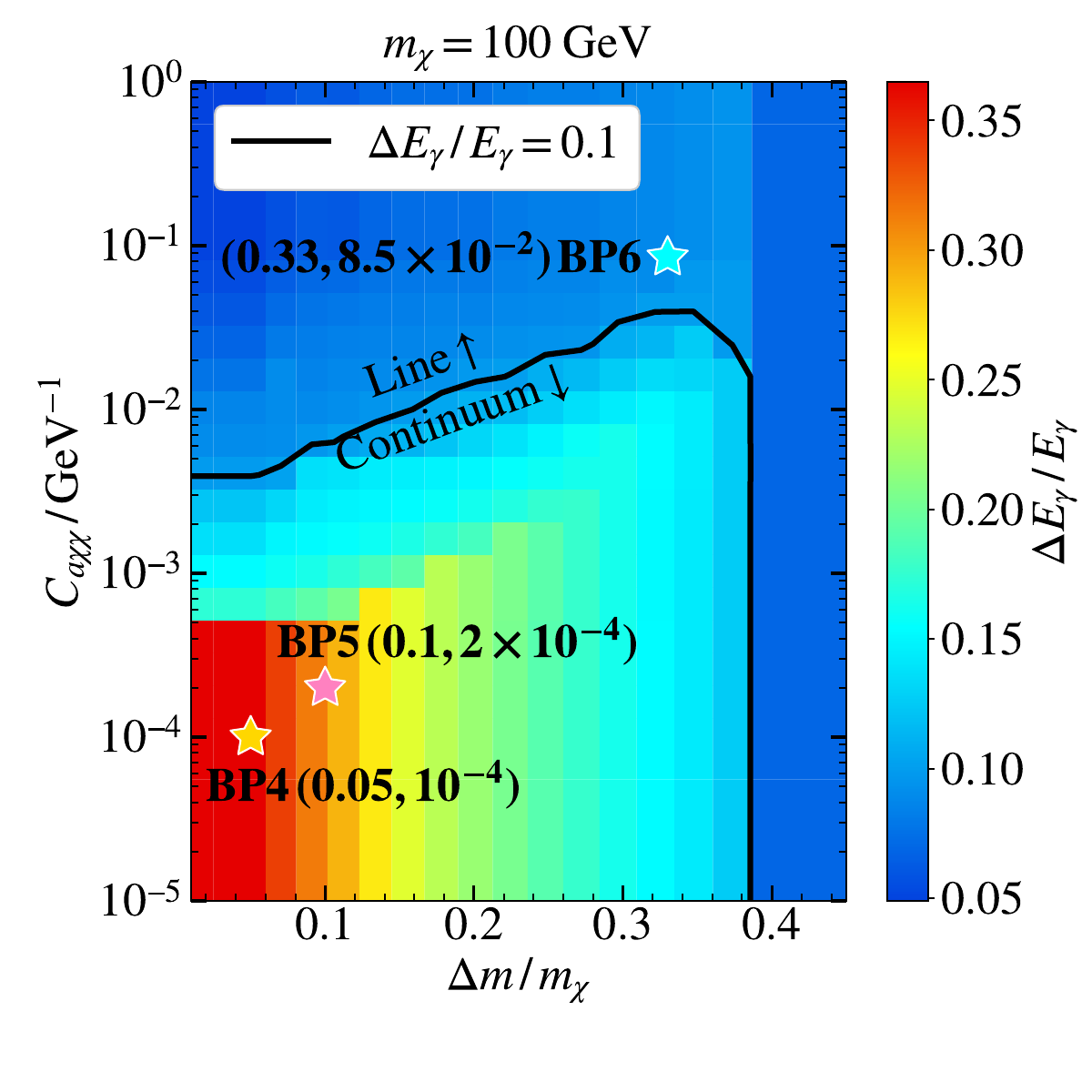}
\includegraphics[width=7.5cm]{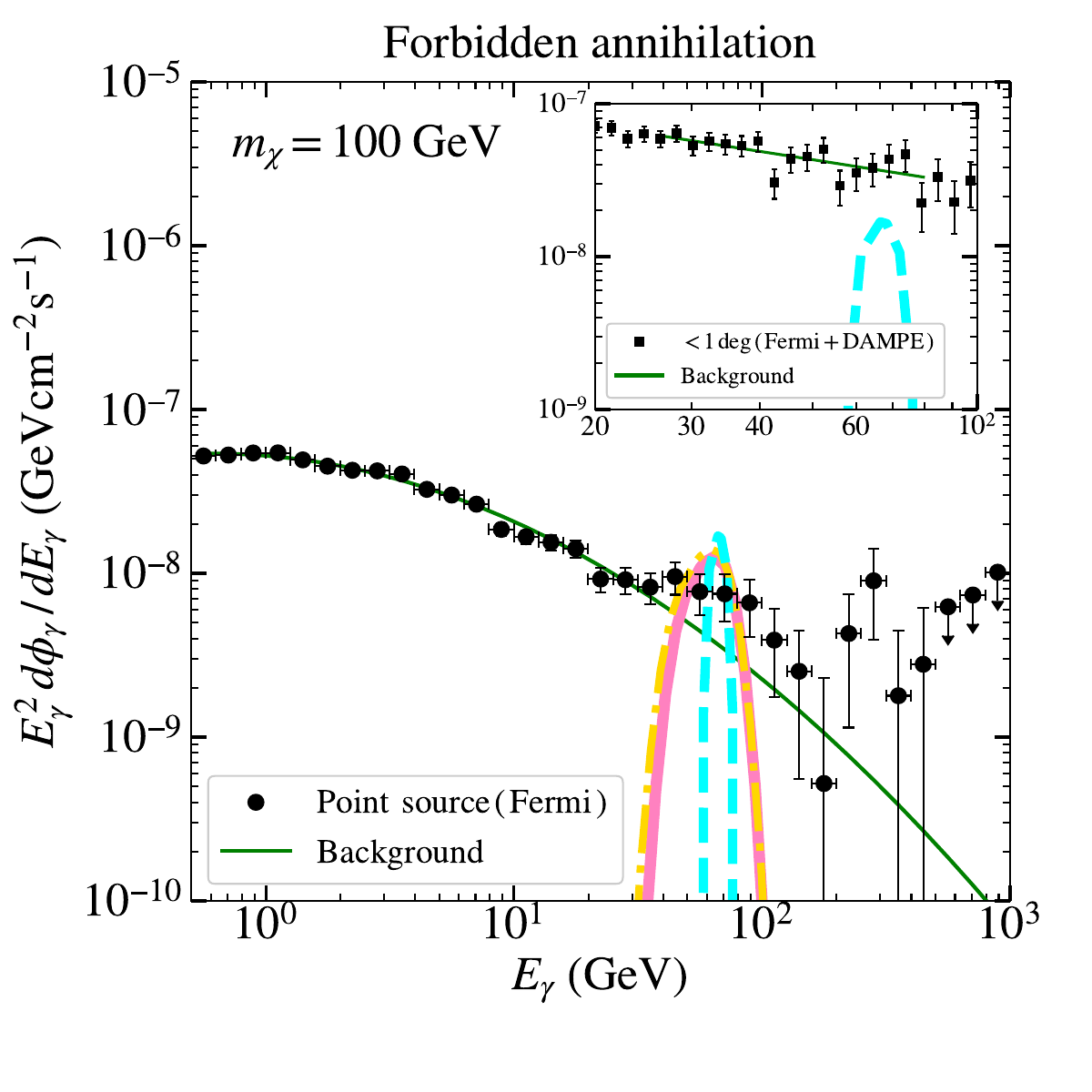}
 \caption{
The $\Delta E_\gamma/E_\gamma$ maps (left panels) and energy spectra (right panels) in $a$-resonance (upper right), and forbidden annihilation (lower right) scenarios. 
The energy resolution of Fermi and DAMPE is $\Delta E_\gamma/ E_\gamma=0.1$ (black line) to distinguish a line or continuum spectrum (see the arrows in the left panels).
For the demonstration of fluxes in the right panel, six benchmark stars are chosen with numbers given in Eq.~\eqref{eq:BP123} and Eq.~\eqref{eq:BP456}.
In the right panels, the black cycles represent the Fermi-LAT spectral energy distributions (SEDs) obtained by standard likelihood methods and black squares in the inset plots represent the SEDs derived by the aperture photometry method for the Fermi+DAMPE data within 1\degree. (In the line spectrum analysis, we do not directly use the SEDs derived by aperture photometry method. See Sec.\ref{sec:line analysis} for more details.)
 }
 \label{fig:flux_p_r}
\end{figure*}

Fig.~\ref{fig:flux_p_r} illustrates the $\Delta E_\gamma/ E_\gamma$ map (left panels) and predicted gamma-ray spectra (right panels) in two more complicate scenarios: resonance annihilation (upper panels) and forbidden DM annihilation (lower panels). 
These cases exhibit either line or continuum spectra for different choice of model parameter configuration. 
The map of $\Delta E_\gamma/ E_\gamma$ is projected to the ($R_{a\chi}$, $C_{a\chi\chi}/\gev^{-1}$) plane for resonance scenario, but 
the ($\Delta m/m_\chi$, $C_{a\chi\chi}/\gev^{-1}$) plane for forbidden annihilation scenario. For the former one, the flux spectrum approximates a triangular shape, and  $\Delta E_\gamma$  is calculated as the integrated flux divided by the peak flux.
For the later one, the flux spectrum approximates a box shape, where $\Delta E_\gamma$ is defined as the integrated flux divided by twice the peak flux. 
The solid black line indicates $\Delta E_\gamma/ E_\gamma\ge 0.1$ (the energy resolution of gamma-ray telescopes) to split the parameter spaces for a monochromatic line (above the black line) and for a continuum spectrum (below the black  line). 
Ideally, different detection methods for different regions can be employed to optimize the signal.

In Fig.~\ref{fig:flux_p_r}, the left panels highlight six benchmark points (\textbf{BP1} to \textbf{BP6}) illustrating their respective parameters.
Their corresponding predicted spectra appear in the right panels.
In the resonance scenario, the spectral shapes of \textbf{BP1} (indicated by the yellow star and line) and \textbf{BP2} (indicated by the red star and line) exhibit a continuum spectrum, whereas \textbf{BP3} (indicated by the blue star and line) shows a line spectrum.
For a fixed $m_\chi=50\gev$, the explicit parameter configurations $\{R_{a\chi}, C_{a\chi\chi}/\gev^{-1}\}$ of these three benchmarks are 
\begin{equation}\label{eq:BP123}
    \textbf{BP1}:\{0.045,5\times 10^{-11}\}, \quad
    \textbf{BP2}:\{0.14,5\times 10^{-11}\}, \quad
    \textbf{BP3}:\{0.65,4\times 10^{-6}\}.
\end{equation} 
Similarly, in the forbidden DM scenario, \textbf{BP4} (gold), \textbf{BP5} (pink), and \textbf{BP6} (cyan) exhibit spectra in the lower panels. 
For the forbidden DM scenario with $m_\chi=100\gev$, \textbf{BP4} and \textbf{BP5} show continuum spectra, whereas \textbf{BP6} predicts a spectral line, with parameters:  
\begin{equation}\label{eq:BP456}
    \textbf{BP4}:\{0.05,10^{-4}\}, \quad
    \textbf{BP5}:\{0.1,2\times 10^{-4}\}, \quad
    \textbf{BP6}:\{0.33,8.5\times 10^{-2}\}.
\end{equation} 
In the inset figures of the right panels, spectra for \textbf{BP3} and \textbf{BP6} are depicted along with combined upper limits from Fermi and DAMPE line spectrum analysis (see the analysis in Sec.~\ref{sec:analysis}).

In the resonance annihilation scenario and the forbidden annihilation scenario, gamma-ray fluxes decrease as $R_{a\chi}$ and $\Delta m/m_\chi$ increase, respectively, due to a reduction in the annihilation cross-section caused by phase-space suppression. 
A large values of $R_{a\chi}$ and $\Delta m/m_\chi$ can cease annihilation, requiring DM particles to be accelerated to higher speeds for compensation. 
Consequently, such DM particles can only annihilate nearer to the BH capture radius $2 R_s$, where a lower DM density can result in fewer gamma-ray fluxes. This suppression can be realized particularly in the forbidden annihilation scenario; Even if taking $C_{a\chi\chi} m_\chi=4\pi$, gamma-ray fluxes from DM annihilation with $\Delta m/m_\chi>0.35$ in this scenario are lower than those from background sources.



\section{Data analysis}
\label{sec:analysis}
\subsection{Constraints from Fermi-LAT SED}
Launched on June 24, 2008, the Fermi-LAT is a high-performance gamma-ray telescope with the energy range from 30 MeV to 1 TeV~\citep{2021AjelloApJS}.
Here we use the nearly 15 years of Fermi-LAT P8R3 data from October 27, 2008 to December 18, 2023 to obtain the spectral energy distribution (SED) of Sgr A$^{\star}$.
In the Fourth Fermi-LAT source catalog~\citep{2020AbdollahiApJS} (4FGL), the point source 4FGL J1745.6-2859 is associated with Sgr A$^{\star}$.
We select the {\tt SOURCE} class events from 500 MeV to 1 TeV within the 14\degree $\times$ 14\degree~region centered at the direction of 4FGL J1745.6-2859 and divide events into 33 evenly spaced logarithmic energy bins. 
We exclude events with zenith angles larger than 90\degree~to reduce the contamination from the Earth’s limb and extract good time intervals with the quality-filter cut {\tt (DATA\_QUAL>0 \&\& LAT\_CONFIG==1)} .
To calculate the SED, we perform the standard binned likelihood analysis\footnote{\url{https://fermi.gsfc.nasa.gov/ssc/data/analysis/scitools/binned_likelihood_tutorial.html}} in each energy bin with the {\tt Fermitools} package\footnote{\url{https://github.com/fermi-lat}}.
The Galactic diffuse emission template ({\tt gll\_iem\_v07.fits}), the isotropic diffuse spectral model ({\tt iso\_P8R3\_SOURCE\_V3\_v1.txt}) and the 4FGL ({\tt gll\_psc\_v32.fit}) are adopted in our work.
The spectral energy distribution (SED) of 4FGL J1745.6-2859 we obtained from the Fermi-LAT gamma-ray observation is plotted in Fig. \ref{fig:flux_p_r}.

To investigate the presence of a DM signal, we refit the measured SED with the null model ($H{_0}$, without DM signal) and the signal model ($H{_1}$, with DM signal) utilizing the $\chi^2$ method\footnote{The upper limits in the SED are not considered in this $\chi^2$ analysis.}.
We adopt the log-parabola function as the $H{_0}$ model, which is the default spectral model of 4FGL J1745.6-2859 in 4FGL:
\begin{equation}
\label{logp}
  \frac{d N}{d E}
=
  N_0 \left( \frac E {E_{b}} \right)^{-\alpha-\beta \ln (E / E_{b})},
\end{equation}
which $E_{b}$ is fixed at 6499.68 MeV as default in 4FGL; $N_0$, $\alpha$ and $\beta$ are set as free parameters in the refitting process.
The $H{_1}$ model is given by the $H_0$ model with the addition of the DM signal.
To quantify the significance of potential DM signals, we define the test statistic (TS) as $ {\rm TS} = {\chi^2_{H_0} - \chi^2_{H_1}}$.
Then we further set upper limits on DM model parameters.
The 95\% confidence level upper limits are obtained at which the $\chi^2_{H_1}$ value is larger by $6.18$ than the $\chi^2_{H_0}$ value (for the $\chi^2$ distribution of 2 degrees of freedom).

\subsection{Constraints of sharp spectral structures from the combined Fermi-LAT and DAMPE data}
\label{sec:line analysis}


Thanks to the large statistics, the Fermi-LAT data are widely used in the search of the line-like~\citep{Abdo:2010nc,Fermi-LAT:2015kyq,Bringmann:2012vr,Liang:2016pvm,Foster:2022nva,DeLaTorreLuque:2023fyg,Fan:2024rcr} and box-like~\citep{Ibarra:2012dw,Li:2018rqo,Shen:2021fie} structures.
We use the 15-yr P8R3\_V3 {\tt CLEAN} Fermi-LAT data~\citep{2021AjelloApJS} between 2008 August 4 and 2023 August 3 (Fermi Mission Elapsed Time (MET) from 239557417 to 712715727) to search for sharp spectral structures.
To improve the sensitivity of sharp structures, we only adopt the {\tt evtype=896} events, which removes $\approx 1/4$ of the data with the worst reconstructed energy quality.
The same zenith cut and data-quality filter cut as the previous subsection are applied with {\tt Fermitools}.

We also adopt the photon data set from DAMPE, which has an extremely high energy resolution and is beneficial to the line-like structure search~\citep{DAMPE:2017cev,DAMPE:2021hsz}.
As a pair-converting gamma-ray telescope, DAMPE can detect photons from $\sim 2~\rm GeV$ to $10~\rm TeV$~\citep{DAMPE:2017cev}.
In the analysis, we use the public 6-yr DAMPE photon events~\citep{DAMPE:2017fxt} collected from 2016 January 1 to 2021 December 31 (DAMPE MET from 94608001 to 283996802).\footnote{Data version v6.0.3 is available in \url{https://dampe.nssdc.ac.cn/dampe/dataquerysc.php}.}
Only the High-Energy-Trigger (HET) events ({\tt evtype=1}) are selected and those collected during the South Atlantic Anomaly (SAA) region or strong solar flares has been excluded.
The {\tt DmpST} package~\citep{Duan:2019wns} is employed in the analysis.

The Fermi-LAT data and the DAMPE data between 3~GeV and 1~TeV are combined to set constraints.
The region of interest (ROI) is defined to be the $1^\circ$ circular region centering at the Sgr~A$^*$ ($\alpha=266.4168^\circ$, $\delta=-29.0078^\circ$~\citep{Xu:2022SgrA}).
We perform an unbinned likelihood analysis with the sliding window technique.
In the window for the sharp structure with the center energy of $E_{\rm s}$, only photons from $0.5E_{\rm s}$ to $1.5E_{\rm s}$ are used in the fittings.
The total likelihood is 
\begin{equation}
    L_{\rm tot} = L_{\rm wsys}^{\rm F}(N_{\rm s} C^{\rm F}_{\rm psf}, N_{\rm sys}^{\rm F}, \Theta_{\rm b}) \times L_{\rm wsys}^{\rm D}(N_{\rm s} C^{\rm D}_{\rm psf}, N_{\rm sys}^{\rm D}, \Theta_{\rm b}),
\end{equation}
where $N_{\rm s}$ is the normalization of the signal and $\Theta_{\rm b} \equiv (N_{\rm b}, \Gamma_{\rm b})$ is the nuisance parameters (normalization and spectral index) of the power-law background.
$L_{\rm wsys}^{\rm F}$ and $L_{\rm wsys}^{\rm D}$ are the unbinned likelihood functions considering the systematic uncertainties for \fermi-LAT and DAMPE respectively, whose definition can be found in Eq.(1) and Eq.(5) in~\citep{DAMPE:2021hsz}.
The systematic fractional signal $|\delta f_{\rm sys}|$, the ratio of the false signal counts to the effective background counts, is $1.5\%$~\citep{Fermi-LAT:2015kyq,Cheng:2023chi} ($2.0\%$~\citep{DAMPE:2021hsz}) for \fermi-LAT (DAMPE).
$N_{\rm sys}^{\rm F}$ and $N_{\rm sys}^{\rm D}$ are the normalization of the false signal for \fermi-LAT and DAMPE, respectively.
To compensate for the photons from the center source but spilling over the ROI, we adopt the method from~\citep{Liu:2022air} by multiplying the $1^\circ$ Point Spread Function (PSF) containments of {\emph Fermi}-LAT ($C_{\rm psf}^{\rm F}(E_{\rm s})$) and DAMPE ($C_{\rm psf}^{\rm D}(E_{\rm s})$) to the normalization of line.

To quantify the significance of the signal and establish the constraint, the likelihood ratio test~\citep{Mattox:1996} is adopted.
The null hypothesis only contains the power-law background $F_{\rm b}=N_{\rm b}E^{-\Gamma_b}$, while the alternative model comprises of the background and the signal convolved with the weighted energy dispersion function.
The TS is defined to be ${\rm TS}_{\rm line} \equiv -2\ln ( \hat{L}_{\rm tot,null}/ \hat{L}_{\rm tot,sig} )$, where $\hat{L}_{\rm tot,null}$ and $\hat{L}_{\rm tot,sig}$ are the maximum likelihood values of the null and alternative models respectively.
We do not detect any significant (${\rm TS}_{\rm line}<10$) sharp structures with central energy between 6~GeV and 500~GeV using the combined data set.
The 95\% confidence level constraint is set by increasing the signal normalization from the best-fit value until the log-likelihood changes by $1.35$~\citep{Chernoff:1954}.
The constraint on the normalization is then converted to that on the DM annihilation cross-sections $\left< \sigma v \right>$.

\section{Results}
\label{sec:result}

We set the 95\% confidence level upper limits of the gamma-ray data in three different DM annihilation scenarios in the DM mass range of ($1 \gev<m_\chi<200 \gev$).
The limits for scenarios $m_\chi > m_a$ is solely determined via the Fermi continuum spectrum analysis. 
Conversely, as outlined in Sec.~\ref{sec:GaFlux}, resonance and forbidden annihilation yield both continuum and line-shaped spectra depending on the parameter configurations. 
Therefore, we conduct the Fermi-LAT continuum-spectrum analysis for all three scenarios, 
while also including line-shaped spectrum analysis from DAMPE and Fermi telescopes specifically for resonance and forbidden annihilation.
To comprehensively investigate these two scenarios, we categorize our search strategies into continuum and monochromatic line spectrum analyses. 
The former is explored in Sec.~\ref{sec:continuum}, while the latter is discussed in Sec.~\ref{sec:line}.

\subsection{Continuum spectrum}
\label{sec:continuum}

\begin{figure*}[ht]
\centering
\includegraphics[width=7.5cm]{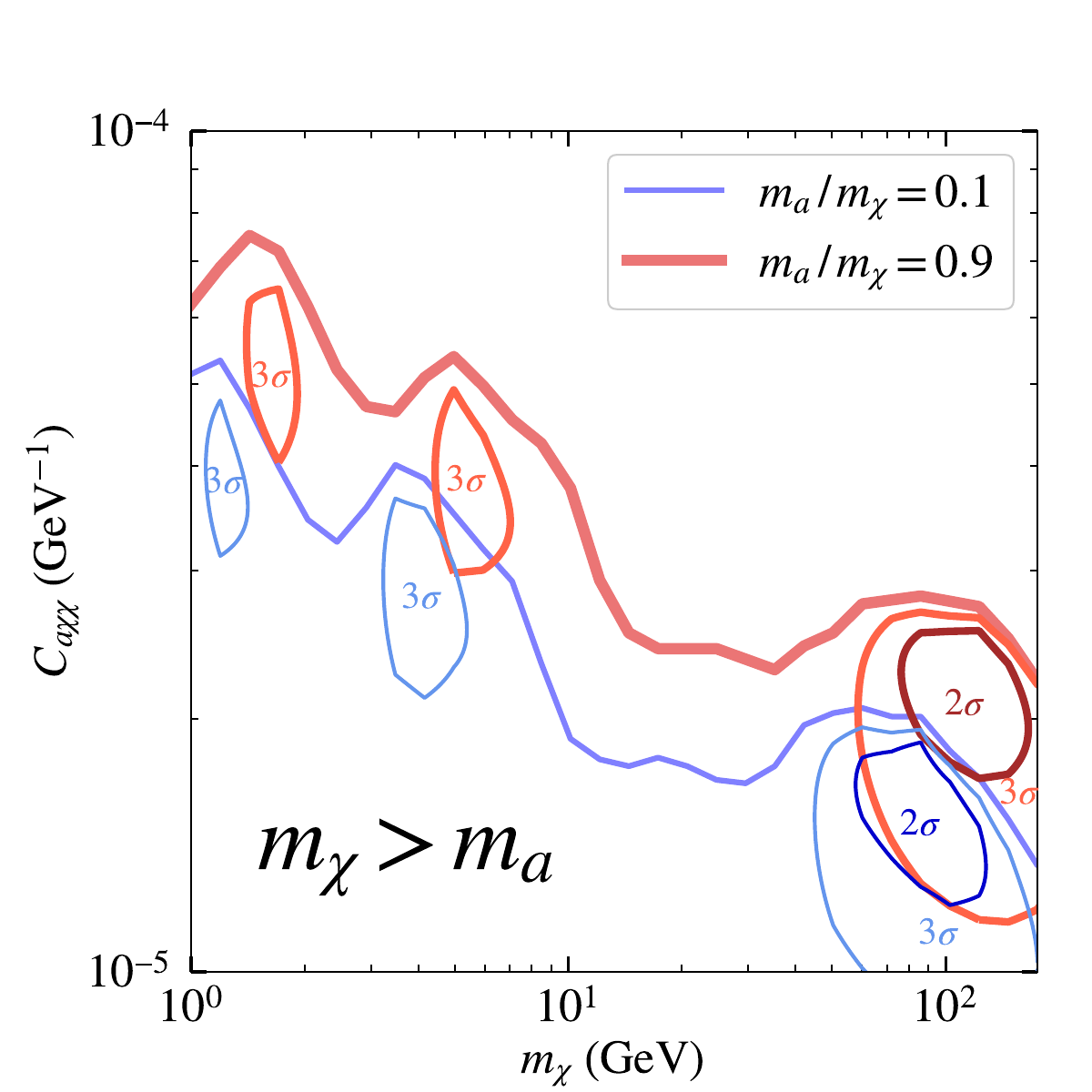}
\includegraphics[width=7.5cm]{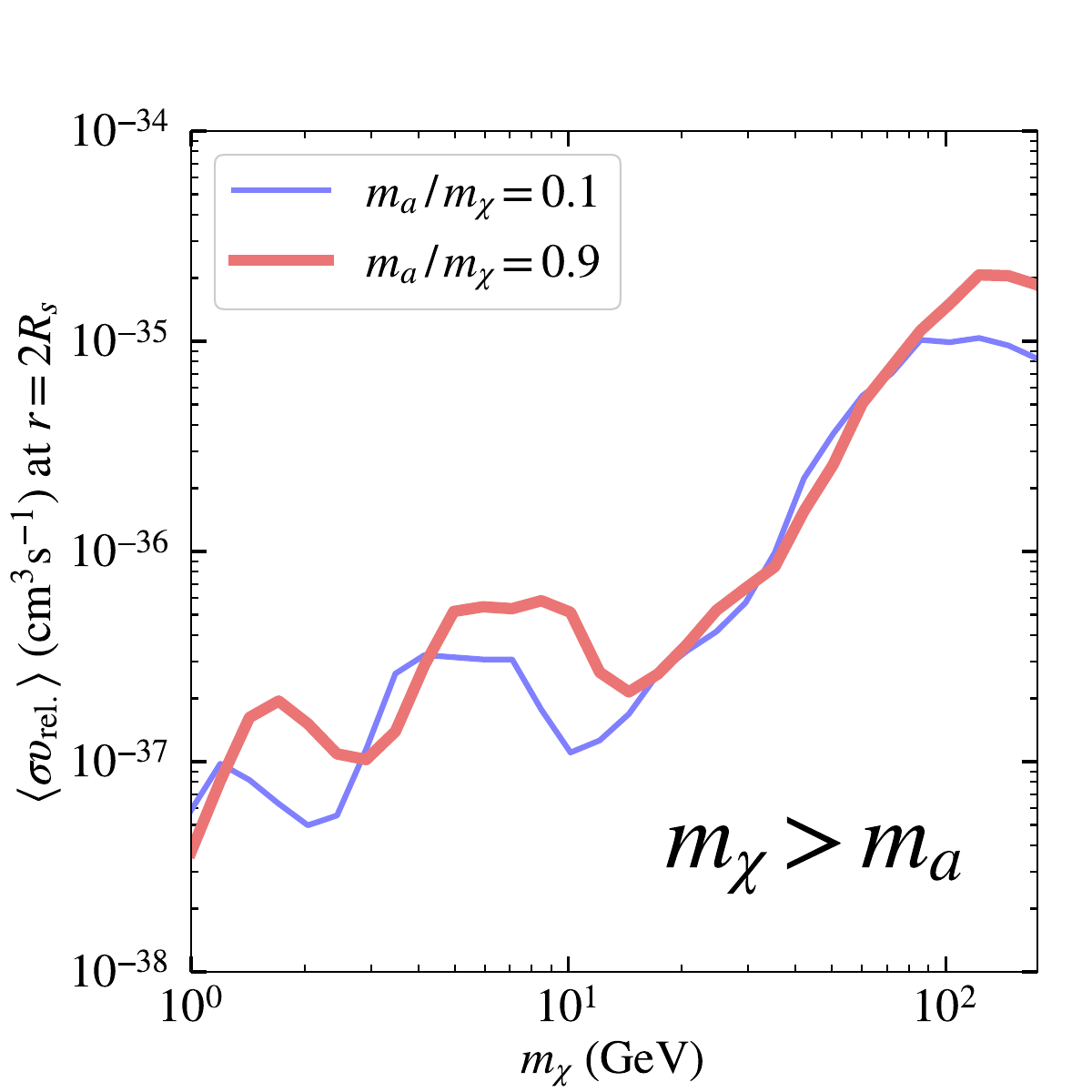}
\includegraphics[width=7.5cm]{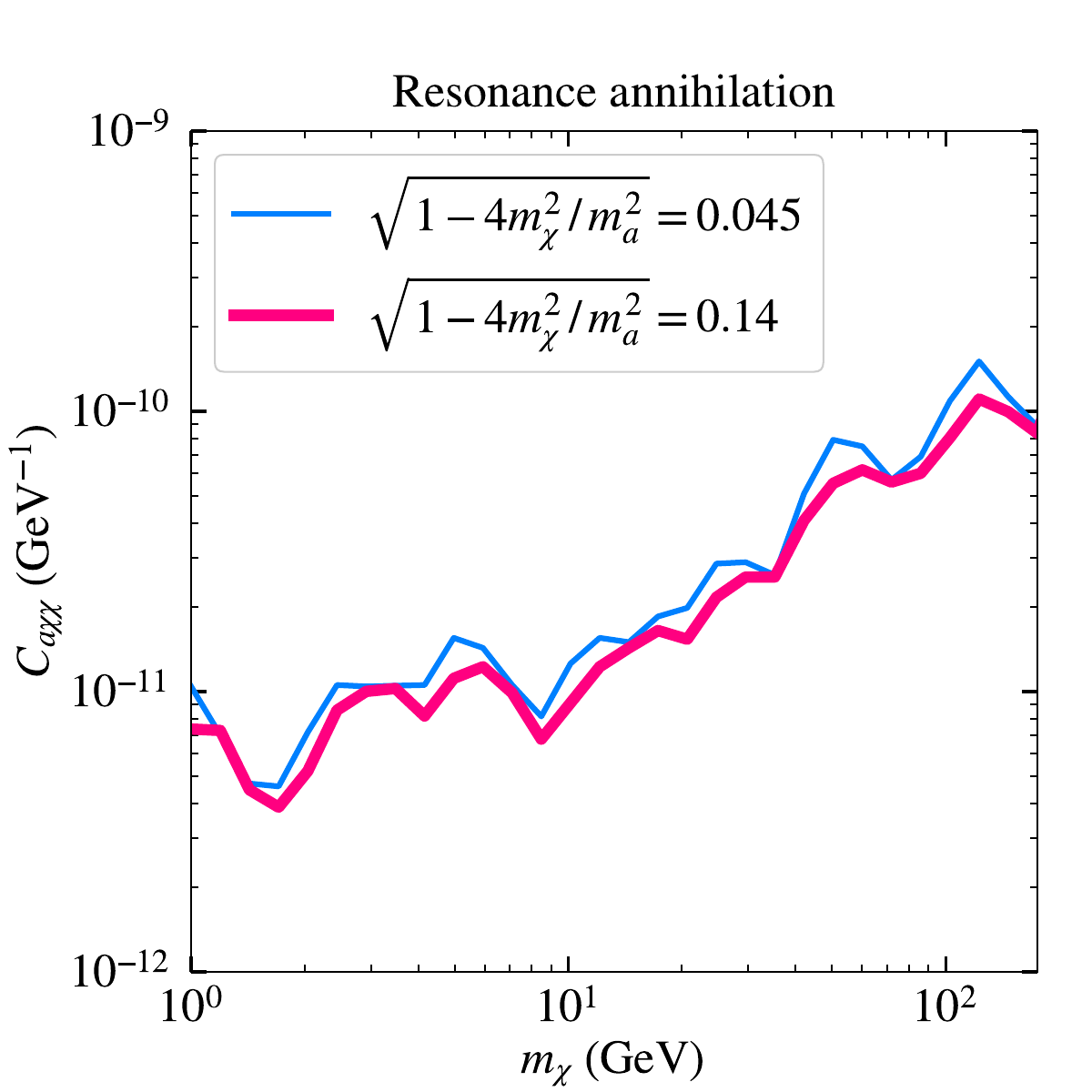}
\includegraphics[width=7.5cm]{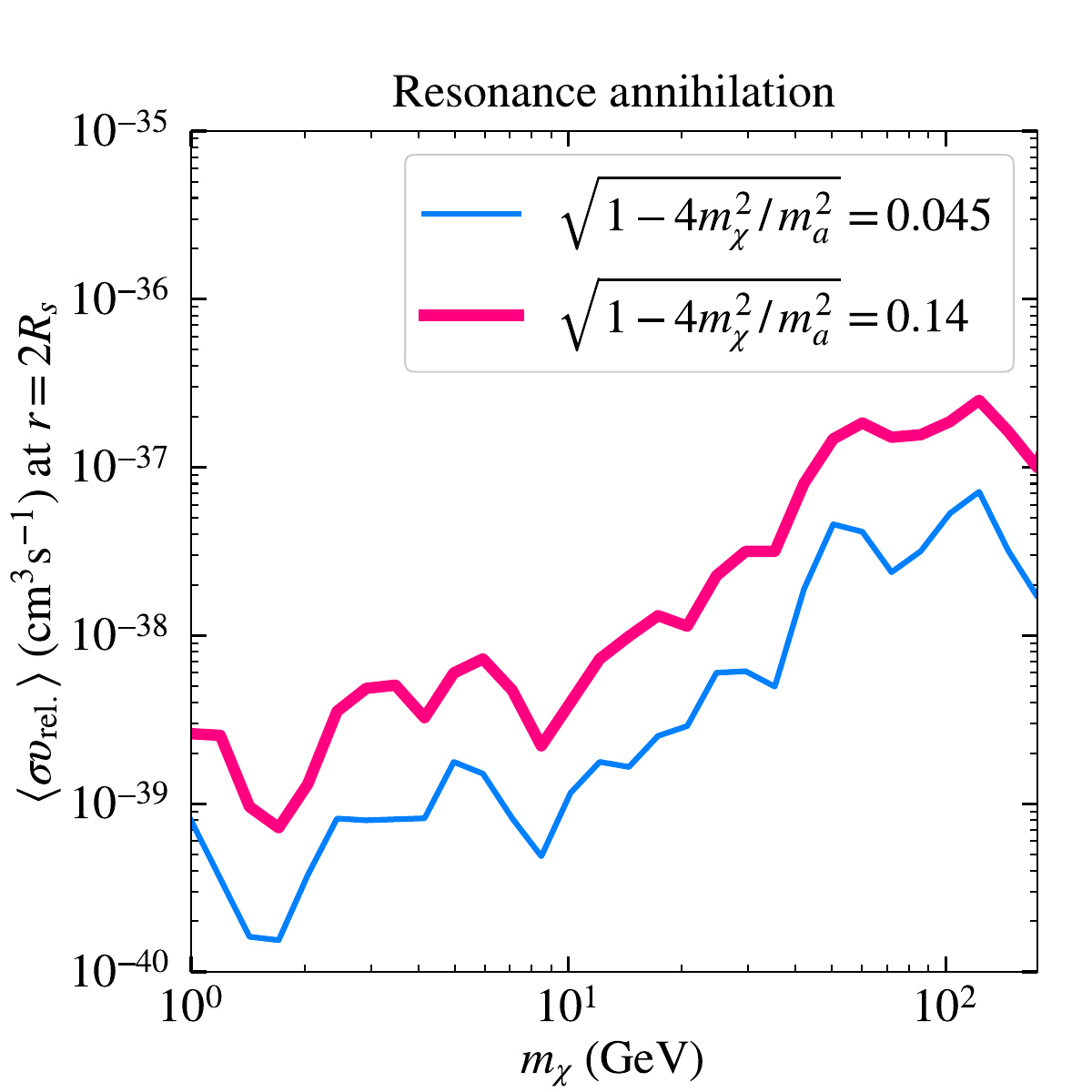}
\includegraphics[width=7.5cm]{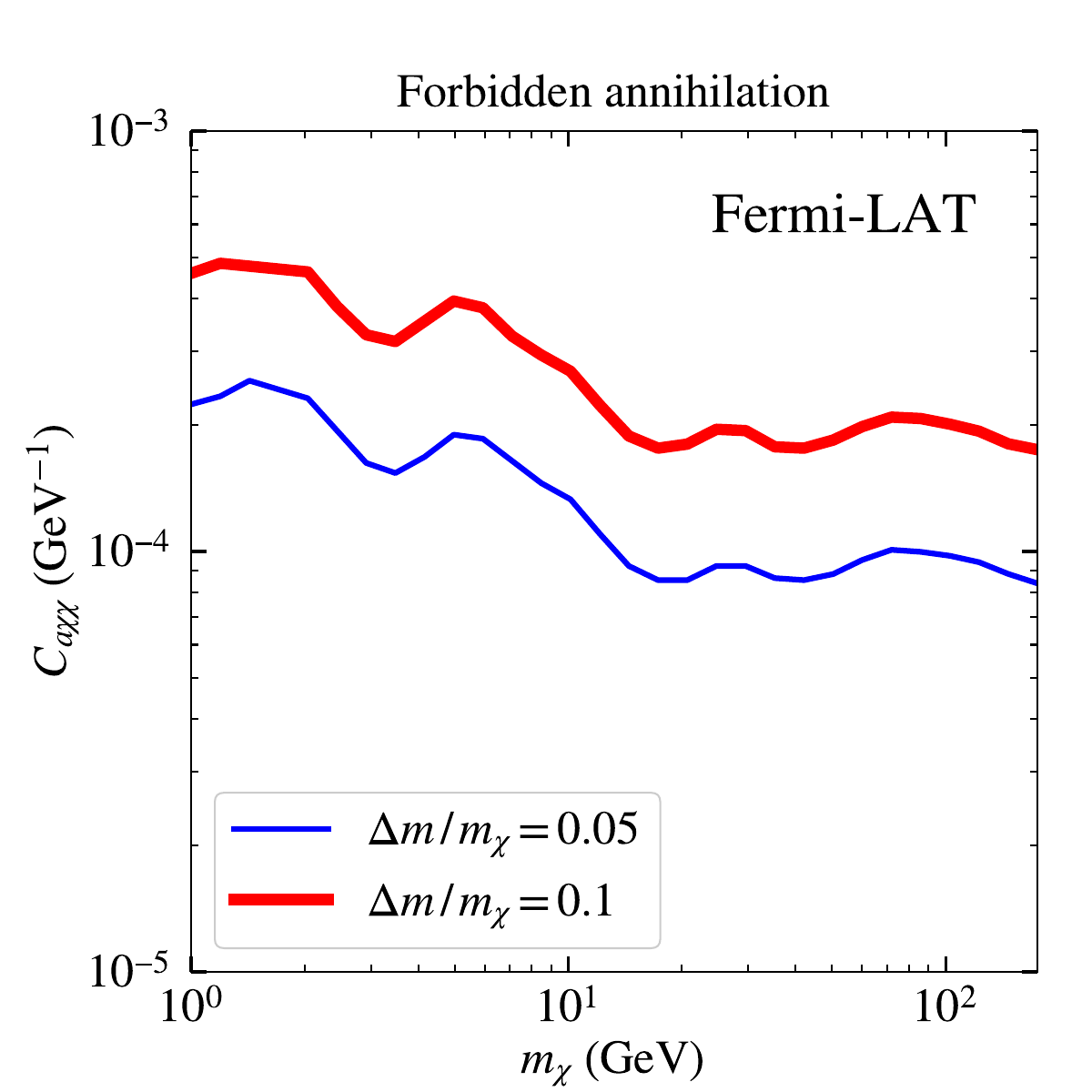}
\includegraphics[width=7.5cm]{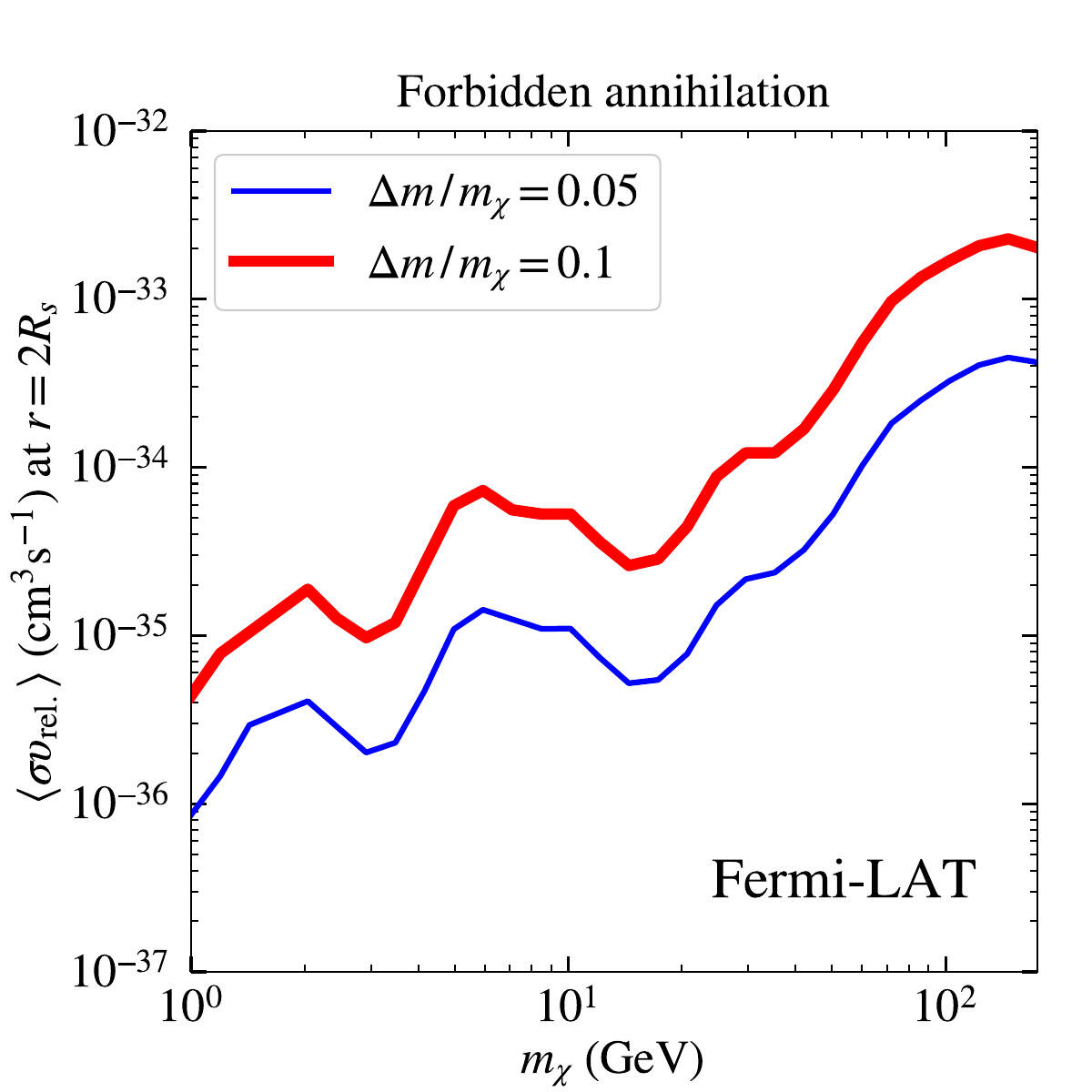}
 \caption{The $95\%$ upper limits for $C_{a\chi\chi}$ (left panels) and $\sv$ at $r=2R_s$ (right panels), based on Fermi continuum spectrum analysis. 
The top, middle, and bottom panels represent scenarios where $m_{\chi} > m_{a}$, resonance annihilation, and forbidden annihilations, accompanied by two benchmarks (blue and red lines). 
In the $m_{\chi} > m_{a}$ scenario (top left panel), a better fit is observed compared to the log-parabola law background, with the contours of $2\sigma$ (inner) and $3\sigma$ (outer) provided as reference. }
 \label{fig:forbidden_cont}
\end{figure*}

In Fig.~\ref{fig:forbidden_cont}, we establish upper limits for $C_{a\chi\chi}$ (left panels) and $\sv$ at $r=2R_s$ (right panels) with a $95\%$ confidence level using Fermi continuum spectrum analysis. 
The top, middle, and bottom panels present scenarios involving $m_{\chi} > m_{a}$, resonance annihilation, and forbidden annihilation cases, each accompanied by blue and red benchmark lines. 
Wiggles in the upper limits result from the statistical fluctuation of observation data. 
Due to the annihilation effective coupling $m_\chi C_{a\chi\chi}$ proportional to $m_\chi$ and gamma-ray fluxes proportional to $\rho_\chi^2/m_\chi^2$, 
the trends in $C_{a\chi\chi}$ and $\sv$ limits may differ. 
Generally, $\sv$ limits are stronger for lighter DM but weaker for heavier DM. 
In scenarios where $\chi\overline{\chi}\to a a$ governs annihilations, such as $m_{\chi} > m_{a}$ and forbidden annihilation cases, 
the annihilation cross-section [Eq.~\eqref{eq:xsecxxaa}] is proportional to $C^4_{a\chi\chi} m^2_{\chi}$. 
Conversely, for $\chi\overline{\chi}\to 2\gamma$ with $a$-resonance condition $m_a\approx 2 m_\chi$ [Eq.~\eqref{eq:xsec_vrel_reson}], 
$\sv$ only depends on $C_{a\chi\chi}^{2}$ 
if the ALP only decays to a photon pair and a DM pair. 
However, $\sv$ is proportional to $m_{\chi}^{-1}$ if the decay channels $a\to ZZ$ and $a\to WW$ open, with a separation at $m_\chi\sim 100\gev$. 
Therefore, the trend of the $C_{a\chi\chi}$ limits for the resonance scenario differs from the other two.

For the $m_{\chi} > m_{a}$ scenario, we can see that a small $m_a/m_\chi$ leads to a strong limit. 
It is worth mentioning that the log-parabola background does not fit the Fermi-LAT SED well at $10\gev<E_\gamma<100\gev$ in $3\sigma$ significance.  
Hence, we draw $2\sigma$ (inner) and $3\sigma$ (outer) contours to demonstrate the signal regions. 
However, because the spectra of the resonance and forbidden scenarios are narrower, they cannot explain this signal, and then no contour is shown.

In the resonance case, for the range $1 \gev < m_\chi < 200 \gev$, allowed $C_{a\chi\chi}$ lies between $7\times 10^{-12}/\gev$ and $10^{-10}/ \gev$. The corresponding DM annihilation cross-sections are $8 \times 10^{-40} \lesssim \sv/({\rm cm}^3/{\rm s}) \lesssim 2\times 10^{-38}$ for $R_{a\chi}=0.045$ (blue solid line) and $3 \times 10^{-39} \lesssim \sv/({\rm cm}^3/{\rm s}) \lesssim  10^{-37} $ for $R_{a\chi}=0.14$ (red solid line). 
Moreover, $\sv$ with $R_{a\chi}=0.14$ exceeds that for $R_{a\chi}=0.045$ with the similar $C_{a\chi\chi}$. 
Despite being unable to explain the $10\gev<E_\gamma<100\gev$ excess, the upper limits around $m_\chi\sim 100\gev$ increase rapidly.


In the forbidden annihilation scenario, a smaller $\Delta m/m_\chi$ imposes a more stringent constraint. For the range $1 \gev < m_\chi < 200 \gev$, allowed $C_{a\chi\chi}$ lies between $2\times 10^{-4}/\gev$ and $10^{-4}/ \gev$ for $\Delta m/m_\chi=0.05$. This constraint is twice as strong as that of $\Delta m / m_\chi = 0.1$.
The corresponding DM annihilation cross-sections are $10^{-36} \lesssim \sv/({\rm cm}^3/{\rm s}) \lesssim 5\times 10^{-34}$ for $\Delta m/m_\chi=0.05$ (blue solid line) and $4 \times 10^{-36} \lesssim \sv/({\rm cm}^3/{\rm s}) \lesssim  2\times 10^{-33} $ for $\Delta m/m_\chi=0.1$ (red solid line).

\subsection{Line spectrum}
\label{sec:line}

\begin{figure*}[ht]
\centering
\includegraphics[width=8.5cm]{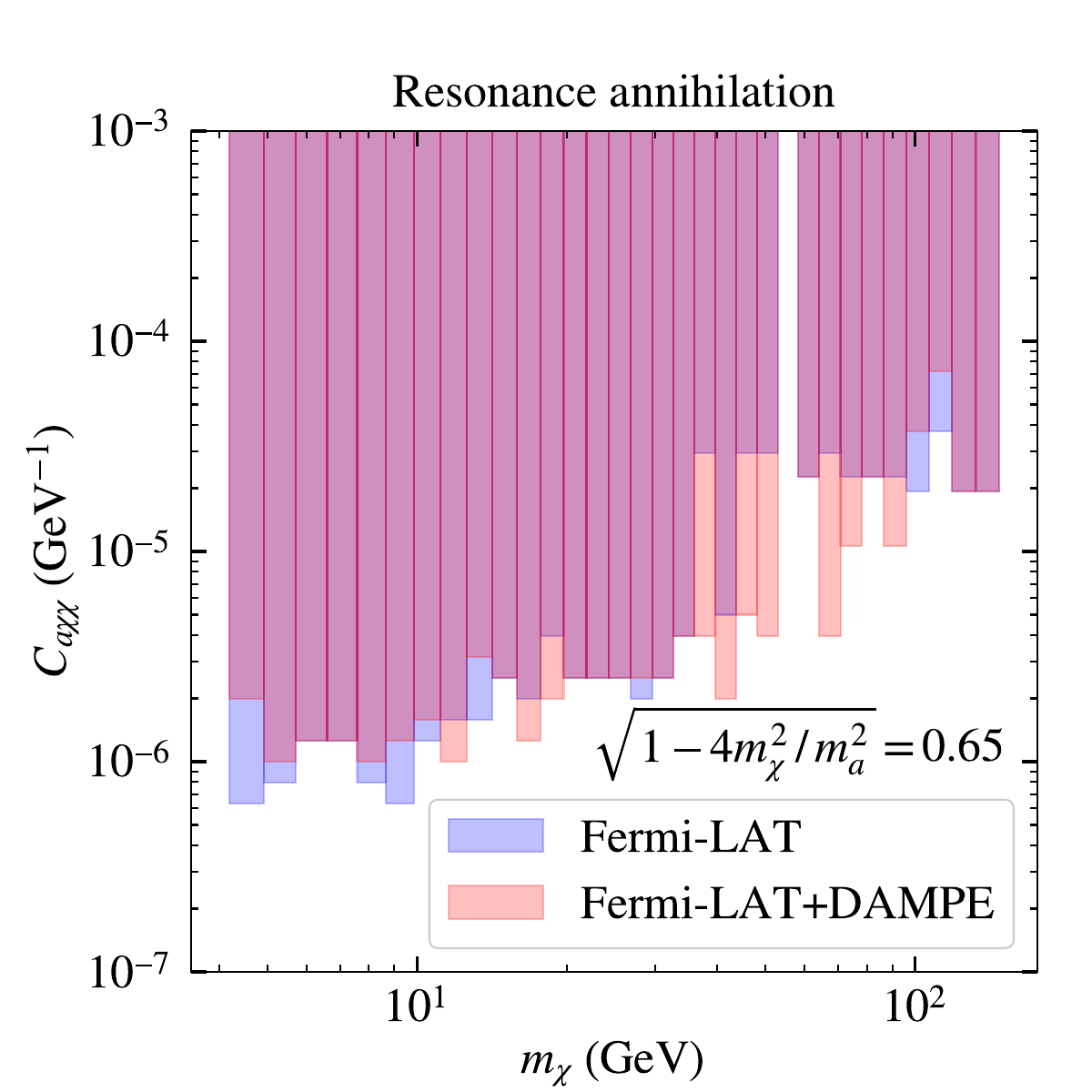}
\includegraphics[width=8.5cm]{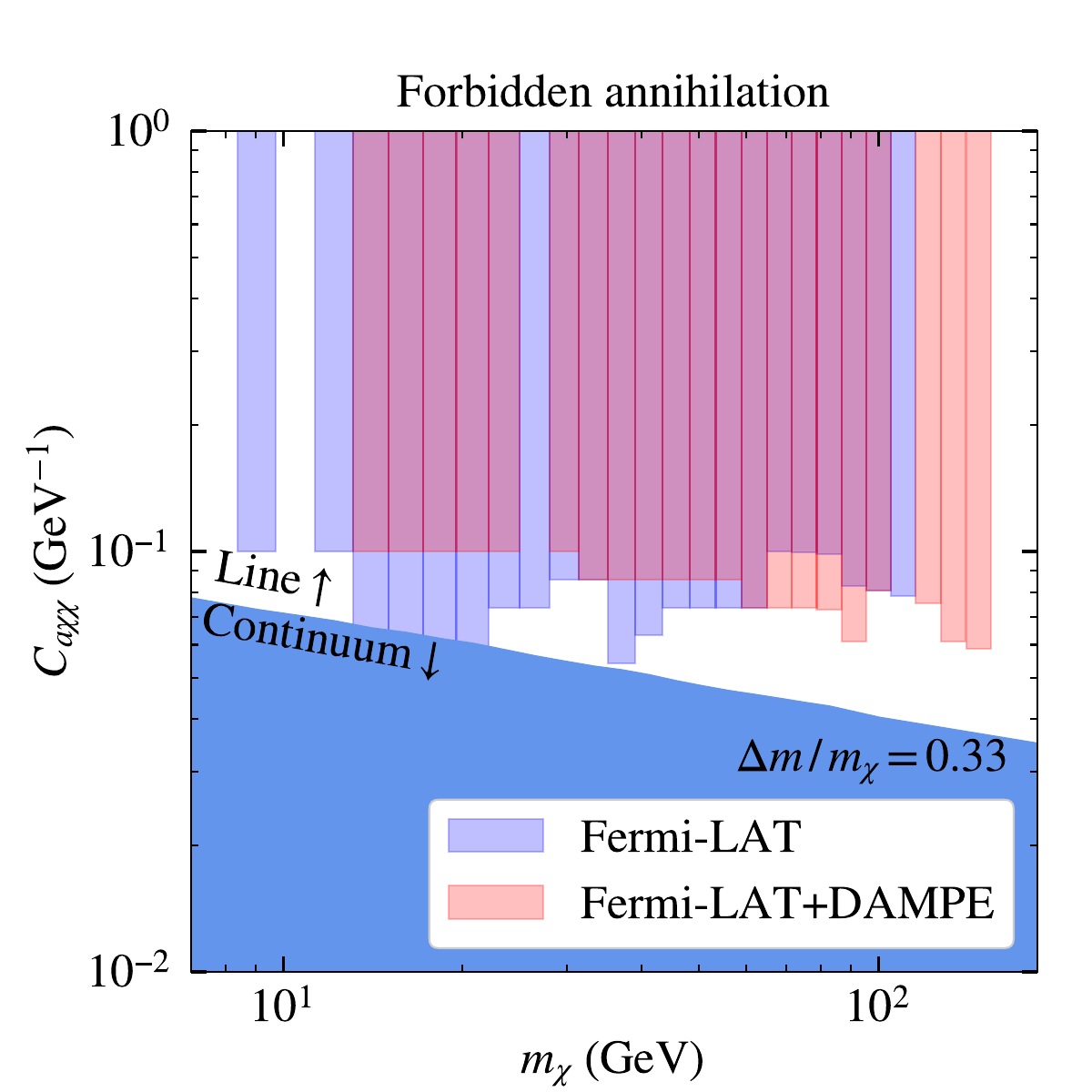}

 \caption{The upper limits of $C_{a\chi\chi}$ for resonance (left panel) and forbidden (right panel) annihilation in the line spectrum scenario. 
The left panel corresponds to $R_{a\chi}=0.65$, while the right panel corresponds to $\Delta m/m_{\chi}=0.33$. 
Pink bars and light blue bars indicate the $95\%$ exclusion limits from Fermi+DAMPE line spectrum and Fermi continuum spectrum analyses, respectively. 
Dark blue regions in the right panel denote parameter spaces where neither Fermi nor DAMPE can identify the annihilation spectra as a spectral line due to their minimum energy resolution. 
 }
 \label{fig:forbidden_line}
\end{figure*}



Fig.~\ref{fig:forbidden_line} illustrates the $95\%$ upper limits on line spectra under two scenarios: $R_{a\chi}=0.65$ for the resonance scenario in the left panel and $\Delta m/m_\chi=0.33$ for the forbidden DM scenario in the right panel. 
The $95\%$ exclusion regions from line spectrum analysis (using Fermi+DAMPE data) and continuum spectrum analysis (using Fermi data) are shown as pink and light blue bars, respectively. 
In the left panel, the parameter $R_{a\chi}=0.65$ is chosen such that the resulting spectra are spectral lines, independent with the coupling size (see the upper left panel of Fig.~\ref{fig:flux_p_r}).
While, in the right panels, dark blue areas indicate regions of parameter space where neither Fermi nor DAMPE can distinguish annihilation spectra as line-shaped due to their limited energy resolution (here we roughly take the threshold value as $\Delta E_\gamma / E_\gamma > 0.1$, which is the corresponding energy resolution (68\%) at 1 GeV for the Fermi-LAT).

Line-shape spectrum analysis, although generally providing weaker constraints compared to continuum spectrum analysis, still offers valuable complementary tests for several $m_\chi$ bins. 
Particularly, at the region $35\gev \lesssim E_\gamma\lesssim 100\gev$, the Fermi SED fit poorly with the log-parabola background model. 
Therefore, In resonance scenarios (where one annihilation yields two photons), 
line spectrum analysis imposes stronger constraints at the region $35\gev \lesssim m_\chi \lesssim 100\gev$. 
Similarly, in forbidden DM scenarios (where one annihilation yields four photons), 
line spectrum analysis also imposes stronger constraints at the region $70\gev \lesssim m_\chi \lesssim 200\gev$.  
In addition, we find that current Fermi+DAMPE data, using the line spectrum analysis, can only probe the parameter space with $R_{a\chi}<0.65$ for the resonance scenario and $\Delta m/m_\chi<0.35$ the forbidden DM scenario. 
For $m_\chi\approx 10\gev$, the $95\%$ exclusions correspond to the cross section at $2R_s$, $\sv\gtrsim \mathcal{O}(10^{-31})$~cm$^3$s$^{-1}$  and $\sv\gtrsim \mathcal{O}(10^{-29})$~cm$^3$s$^{-1}$, in the resonance and the forbidden DM scenario, respectively.

\section{Summary and Discussion}
\label{sec:conclusion}
The current indirect detections mostly focus on detecting DM annihilation assuming $s$-wave behavior and DM velocities below $10^{-3}c$. 
Despite the sensitivity of these instruments may not see significant improvements in the near future, there remain areas of parameter space where annihilation dynamics exhibit velocity dependencies, such as $p$-wave annihilation, resonance annihilation with peaks at velocities exceeding $10^{-3} c$, or forbidden annihilation requiring semi-relativistic velocities. 
These annihilation processes feature velocity suppression in their cross-sections, requiring acceleration for enhanced detectability. 
Given the considerable DM abundance accumulated by BH accretion, annihilation signals (if velocity-dependent) could be amplified by the gravitational potential of SMBH. 
In this study, we revisit the accretion-enhanced annihilation mechanism for the SMBH Sgr A$^{\star}$ in the galactic center by utilizing the gamma-ray data from Fermi and DAMPE telescopes.

This work examines a simplified DM model using an electroweak ALP portal framework. 
We focused on a Dirac fermionic DM with a mass above GeV and a pseudoscalar ALP. 
In this model, DM particles can annihilate into photons through either $s$-wave contribution via $a$-resonance or $p$-wave contribution ($\chi\overline{\chi}\to a a$). These annihilation channels naturally occur under specific mass conditions: 
$m_\chi > m_a$ for $p$-wave annihilation, $m_a \approx 2 m_\chi$ for $a$-resonant annihilation, and $m_\chi\lesssim m_a$ for the forbidden DM channel.

When considering the accretion-enhanced annihilation mechanism within the DM spike, we observe that the DM annihilation cross-section $\sv$ and the photon yield spectrum $\dnde$ both depend on relative velocity and the annihilation position to the BH.
We accurately calculate the expected gamma-ray fluxes induced by DM annihilation near Sgr A$^{\star}$, considering both velocity and positional dependencies in $\sv$ and $\dnde$.
After being accelerated by accretion, the initially monochromatic photon line spectrum can broaden to a wider range, potentially detectable by the Fermi gamma-ray telescope within the energy range of $500\mev$ to $1$ TeV.

We find that only the $m_\chi>m_a$ scenario, with DM mass around $\mathcal{O}(100\gev)$ and $C_{a\chi\chi}/\gev^{-1}\sim 2\times 10^{-5}$, can improve the test statistics of the Fermi data by $3\sigma$ when compared to a log-parabola background. 
This improvement is due to the need for an additional contribution with an energy spread to cover the $35\gev\lesssim E_\gamma\lesssim 100\gev$ range, which is not sufficiently covered by the narrower spectra of the $a$-resonance and forbidden annihilation. 
The spectrum of resonance annihilation and forbidden DM annihilation are the sharpest among the scenarios, while annihilation events farther from the BH may lead to forbidden channels or a monochromatic line spectrum.
Therefore, in cases of resonance annihilation and forbidden DM annihilation within the spectral line parameter space (see Fig.~\ref{fig:flux_p_r}), the resulting photon spectrum maintains line spectrum characteristics and meets the $\Delta E_\gamma/E_\gamma<0.1$ criteria required by the energy resolution of Fermi and DAMPE line spectrum analysis.

In this study, we examined the predicted gamma-ray fluxes of three annihilation mechanisms using Fermi 15-year data. We note that the resonance annihilation and forbidden annihilation mechanism yields both continuum and line-like gamma-ray spectra. 
When $\Delta m/m_\chi < 0.35$ and $C_{a\chi\chi}/\gev^{-1} \gtrsim 5\times 10^{-2}$ for the forbidden annihilation, or $R_{a\chi} > 0.5$ and $C_{a\chi\chi}/\gev^{-1} \gtrsim 10^{-8}$ for the resonance annihilation, 
the photon spectra look like a gamma-ray line, prompting us to use both Fermi and DAMPE data to investigate this parameter space. 
Quantitatively, we summarize the upper limits of $C_{a\chi\chi}$ as follows:
\begin{itemize}
    \item  In the scenario where $m_\chi > m_a$, the upper limit at $m_\chi \simeq 10\gev$ is $C_{a\chi\chi}/\gev^{-1} \lesssim \mathcal{O}(3 \times 10^{-5})$, slightly dependent on the ratio $m_a/m_\chi$. The Fermi data, when compared to a log-parabola background, prefers this scenario with parameters around $m_\chi \approx 100\gev$ and $C_{a\chi\chi}/\gev^{-1} \approx 2 \times 10^{-5}$, corresponding to a cross-section $\sv$ at $2 R_s$ of approximately $\mathcal{O}(10^{-35})~\text{cm}^3\text{s}^{-1}$ at a $2\sigma$ significance level.

    \item In the resonance annihilation scenario and forbidden annihilation scenario, the limits vary with different values of $m_\chi$, $m_{a}$ and $C_{a\chi\chi}$. 
    The distinction between continuum and line spectra is shown in the left panel of Fig.~\ref{fig:flux_p_r}. 
    Due to the nature of resonance annihilation and forbidden annihilation, when there is a large $R_{a\chi}$ or $\Delta m/m_\chi$, a larger $C_{a\chi\chi}$ is required, which can modify the spike profile.

    \item For the continuum spectrum in the $a$-resonance scenario, varying $R_{a\chi} \equiv \sqrt{1-4m_\chi^2/m_a^2}$ between 0.14 and 0.045 does not affect the upper limit of $C_{a\chi\chi}/\gev^{-1} \lesssim \mathcal{O}(10^{-11})$ at $m_\chi \simeq 10\gev$. 
    This is the most stringent limit among the scenarios. 
    For the continuum spectrum in the forbidden annihilation scenario, 
    the allowed $C_{a\chi\chi}$ at $m_\chi \simeq 10\gev$ is $C_{a\chi\chi}/\gev^{-1} \lesssim \mathcal{O}(3 \times 10^{-4})$ for $\Delta m/m_\chi = 0.1$, 
    but $C_{a\chi\chi}/\gev^{-1} \lesssim \mathcal{O}(10^{-4})$ for $\Delta m/m_\chi = 0.05$. 
    The associated annihilation cross section at $2R_s$ is around $10^{-34}\text{cm}^3\text{s}^{-1}$.

    \item For the line spectrum in both the $a$-resonance annihilation scenario and the forbidden annihilation scenario, a higher value of $C_{a\chi\chi}$ is required. In the resonance annihilation scenario, $C_{a\chi\chi}/\gev^{-1} \gtrsim \mathcal{O}(4 \times 10^{-6})$ at $m_\chi \simeq 50\gev$ for a benchmark $R_{a\chi} = 0.65$ by using Fermi and DAMPE data. 
    The corresponding annihilation cross-section at $2R_s$ is around $10^{-30}\text{cm}^3\text{s}^{-1}$. 
    In the case of forbidden annihilation scenario,  
    $C_{a\chi\chi}/\gev^{-1} \gtrsim \mathcal{O}(8.5 \times 10^{-2})$ at $m_\chi \simeq 100\gev$ for a benchmark $\Delta m/m_\chi = 0.33$ and its corresponding annihilation cross-section at $2R_s$ is also around $10^{-27}\text{cm}^3\text{s}^{-1}$.   

 

\end{itemize}

Finally, we aim to compare our findings with existing researches in the field. 
Previous studies~\cite{Chiang:2019zjj,Alvarez:2020fyo,Liu:2022air,Balaji:2023hmy} focus on $s$-wave annihilation. 
In contrast, only $a$-resonance annihilation in our work is under $s$-wave condition, highlighting scenario where annihilation cross-sections are enhanced near the BH by reaching the pole velocity. 
Ref.~\cite{Cheng:2023dau} also examines resonance annihilation through a Higgs portal DM model, 
while annihilation into photon pairs is not the tree-level contribution. 
Some studies investigate $p$-wave annihilation near the Sgr A*~\cite{Shelton:2015aqa,Sandick:2016zeg,Johnson:2019hsm,Cheng:2022esn} and in the center of M87 \cite{Christy:2023tdv}.
Only Ref.~\cite{Johnson:2019hsm} examines the axion-portal DM model, but we derive constraints using the halo slope given by S2 data rather than $\gamma\ge 1$. 
In Appendix~\ref{app:spike_density}, we show the impact of $\gamma$ on spike halo density and the upper limits of $C_{a\chi\chi}$ in the $a$-resonance scenario.
Furthermore, Ref.~\cite{Johnson:2019hsm} claims that constraints derived from the continuum analyses are more stringent than those from line analyses. 
However, our analysis demonstrates that integrating both line and continuum analyses using data from Fermi and DAMPE can offer complementary limits.
Ref.~\cite{Shelton:2015aqa} also discusses various search strategies for lines or continuum signals depending on parameter space, 
but they focus on the Higgs-portal model. 
In contrast, we demonstrate that the axion-portal model supports a monochromatic line in both $a$-resonance and forbidden DM scenarios.

\section*{Acknowledgments}
We thank an anonymous referee for useful comments on gravitational potential.  
YST, MWY, ZQG and ZQX are supported by the National Key Research and Development Program of China
(No. 2022YFF0503304), and the Project for Young Scientists in Basic Research of the Chinese Academy of Sciences 
(No. YSBR-092). 
CTL and XYL are supported by the Special funds for postdoctoral overseas recruitment, Ministry of Education of China (No.~164080H0262403). 
This research has made use of data obtained from the High Energy Astrophysics Science Archive Research Center (HEASARC), provided by NASA’s Goddard Space Flight Center. 
This research has made use of the data resources from DArk Matter Particle Explorer (DAMPE) satellite mission supported by Strategic Priority Program on Space Science, and China data service provided by National Space Science Data Center of China. 

\appendix



\section{Potential profile}
\label{app:Potential profile}

\begin{figure*}[ht]
\centering
\includegraphics[width=8.5cm]{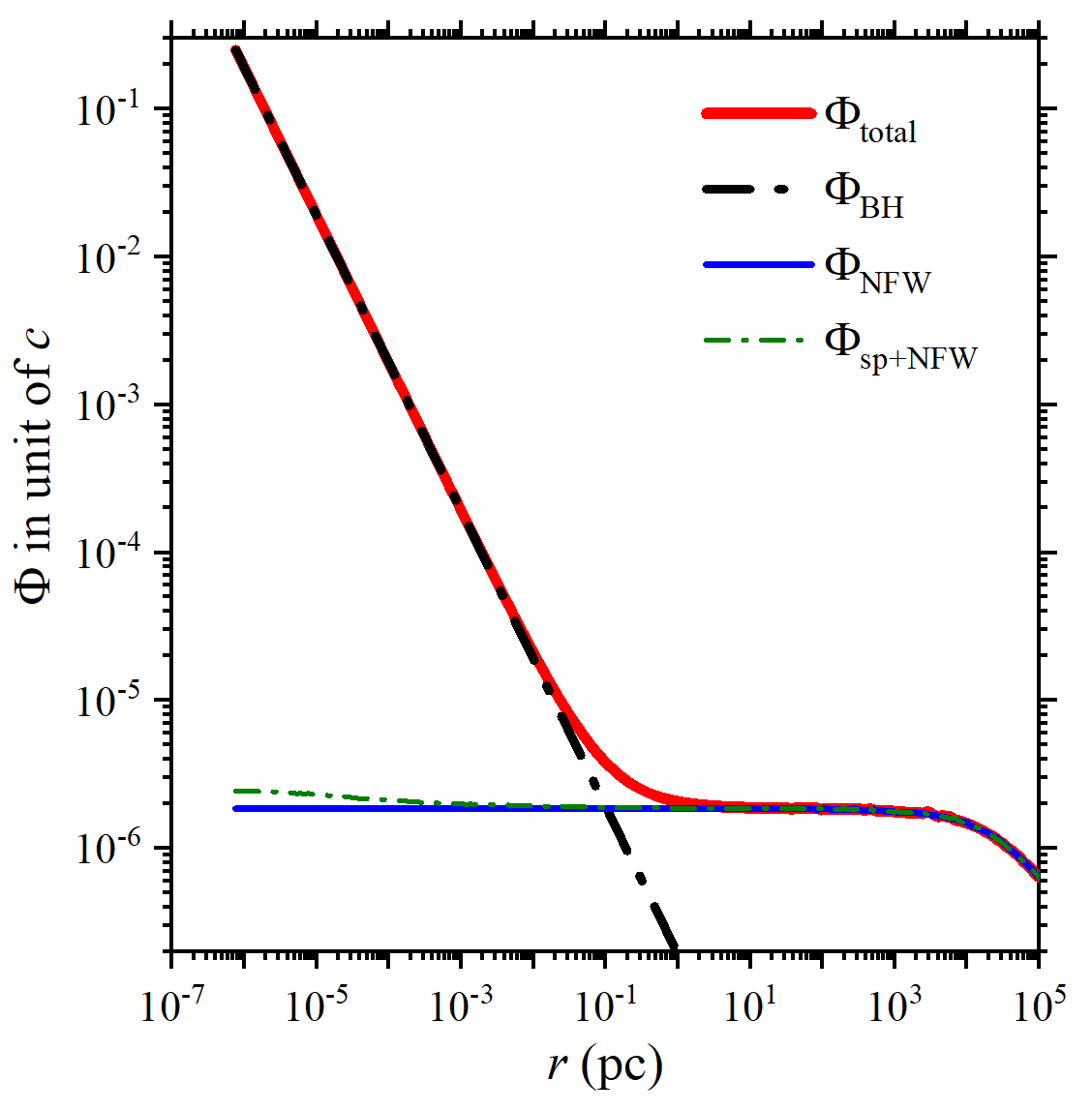}
\caption{ The total gravitational potential $\Phi_{\rm total}$ (illustrated as a red solid line) containing contributions from both BH (black dot-dashed line) and DM (green dot-dashed line, NFW with spike profile). 
The blue solid line represents the gravitational potential only with the NFW profile contribution.}
 \label{potential profile}
\end{figure*}

In Fig.~\ref{potential profile}, we illustrate the potential functions from BH, DM (NFW with spike profile), 
NFW profile, and the sum, respectively represented by black dot-dashed line, green dot-dashed line, blue solid line, and red solid line. 
In the innermost region, the BH gravitational potential dominates, while the NFW DM gravitational potential takes over if $r>10^{-1} \pc$.

\section{DM self-interaction}
\label{app:self_interaction}
Considering Dirac fermion-type DM, the DM self interaction has the following three channels, $\chi\bar{\chi}\to\chi\bar{\chi}$, $\chi\chi\to\chi\chi$ and $\bar{\chi}\bar{\chi}\to\bar{\chi}\bar{\chi}$. For the channel $\chi\bar{\chi}\to\chi\bar{\chi}$, the cross-section in this model is
\begin{equation}
\begin{aligned} \sigma(\chi\bar{\chi}\to\chi\bar{\chi})=\frac{(C_{a\chi\chi} m_\chi)^4}{16\pi s^2 \beta_\chi^2}\left[\frac{s\beta_\chi^2(s \beta_\chi^2+2m_a^2)}{s \beta_\chi^2+m_a^2}+\frac{s^3 \beta_\chi^2}{(s-m_a^2)^2+m_a^2\Gamma_a^2}\right.\\
\left. +2m_a^2\ln(\frac{m_a^2}{s \beta_\chi^2+m_a^2})-
\frac{s m_a^2\ln(\frac{m_a^2}{s \beta_\chi^2+m_a^2})-s^2 \beta_\chi^2}{\sqrt{(s-m_a^2)^2+m_a^2\Gamma_a^2}}\right],
\end{aligned}
\end{equation}
for the channel $\chi\chi\to\chi\chi$ and $\bar{\chi}\bar{\chi}\to\bar{\chi}\bar{\chi}$, the cross-section is
\begin{equation}
\begin{aligned} \sigma(\chi\chi\to\chi\chi)=\sigma(\bar{\chi}\bar{\chi}\to\bar{\chi}\bar{\chi})=\frac{(C_{a\chi\chi} m_\chi)^4(5m_a^2+3s \beta_\chi^2)(1-\frac{m_a^2}{s\beta_\chi^2+m_a^2}+\frac{2m_a^2\ln{\frac{m_a^2}{s\beta_\chi^2+m_a^2}}}{s\beta_\chi^2+2m_a^2})}{32\pi s^2 \beta_\chi^2}.
\end{aligned}
\end{equation}
Here $\beta_\chi = \sqrt{1-\frac{4m_\chi^2}{s}}$ and $s\equiv E_{\rm cm}^2$.

\section{Phase Space Distribution}
\label{app:phase_space}

To simplify the calculation, we express the velocity ($v_1, v_2$) and angular momentum ($L_1, L_2$) of two DM particles as functions of the center of mass velocity ($v_{\rm CM}$), relative velocity ($v_{\rm rel.}$), and angles,
\begin{equation}
    \begin{aligned}
        v_1=|\mathbf{v_1}|=\sqrt{v_{\rm CM}^2+\frac{v_{\rm rel.}^2}{4}-v_{\rm CM} v_{\rm rel.} \cos\theta},\\
        v_2=|\mathbf{v_2}|=\sqrt{v_{\rm CM}^2+\frac{v_{\rm rel.}^2}{4}+v_{\rm CM} v_{\rm rel.} \cos\theta},
    \end{aligned}
\end{equation}
where the definitions of the notations are in consist with Sec.~\ref{sec:halo}, and  
the two angular momentum are 
\begin{equation}
    \begin{aligned}
        L_1^2=&\left|\mathbf{r}\times\left(\mathbf{v_{\rm CM}}-\frac{\mathbf{v_{\rm rel.}}}{2}\right)\right|^2\\
        =&r^2\left[\frac{v_{\rm rel.}^2}{4}\sin^2\theta \sin^2 \phi+\{-\frac{v_{\rm rel.}}{2}\cos\alpha\sin\theta {\rm cos}\phi\right.\\
        &\left. -\sin\alpha (v_{\rm CM}-\frac{v_{\rm rel.}}{2}\cos\theta)\}^2\right],\\
        L_2^2&=|\mathbf{r}\times(\mathbf{v_{\rm CM}}+\frac{\mathbf{v_{\rm rel.}}}{2})|^2\\
        &=r^2[\frac{v_{\rm rel.}^2}{4}{\sin^2\theta\sin^2 \phi}+\{\frac{v_{\rm rel.}}{2}\cos\alpha\sin\theta {\rm cos}\phi\\
        &-\sin\alpha(v_{\rm CM}+\frac{v_{\rm rel.}}{2}\cos\theta)\}^2]. 
    \end{aligned}
\end{equation}

\section{The ALP partial decay widths}
\label{app:Gamma_a}
\begin{figure*}[ht]
\centering
\includegraphics[width=8.5cm]{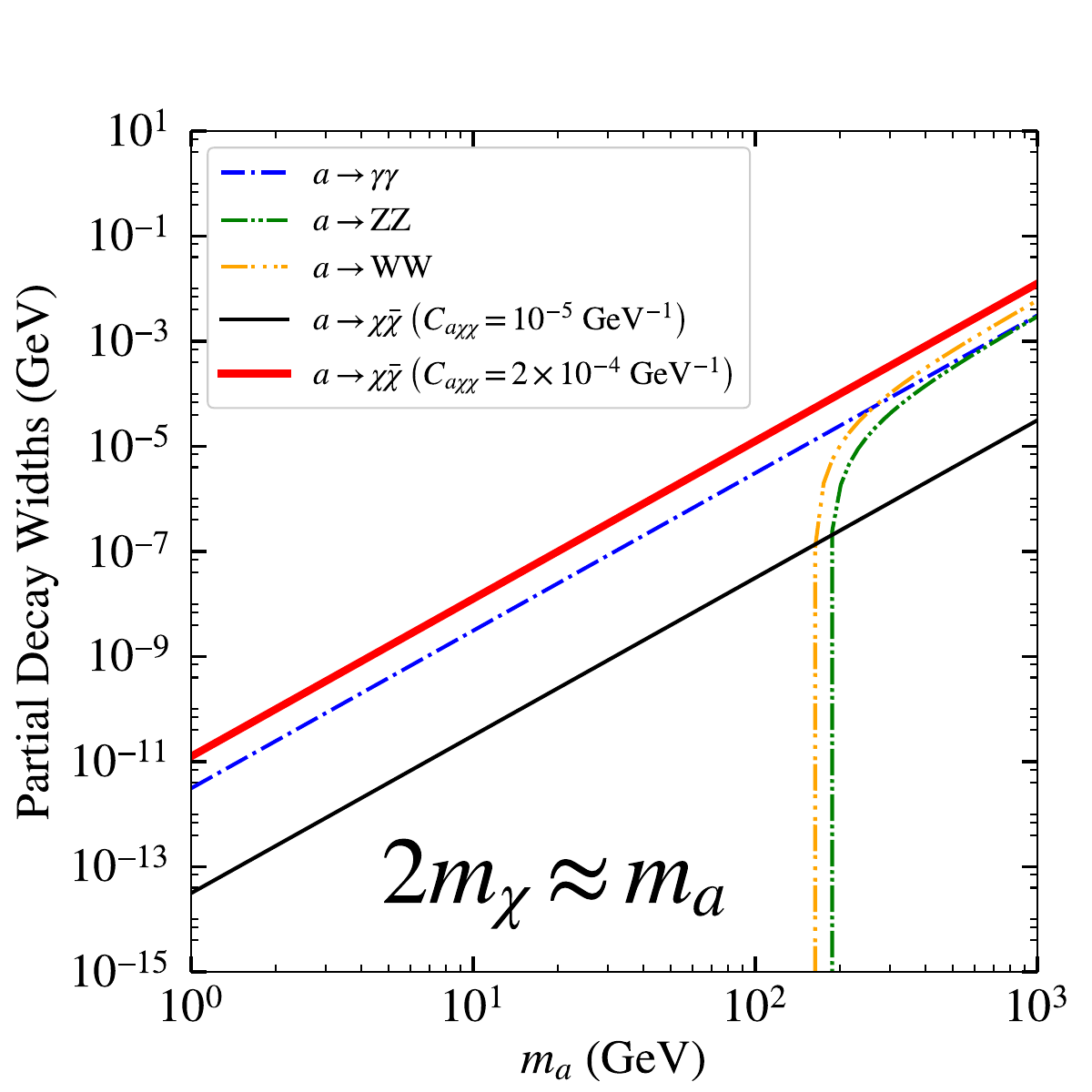}
\caption{The ALP partial decay widths of $a\to\gamma\gamma$, $a\to Z Z$, $a\to W^+ W^-$(dash lines), and $a\to \chi\overline{\chi}$ with the assumption of $C_{BB}= C_{WW}=2.5\times10^{-5}$ GeV$^{-1}$, $2m_{\chi}\approx m_a$($m_{\chi}=m_a/2.001$). The black solid line corresponds to an ALP-dark matter coupling constant $C_{a\chi\chi}= 1\times10^{-5}$ GeV$^{-1}$, while the red solid  line demonstrates the scenario for a significantly larger coupling constant $C_{a\chi\chi}= 2\times10^{-4}$ GeV$^{-1}$. Note that the $C_{a\chi\chi}$ here has been excluded, and it actually requires $10^{-12}$ GeV$^{-1}$.}
 \label{Br}
\end{figure*}

We present the ALP partial decay widths for resonance scenarios in Fig.~\ref{Br}. For the purpose of presentation, we have considered   coupling constant $C_{a\chi\chi}=2\times 10^{-4} /\gev$(red line) and $C_{a\chi\chi}=10^{-5} /\gev$(black line) with $m_a/m_{\chi}=2.001$,  $C_{BB}= C_{WW}=2.5\times10^{-5}$ GeV$^{-1}$. It is worth noting that $C_{a\chi\chi}$ here has been excluded, and it actually requires $10^{-12}$ GeV$^{-1}$. When $m_{\chi}>m_{a}$, only three decay channels $a\to\gamma\gamma$, $a\to Z Z$, and $a\to W^+ W^-$ are possible. For $m_a < 100$ GeV, ALPs mainly decay into a photon pair, making the channel of $a\to\gamma\gamma$ dominant. As $m_a$ increase, the contribution of $a\to Z Z$ and $a\to W^+ W^-$ will become dominant. 
Furthermore, in this resonance annihilation scenario, when $C_{a\chi\chi}=10^{-5} /\gev$, although the decay channel $a\to\chi\overline{\chi}$ is open, its partial decay width remains significantly smaller compared to the other three partial decay widths. However, when $C_{a\chi\chi}=2\times 10^{-4} /\gev$, the decay channel $a\to\chi\overline{\chi}$ become dominant.

\section{The slope of spike density}
\label{app:spike_density}

\begin{figure*}[ht]
\centering
\includegraphics[width=8.5cm]{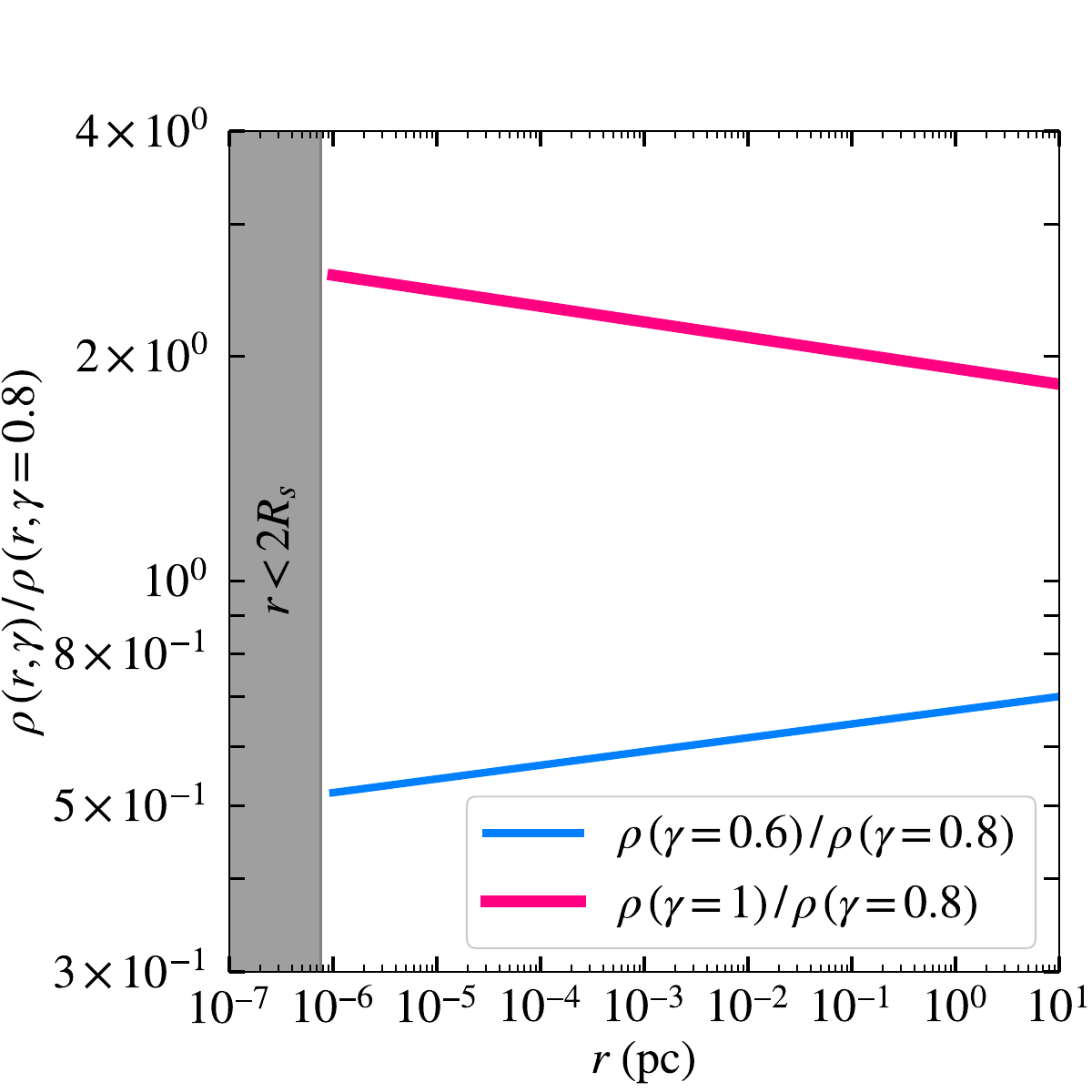}
\includegraphics[width=8.5cm]{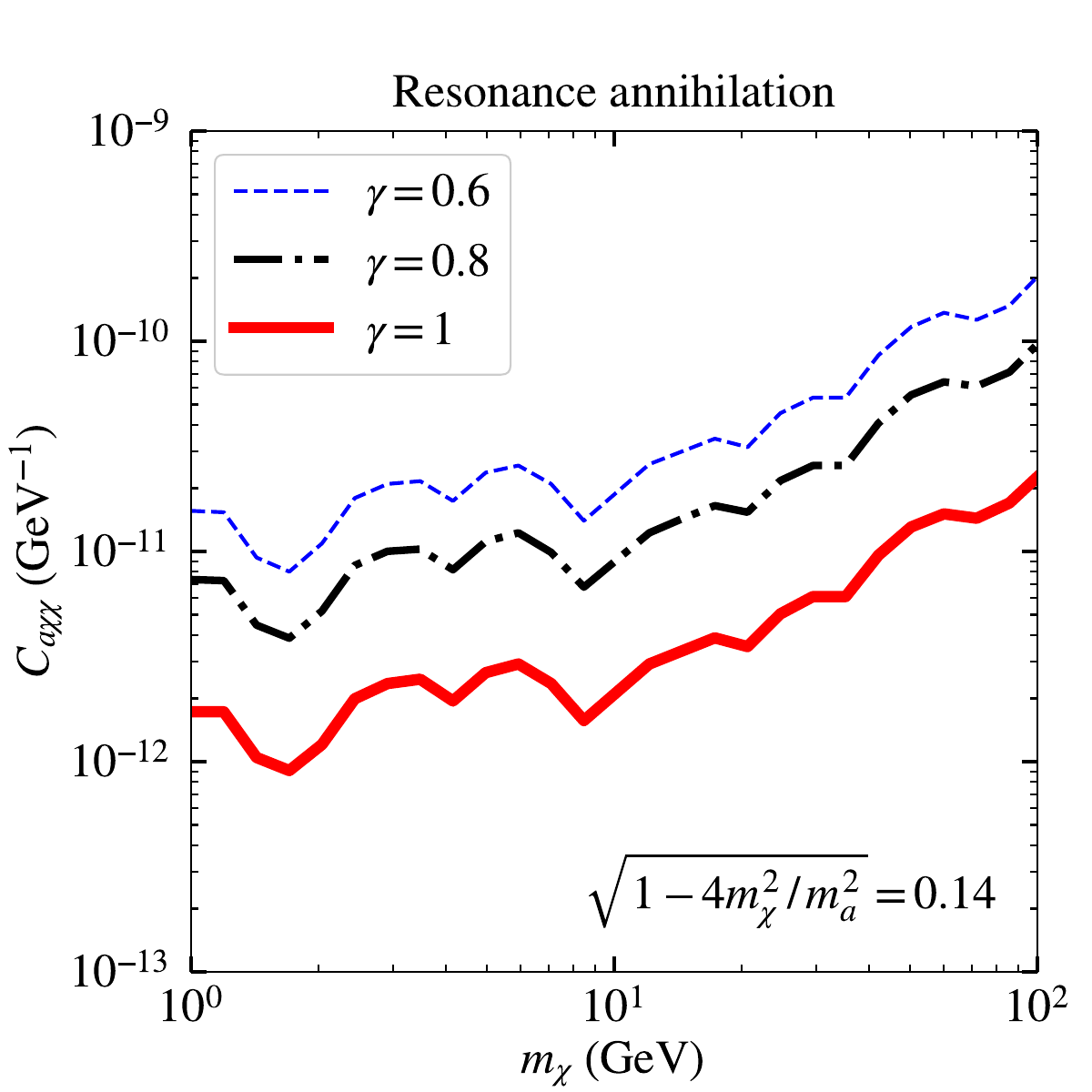}
 \caption{The left panel: the ratio of  the density profiles $\rho(r,\gamma=0.6)/\rho(r,\gamma=0.8)$ (cornflowerblue line) and $\rho(r,\gamma=1)/\rho(r,\gamma=0.8)$ (peach line) as a function of radius. The right panel: the 95\% upper limits for $C_{a\chi\chi}$ for different density profile at $\gamma=0.6$ (blue dash line), $\gamma=0.8$ (black dash-dotted line), $\gamma=1$ (red solid line).}
 \label{fig:compare}
\end{figure*}

In Fig.~\ref{fig:compare}, we demonstrate the impact of $\gamma$ on the spike density profile (left) and the upper limits of $C_{a\chi\chi}$ for the $a$-resonance scenario.   
The spike density profiles, normalized with a default slope of $\gamma=0.8$, are shown in red ($\gamma=1$) and blue ($\gamma=0.6$).
Compared to $\gamma=0.8$, the spike density is twice as large for $\gamma=1$ and smaller by a factor of 0.6 for $\gamma=0.6$.    
In the right panel of Fig.~\ref{fig:compare}, we take $a$-resonance scenario with $R_{a\chi}=0.14$ as an example. 
The slope changes from $\gamma=1$ to $0.6$, weakening the upper limits of $C_{a\chi\chi}$ by a factor of four.

\bibliography{refs}
\end{document}